\documentclass[tighten,twocolumn]{aastex631}

\newcommand{\NIJ}{Radboud University, Department of Astrophysics/IMAPP, P.O. Box 9010, 6500 GL Nijmegen, The Netherlands}

\accepted{to ApJ; January 2, 2026}
\shorttitle{X-band Galactic Center Survey}
\shortauthors{Perez et al.}
\graphicspath{{./}{figures/}}
\usepackage{glossaries}
\usepackage{comment}

\usepackage{tikz}
\usetikzlibrary{shapes.geometric, arrows}

\glsdisablehyper

\begin{document}

\title{On the Deepest Search for Galactic Center Pulsars \\ and an Examination of an Intriguing Millisecond Pulsar Candidate}

\newacronym{alfa}{ALFA}{Arecibo L-Band Feed Array}
\newacronym{dm}{DM}{dispersion measure}
\newacronym{frb}{FRB}{fast radio burst}
\newacronym{fwhm}{FWHM}{full-width at half-maximum}
\newacronym{gbt}{GBT}{Green Bank Telescope}
\newacronym{gc}{GC}{Galactic Center}
\newacronym{hpbw}{HPBW}{half-power beam width}
\newacronym{if}{IF}{intermediate frequency}
\newacronym{igm}{IGM}{intergalactic medium}
\newacronym{ism}{ISM}{interstellar medium}
\newacronym{iism}{IISM}{ionized interstellar medium}
\newacronym{lst}{LST}{local sidereal time}
\newacronym{msp}{MSP}{millisecond pulsar}
\newacronym{nip}{NIP}{non-image Processing}
\newacronym{rfi}{RFI}{radio-frequency interference}
\newacronym{rrat}{RRAT}{rotating radio transients}
\newacronym{rm}{RM}{rotation measure}
\newacronym{ska}{SKA}{Square Kilometre Array}
\newacronym{sefd}{SEFD}{system equivalent flux density}
\newacronym{seti}{SETI}{search for extraterrestrial intelligence}
\newacronym{snr}{S/N}{signal-to-noise ratio}
\newacronym{sps}{SPS}{single pulse search}
\newacronym{tab}{TAB}{tied-array beam}
\newacronym{trapum}{TRAPUM}{TRansients And PUlsars with MeerKAT}
\newacronym{uwl}{UWL}{Ultra Wideband Low}
\newacronym{vlbi}{VLBI}{very long baseline interferometry}
\newacronym{bl}{BL}{Breakthrough Listen}
\newacronym{eti}{ETI}{extra-terrestrial intelligence}

\newcommand{\BI}{Breakthrough Initiatives\xspace}
\newcommand{\BLI}{Breakthrough Listen Initiative\xspace}
\newcommand{\BL}{Breakthrough Listen\xspace}
\newcommand{\SKA}{\textit{Square Kilometre Array}\xspace}
\newcommand{\MK}{\textit{MeerKAT}\xspace}
\newcommand{\VLA}{\textit{Karl G. Jansky Very Large Array}\xspace}
\newcommand{\Parkes}{\textit{Parkes Observatory}\xspace}
\newcommand{\MWA}{\textit{Murchison Widefield Array}\xspace}
\newcommand{\LOFAR}{\textit{LOw Frequency ARray}\xspace}
\newcommand{\ATA}{\textit{Allen Telescope Array}\xspace}
\newcommand{\GBT}{\textit{Green Bank Telescope}\xspace}
\newcommand{\APF}{\textit{Automated Planet Finder}\xspace}
\newcommand{\sgra}{Sgr\,A$^{*}$}
\newcommand{\subtot}[1]{{\textcolor{Cerulean}{#1}}}
\newcommand{\subsubtot}[1]{\textit{\textcolor{OliveGreen}{#1}}}
\newcommand{\tot}[1]{{\textcolor{NavyBlue}{#1}}}
\newcommand{\ie}{i.\,e.\,}
\newcommand{\eg}{e.\,g.\,}

\newcommand{\UCB}{Breakthrough Listen,  University of California, Berkeley, 501 Campbell Hall \#3411, Berkeley, CA 94720, USA}
\newcommand{\SSL}{Space Sciences Laboratory, University of California, Berkeley, Berkeley, CA 94720, USA}
\newcommand{\SETI}{SETI Institute, 339 Bernardo Ave, Suite 200, Mountain View, CA 94043, USA}
\newcommand{\oxford}{Astrophysics, Department of Physics, University of Oxford, Denys Wilkinson Building, Keble Road, Oxford OX1 3RH, UK}

\correspondingauthor{Karen I. Perez}
\email{karen.i.perez@columbia.edu}

\author[0000-0002-6341-4548]{Karen I.~Perez}
\affiliation{Department of Astronomy, Columbia University, 550 West 120th Street, New York, NY 10027, USA}
\affiliation{\SETI}

\author[0000-0002-8604-106X]{Vishal Gajjar}
\affiliation{\SETI}
\affiliation{\UCB}

\author[0000-0002-9870-2742]{Slavko Bogdanov}
\affiliation{Columbia Astrophysics Laboratory, Columbia University, 550 West 120th Street, New York, NY 10027, USA}

\author[0000-0003-4814-2377]{Jules P.~Halpern}
\affiliation{Department of Astronomy, Columbia University, 550 West 120th Street, New York, NY 10027, USA}
\affiliation{Columbia Astrophysics Laboratory, Columbia University, 550 West 120th Street, New York, NY 10027, USA}

\author[0000-0002-6664-965X]{Paul B.~Demorest}
\affiliation{National Radio Astronomy Observatory, P.O. Box O, Socorro, NM 87801, USA}

\author[0000-0003-4823-129X]{Steve Croft}
\affiliation{\oxford}
\affiliation{\UCB}
\affiliation{\SETI}

\author[0000-0002-7042-7566]{Matt Lebofsky}
\affiliation{\UCB}

\author[0000-0001-6950-5072]{David H.\ E.\ MacMahon}
\affiliation{\UCB}

\author[0000-0003-2828-7720]{Andrew P. V. Siemion}
\affiliation{\oxford}
\affiliation{\UCB}
\affiliation{\NIJ}
\affiliation{\SETI}
\newcommand{\KZA}{University of Malta, Institute of Space Sciences and Astronomy}
\affiliation{\KZA}
 
\begin{abstract} 
We report results of one of the most sensitive pulsar surveys to date targeting the innermost region of the Galactic Center (GC) using the Robert C. Byrd Green Bank Telescope (GBT) at X-band (8--12\,GHz) using data from the Breakthrough Listen initiative. In total, we collected 9.5\,hr of data covering the wider $\sim 8 \arcmin$ diameter of the GC bulge, and 11\,hr on the inner 1\farcm4 region between 2021 May and 2023 December. We conducted a comprehensive Fourier-domain periodicity search targeting both canonical pulsars (CPs) and millisecond pulsars (MSPs), using constant and linearly changing acceleration searches to improve sensitivity to compact binaries. Assuming weak scattering, our searches reached luminosity limits of \(L_{\mathrm{\rm min}} \approx 0.14\,\mathrm{mJy\,kpc^2}\) for CPs and \(L_{\mathrm{\rm min}} \approx 0.26\,\mathrm{mJy\,kpc^2}\) for MSPs --- sensitive enough to detect the most luminous pulsars expected in the GC. Among 5,282 signal candidates, we identify an interesting 8.19\,ms MSP candidate (DM of 2775\,pc\,cm\(^{-3}\)), persistent in time and frequency across a 1-hr scan at a flux density of \(S_{\mathrm{\rm min}} \approx 0.007\,\mathrm{mJy}\). We introduce a novel randomization test for evaluating candidate significance against noise fluctuations, including signal persistence via Kolmogorov-Smirnov tests and flux-vs-DM behavior. We are unable to make a definitive claim about the candidate due to a mixed degree of confidence from these tests and, more broadly, its non-detection in subsequent observations. This deepens the ongoing missing pulsar problem in the GC, reinforcing the idea that strong scattering and/or extreme orbital dynamics may obscure pulsar signals in this region.
\end{abstract}

\keywords{Galactic center; Neutron stars; Radio pulsars; Radio transient sources; Binary pulsars; Millisecond pulsars; Magnetars; Surveys}


\section{Introduction} 
\label{sec:intro}
The Galactic Center (GC) hosts the highest number density of stars in the Galaxy (\( \sim 10^{6}\, \rm{pc^{-3}}\) within 1\,pc of Sagittarius A* (Sgr A*); \citealt{Genzel_1996, Macquart_2015, Schodel_2018}), and its line-of-sight offers the largest integrated Galactic star count of any direction in the sky. Therefore, the Breakthrough Listen (BL) initiative is undertaking one of the most extensive and sensitive searches for technosignatures and pulsars directed toward the GC \citep{Gajjar_2021}. Estimates suggest that approximately 10\% of all high-mass stars in the Galaxy reside within $\sim 200$\,pc of the GC \citep{Figer_2004, Lorimer_2004, Deneva2009}, implying a correspondingly large population of neutron stars (NSs) and black holes (BHs). However, despite the expectation of a large population of close-in NSs, no pulsars have been detected within 1\,pc of Sgr A* (an angular offset of 25\farcs2 for $R_0=8.18$\,kpc), with the exception of the GC magnetar SGR\,J1745$-$2900 (2\farcs4, or $< 0.1$\,pc away) \citep{Eatough_2013a, Eatough_2013b}. The next closest known pulsars remain 12--18\arcmin\ away ($\sim 30-40$\,pc in projection) \citep{Johnston2006, Deneva2009, Lower_2024}. Nevertheless, indirect observational evidence suggests the presence of NSs near Sgr A*, including dense clusters of young, massive stars \citep{Ghez_2005}, transient sources which might be low-mass X-ray binaries \citep{Muno_2005, Mori_2021}, a possible pulsar wind nebula \citep{Wang_2006}, radio variables \citep{Zhao_2020}, and counterparts to sources such as X-ray binaries or unassociated $\gamma$-rays \citep{FermiLAT_2017}. 

Constant star formation and recent starburst models predict $10^{2} - 10^{8}$ NSs in the GC, with $10 - 10^{5}$ likely to be active pulsars \citep{Cordes_1997, Nogueras_2019}. Considering their velocity, lifetime, beaming direction, luminosity, and observational detectability, it is estimated that $\sim 100 - 1000$ active pulsars could reside in close orbits ($P_{b} \lesssim 100$\,yr) around Sgr A*, with 1 -- 10 of the most luminous potentially detectable by current telescopes \citep{Cordes_1997,Mezger_1999,Pfahl_2004, Deneva2009, Wharton_2012}. This population could also be increased by recycled pulsars originating from binary interactions such as tidal captures and stellar collisions \citep{Samsing2017}. Observing at higher frequencies helps mitigate some observing challenges, namely scattering (a form of temporal smearing due to multi-path propagation), which scales as \(\nu^{-4}\). Assuming weak scattering, the optimal frequencies for pulsar detection toward the GC are estimated to be up to $\sim 9$\,GHz for canonical pulsars (CPs)  and $\sim 22$\,GHz for millisecond pulsars (MSPs), with an optimal GBT survey range of 8 -- 14\,GHz \citep{Cordes_1991, Rajwade_2017, Macquart_2015} --- a range within which this survey falls. 

Discovering a population of pulsars in this region would offer many benefits in probing the GC, including using their age and spin distributions to constrain the past star formation rate \citep{Lorimer_1993} and gravitational potential in the GC \citep{Cordes_1997}. Additionally, pulsars could probe the scattering region around Sgr A*, refining electron density models for the inner Galaxy \citep{Deneva2009}. Given the extreme stellar density and dynamical environment of the GC, pulsars in exotic configurations --- such as pulsar-black hole binaries --- are also expected to reside there \citep{FaucherGiguere2011,Gajjar_2021}. Perhaps more importantly, discovering and deriving a phase-connected timing solution for a pulsar in a close orbit ($P_b$ $\lesssim$ 1\,yr) around the supermassive black hole (SMBH) Sgr A* would enable 
measurements of the black hole's mass, spin, and quadrupole moment \citep{Wex_Kopeikin_1999, Liu_2012, Liu_2014}, thereby providing a precise test of the no-hair theorem in General Relativity \citep{Liu_2012, Johannsen2016, Psaltis2016} for Kerr black holes. Additionally, a pulsar in a very close orbit around Sgr A* ($P_b$ $\lesssim$ 1\,day) could experience gravitational deflection of its pulses by the SMBH, enabling a probe of the strongly curved spacetime geometry and gravitational field near Sgr A* \citep{Wang_2009a, Wang_2009b,Liu_2012,Stovall2012,DellaMonica2023}.

The lack of CP detections within the innermost 40\,pc may indicate an intrinsic deficit, potentially dominated by a magnetar population \citep{Dexter2014}. In the case of MSPs, the dearth may reflect their inherently lower luminosities compared to ordinary pulsars \citep{Burgay_2013}. Pulsars in the GC are also likely to be more challenging to detect than in typical field surveys due to the region's extreme scattering and complex orbital dynamics. Even in less extreme Galactic environments, marginal candidates have later been confirmed, demonstrating the importance of rigorous candidate vetting in pulsar surveys. 

For instance, PSR\,J2322$-$2650, a 3.5\,ms pulsar discovered in the High Time Resolution Universe (HTRU) survey with an initial ${\rm S/N}\approx12$, was confirmed only after follow-up observations with the Lovell Telescope at the Jodrell Bank Observatory, establishing its binary nature \citep{Spiewak_2018}. Similarly, the coherent re-processing of the Parkes multi-beam pulsar survey (PMPS) data confirmed the intermittent pulsar PSR J1808$-$1517 \citep{Eatough_2013}, and the Einstein@Home project uncovered 24 new pulsars missed in original PMPS analysis \citep{Knispel_2013}. Across past PMPS analyses, many high-quality candidates remain unconfirmed, some of which may be intermittent. More recently, GPU-accelerated coherent reprocessing of HTRU-South Low Latitude survey data recovered dozens of new pulsars by folding candidates to a much lower spectral S/N and exploiting the coherence of folding over the incoherent summing of Fourier components \citep{Sengar_2024}. Given the profound implications of detecting a short-period MSP orbiting Sgr A*, it is important to examine and follow up on any low significance candidates exhibiting pulsar-like properties to determine their true nature, as we have attempted to do in this work.

In this paper, we present results from one of the most sensitive pulsar surveys yet conducted toward the inner 1\farcm4 (3.33\,pc) region of the GC. We also covered the wider $\sim 8 \arcmin$ diameter of the GC bulge. In Section \ref{sec:strategy_and_observations} we detail our observing strategy, data products, and pre-processing techniques, while in Section \ref{sec:periodicity_searches}, we describe our periodicity searches employing \texttt{PRESTO} acceleration algorithms. In Section \ref{sec:results}, we present the results of these searches, including the discovery and characteristics of a particularly intriguing MSP candidate, hereafter referred to as BLPSR. Section \ref{sec:confirm_blpsr} rigorously investigates this candidate using novel statistical tests involving data randomization, quantifying the likelihood of its detection occurring by chance. Section \ref{sec:discussion} analyzes our survey sensitivity and compares it with previous pulsar searches in the GC region. We also discuss follow-up observations of BLPSR, examine potential orbital parameters under the assumption that the candidate is genuine, and explore various astrophysical scenarios that might account for its non-detection in subsequent scans. Finally, Section \ref{sec:future_work} summarizes ongoing and planned future observational efforts.
 
\section{Strategy and Observations}
\label{sec:strategy_and_observations}
\subsection{Observing Strategy}
\label{sec:observing_strategy}
The Breakthrough Listen Galactic Center (BL-GC) survey is an extensive endeavor to search for radio technosignatures, pulsars, transients, spectral lines, and masers in the GC and neighboring Galactic bulge from 0.7 -- 93\,GHz (see \citealt{Gajjar_2021} for full survey description). The 0.7 -- 4\,GHz component of the survey utilizes the Ultra-Wide Band (UWL) receiver on the Parkes Telescope, while the 4 -- 93\,GHz component uses the Green Bank Telescope (GBT). As part of this survey, \cite{Suresh2022} reported null results from ancillary pulsar searches conducted between 4 -- 8\,GHz. In this work, we extend the search to the 8 -- 12\,GHz GBT X-band. The GBT X-band receiver is a circularly polarized, single beam receiver with a $\theta_{\rm {HPBW}} = 1\farcm4$ half-power beam width (HPBW) at $\nu_{c} = 10$\,GHz. At this frequency, the GBT's aperture efficiency is 71\%, with a beam efficiency of 97\%. Unlike traditional single-dish telescopes, the unique off-axis design of the GBT significantly reduces sidelobe contamination\footnote{\url{https://www.gb.nrao.edu/scienceDocs/GBTpg.pdf}}, which is especially important in mitigating Radio Frequency Interference (RFI) and identifying real astrophysical signals (see Section \ref{sec:shortpointings_results}).  

\begin{figure}[!t]
\centering
    \includegraphics[width=0.45\textwidth,]{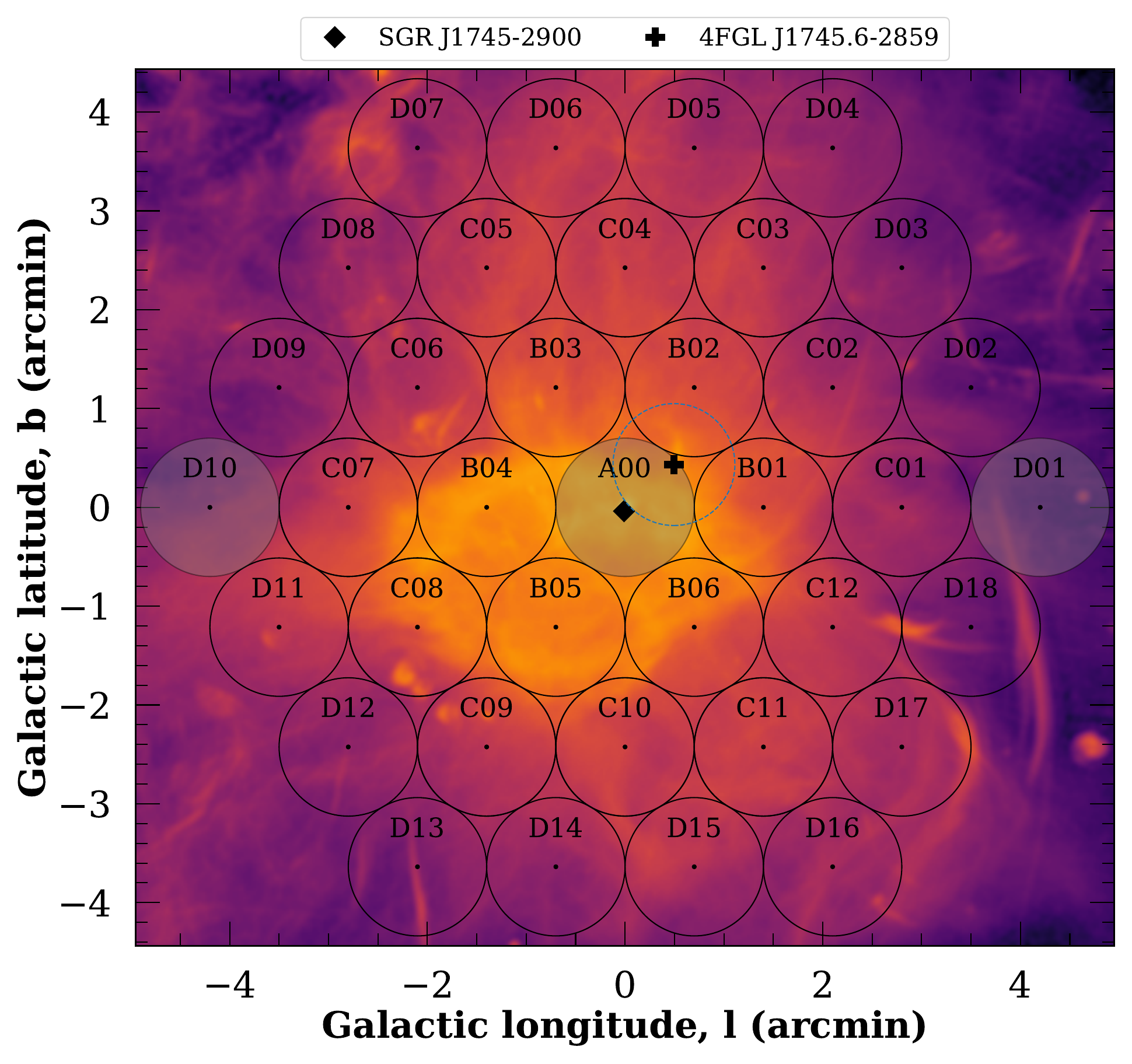}	
    \caption{Pointings at GBT X-band ($1\farcm4$ beam width) for the BL--GC survey with an overlaid 1.28\,GHz MeerKAT total intensity mosaic of the GC \citep{Heywood_2022}. The central pointing, labeled as A00, indicates the deep pointing of the GC ($l=0$\textdegree, $b=0$\textdegree). The black diamond represents the location of the known GC magnetar, J1745$-$2900, near the SMBH Sgr A*. The black cross shows the centroid of the 4FGL J1745.6$-$2859 GC source from the fourth \textit{Fermi}--LAT catalog, with the dashed blue circle showing its 95\% positional uncertainty (a radius of of 37\,\arcsec). The rest of the 36 pointings fully sample the wider $\sim 8 \arcmin$ diameter of the GC bulge region and represent our short 5-min pointings. The gray-shaded pointings show two separate pairs for our ON-OFF observations; A00--D01 and A00--D10. Due to our odd numbered pointings, A00 is the only pointing which is repeated in a pair. The black points in the center of the beams show the beam centers. 
    }
    \label{fig:Xband_gbt_pointings}
\end{figure}

We sample an $\sim 8 \arcmin$ diameter region of the GC and its surroundings with 37 distinct pointings, wherein A00 denotes the GC ($l=0$\textdegree, $b=0$\textdegree) pointing (see Figure \ref{fig:Xband_gbt_pointings}). The black diamond shows the location of the known GC magnetar, J1745$-$2900. The cross shows the 4FGL J1745.6$-$2859 GC source from the fourth \textit{Fermi} Large Area Telescope (LAT) catalog. The source is $\sim \,0\fdg01$ from Sgr A*, and believed to be its $\gamma$-ray counterpart \citep{Cafardo_2021}; it is the only \textit{Fermi}-LAT catalog source within 4\arcmin\, of the GC. The 1.28\,GHz MeerKAT total intensity mosaic is overlaid in the background to illustrate the GC region covered \citep{Heywood_2022}. 

All pointings are arranged in three concentric hexagonal rings, labeled A, B, C, and D, with 1, 6, 12, and 18 pointings per ring, respectively. These pointings were conducted with three short, ON--OFF 5-minute duration cadences per pointing, for a total of 9.5\,hr on target across all pointings. To optimize observing efficiency and reject RFI via position switching, we conducted alternating observations of pairs of pointings. Pointing pairs were chosen such that the beam centers of pairs of pointings were separated by at least $2 \theta_{\rm {HPBW}}$ on the sky (see the gray-shaded pairs A00--D01 and A00--D10 in Figure \ref{fig:Xband_gbt_pointings} for example). All other pointing pairs include: D03--D18, D02--D17, C01--D16, C02--C11, C03--C12, D05--B01, C04--B06, B02--C10, C05--B05, B03--C09, B04--D13, C08--D15, C07--D14, C06--D12, D08--D11, D06--D09, and D07--D04. Table \ref{tab:shortpointings_log} summarizes all of their coordinates.

We also conducted 11\,hr of deep 1- and 2-hr pointings of only the central most region (A00). These observations were conducted over two epochs, 1.5\,yr apart. Because interstellar scintillation causes signal intermittency at various time scales depending on the degree of scattering (\citealt{Cordes_1991, Cordes_1997}), this strategy helps us maximize our chance of detecting narrowband technosignatures \citep{Brzycki_2024_GC_SETI}  and other astrophysical sources such as pulsars that might have lower duty cycles.

\subsection{Observations}
\label{sec:observations_description}

\subsubsection{Short Pointings}
\label{sec:shortpointings_obs}
We observed the 37 short pointings (see Table \ref{tab:shortpointings_log}) from 2021 May 14 to 2022 June 6 with the GBT using the Breakthrough Listen Digital Backend (BLDB; \citealt{MacMahon_BLDR}) at X-band (7.50--11.25\,GHz). Each pointing consisted of three 5-min scans. Baseband voltages were recorded and converted into \textit{mid-time} resolution total intensity \texttt{filterbank} products (8.2\,GB/file) to search for canonical pulsars. For these mid-time resolution data products, the number of observing channels was 10,240, with a channel bandwidth of $366.21$\,kHz, and a sampling time of $349.53\,\mu$s.

\subsubsection{Long Pointings}
\label{sec:longpointings_obs}
The 11\,hr of deep observations on the A00 pointing were conducted on 2022 June 10 (Epoch 1) and 2023 December 9 (Epoch 2). Table \ref{tab:A00_obs_log} summarizes our long A00 pointing observations, which were conducted over two epochs. The table lists the starting Modified Julian Date (MJD) for each scan. Epoch 1 consists of five 1-hr scans (totaling 5\,hr, 3.1\,TB/scan), and Epoch 2 comprises three 2-hr scans (totaling 6\,hr, 8\,TB/scan). The increased integration time later adapted during epoch 2 was to improve our sensitivity. Our observations were constrained to $\sim 6$\,hr per day --- the maximum duration the GC remains above the horizon. To verify our data, we recorded a 5-minute scan on test pulsar J1744$-$1134 during Epoch 1 and on B1929+10 during Epoch 2. 

During Fall 2023 the GBT X-band receiver was replaced with an upgraded instrument, which added an extra 1\,GHz of bandwidth, thus covering 7.50--12.38\,GHz. Epoch 2 used the upgraded receiver, greatly improving our sensitivity for detecting MSPs in particular. Baseband voltages were recorded and converted into high-time resolution total intensity \texttt{filterbank} products. These 11\,hr of observations were used to conduct a deep search for MSPs --- an ancillary science goal of the BL--GC survey, which was primarily designed as a technosignature search \citep{Gajjar_2021}. Hence, for both epochs, we also produced \textit{high-time} resolution data products with time and frequency resolutions of $43.69\,\mu$s and $91.55$\,kHz, respectively, resulting in a total of 40,960 frequency channels during Epoch 1 and 53,248 channels during Epoch 2. A detailed summary of the BL reduction pipeline and discussion of various data products is provided by \cite{Lebofsky_2019}.

\begin{deluxetable*}{cCccCCCC}[t]
\tablecolumns{8}
\tablewidth{0pt}
\tablecaption{Summary of GBT X-band Short Observations.
\label{tab:shortpointings_log}}
\tablehead{
\colhead{Pointing} & \colhead{Date} & \colhead{R.A. (J2000)} & \colhead{Decl. (J2000)}  & \colhead{$l$} & \colhead{$b$} & \colhead{Total exp\tablenotemark{a}} & \colhead{\# of candidates\tablenotemark{b}} \\
\colhead{} & \colhead{(MJD)} & \colhead{(h:m:s)} & \colhead{(d:m:s)} & \colhead{(deg)} & \colhead{(deg)} & \colhead{(hr)}
      }
    \startdata
    \hline
    \multicolumn{8}{c}{Short Pointings} \\
    \hline
    A00\tablenotemark{c} & 59348.258 & 17:45:40.04 & $-$29:00:28.10 & 359.944 & -0.046 & 0.5 & 26  \\
    B01 & 59348.412 & 17:45:43.38 & $-$28:59:16.40 & 359.968 & -0.046 & 0.25 & 5 \\
    B02 & 59736.187 & 17:45:36.98 & $-$28:59:14.35 & 359.956 & -0.026 & 0.25 & 7 \\
    B03 & 59736.231 & 17:45:33.64 & $-$29:00:26.04 & 359.933 & -0.026 & 0.25 & 20 \\
    B04 & 59736.252 & 17:45:36.70 & $-$29:01:39.80 & 359.921 & -0.046 & 0.25 & 20 \\
    B05 & 59736.212 & 17:45:43.11 & $-$29:01:41.85 & 359.933 & -0.066 & 0.25 & 11 \\
    B06 & 59348.412 & 17:45:46.44 & $-$29:00:30.14 & 359.956 & -0.066 & 0.25 & 9 \\
    C01 & 59348.344 & 17:45:46.71 & $-$28:58:04.69 & 359.991 & -0.046 & 0.25 & 5 \\
    C02 & 59348.366 & 17:45:40.31 & $-$28:58:02.65 & 359.979 & -0.026 & 0.25 & 8 \\
    C03 & 59348.387 & 17:45:33.91 & $-$28:58:00.60 & 359.968 & -0.006 & 0.25 & 8 \\
    C04 & 59348.430 & 17:45:30.57 & $-$28:59:12.29 & 359.944 & -0.006 & 0.25 & 6 \\
    C05 & 59736.209 & 17:45:27.24 & $-$29:00:23.97 & 359.921 & -0.006 & 0.25 & 10 \\
    C06 & 59736.317 & 17:45:30.30 & $-$29:01:37.73 & 359.909 & -0.026 & 0.25 & 16 \\
    C07 & 59736.295 & 17:45:33.37 & $-$29:02:51.49 & 359.898 & -0.046 & 0.25 & 16 \\
    C08 & 59736.274 & 17:45:39.77 & $-$29:02:53.55 & 359.9909 & -0.066 & 0.25 & 5 \\
    C09 & 59736.234 & 17:45:46.17 & $-$29:02:55.59 & 359.921 & -0.087 & 0.25 & 6 \\
    C10 & 59736.191 & 17:45:49.51 & $-$29:01:43.87 & 359.944 & -0.087 & 0.25 & 8 \\
    C11 & 59348.369 & 17:45:52.84 & $-$29:00:32.15 & 359.968 & -0.087 & 0.25 & 7 \\
    C12 & 59348.391 & 17:45:49.78 & $-$28:59:18.42 & 359.979 & -0.066 & 0.25 & 8 \\
    D01 & 59348.283 & 17:45:50.04 & $-$28:56:52.97 & 0.014 & -0.046 & 0.25 & 7 \\
    D02 & 59348.323 & 17:45:43.64 & $-$28:56:50.95 & 0.003 & -0.026 & 0.25 & 4 \\
    D03 & 59348.301 & 17:45:37.25 & $-$28:56:48.90 & 359.991 & -0.006 & 0.25 & 10 \\
    D04 & 59736.385 & 17:45:30.85 & $-$28:56:46.84 & 359.979 & 0.014 & 0.25 & 28 \\
    D05 & 59348.409 & 17:45:27.51 & $-$28:57:58.52 & 359.956 & 0.014 & 0.25 & 9 \\
    D06 & 59736.360 & 17:45:24.18 & $-$28:59:10.20 & 359.933 & 0.014 & 0.25 & 26 \\
    D07 & 59736.381 & 17:45:20.84 & $-$29:00:21.88 & 359.909 & 0.014 & 0.25 & 17 \\
    D08 & 59736.338 & 17:45:23.90 & $-$29:01:35.65 & 359.898 & -0.006 & 0.25 & 24 \\
    D09 & 59736.363 & 17:45:26.96 & $-$29:02:49.42 & 359.886 & -0.026 & 0.25 & 20 \\ 
    D10 & 59348.262 & 17:45:30.03 & $-$29:04:03.18 & 359.874 & -0.046 & 0.25 & 16 \\
    D11 & 59736.342 & 17:45:36.43 & $-$29:04:05.25 & 359.886 & -0.066 & 0.25 & 9 \\
    D12 & 59736.320 & 17:45:42.84 & $-$29:04:07.29 & 359.898 & -0.087 & 0.25 & 11 \\
    D13 & 59736.256 & 17:45:49.24 & $-$29:04:09.32 & 359.909 & -0.107 & 0.25 & 17 \\
    D14 & 59736.299 & 17:45:52.58 & $-$29:02:57.60 & 359.933 & -0.107 & 0.25 & 13 \\
    D15 & 59736.277 &17:45:55.91 & $-$29:01:45.88 & 359.956 & -0.107 & 0.25 & 18 \\
    D16 & 59348.348 & 17:45:59.24 & $-$29:00:34.15 & 359.979 & -0.107 & 0.25 & 6 \\
    D17 & 59348.326 & 17:45:56.18 & $-$28:59:20.43 & 359.991 & -0.087 & 0.25 & 12 \\
    D18 & 59348.305 & 17:45:53.11 & $-$28:58:06.70 & 0.003 & -0.066 & 0.25 & 7
    \enddata
    \tablenotetext{a}{Each pointing was observed for 15 minutes, split into three 5-minute scans, alternating between pairs of pointings as part of the technosignature survey strategy (see Section \ref{sec:observing_strategy}).}
    \tablenotetext{b}{Number of candidate signals detected per pointing for $z_{\rm max}=0$.} 
    \tablenotetext{c}{The A00 short pointing was observed for 30 minutes due to it being part of two pairs of pointings.}
\end{deluxetable*}

\subsection{Data Pre-processing}
\label{sect:preprocessing}
\begin{figure*}[!t]
\centering
    \includegraphics[width=1.0\linewidth,]{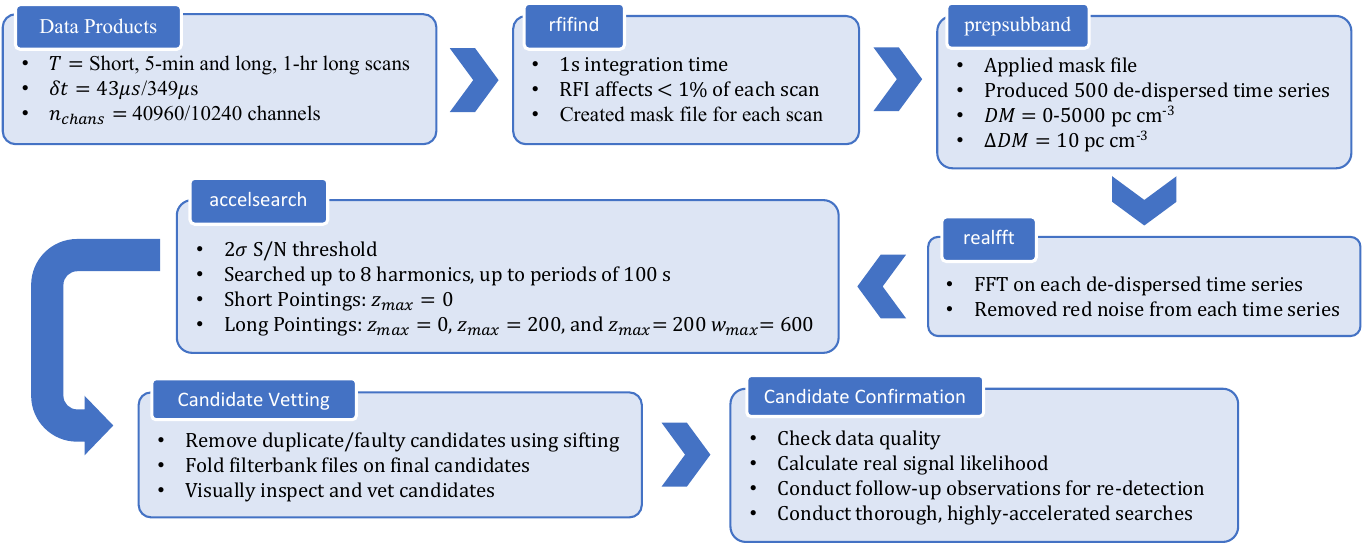}	
   \caption{Flowchart of PRESTO commands run for our survey search.}
    \label{fig:presto_flowchart}
\end{figure*}

All scans were saved and processed on our local workstation (processor model AMD Ryzen Threadripper Pro 5995WX with 64 cores and 128 threads) and standard single-process \texttt{PRESTO}\footnote{Available for download from \url{https://github.com/scottransom/presto}} \citep{2001PhDT.......123R} routines were run as outlined below (see Figure~\ref{fig:presto_flowchart}). We examined our data for narrow-band RFI signals using the \texttt{rfifind} package with an integration time of 1\,s and created an RFI mask for each scan. We find RFI affects $\sim 1.5$\% of each short scan and $\sim  0.03 - 0.1$\% of each long scan, which can be attributed to the low-RFI environment at such high frequencies. This resulted in an effective bandwidth of 3.694\,GHz for the short scans, and 3.749\,GHz and 4.875\,GHz for the long scans from Epochs 1 and 2, respectively. We used the RFI masks with the \texttt{prepsubband} package to generate 500 de-dispersed time series across 0--5000\,pc\,cm$^{-3}$, with a dispersion measure (DM) step of 10\,pc\,cm$^{-3}$ using 256 and 4096 sub-bands for the short and long pointings, respectively. This DM step size was chosen using the \texttt{DDplan.py} code to optimize the sensitivity of pulsar searching. We opted for 4096 sub-bands for the long pointings to minimize the time delay smearing caused by the frequency dependence of radio waves in the interstellar plasma and thereby ensure a thorough search for candidates. Using sub-band widths of 0.92\,MHz for Epoch 1 and 1.19\,MHz for Epoch 2, we calculated negligible dispersive delays of 16\,$\mu$s and 18\,$\mu$s per sub-band, respectively. To reduce computational cost while retaining sensitivity, we also downsampled by a factor of 4.

We then applied a fast Fourier transform to each de-dispersed time series with the \texttt{realfft} package. To reduce the effects of red noise and help flatten the power spectrum, we applied the \texttt{rednoise} routine to each spectra by subtracting the low-frequency noise baseline. This involves measuring the median power in blocks of Fourier frequency bins and scaling it by log2, which converts it to an equivalent mean power, assuming exponential noise distribution. The block size increases with frequency using adaptive windowing, starting with 6 bins at low frequencies (\texttt{-startwidth 6}) and growing to 100 bins (\texttt{-endwidth 100}) by 6\,Hz (\texttt{-endfreq 6}). For frequencies above 6\,Hz where the spectrum flattens, the block size remains fixed at 100 bins \citep{Presto_Ransom_2011, Lazarus_2015, Torne2021, Suresh2022}. After de-dispersing, the data was barycentered to remove the effects from the rotation of the Earth and its motion around the Sun.  

\section{Periodicity Searches}
\label{sec:periodicity_searches}
Detecting pulsars in binary systems, especially those orbiting compact, massive objects such as Sgr A*, is challenging due to the binary orbital motion and acceleration of the pulsar. This introduces a time-dependent Doppler drift which causes the apparent pulsar spin frequency to change with time, causing the signal harmonics to smear across neighboring frequency bins in the power spectrum. Fourier-domain acceleration searches account for this smearing by assuming a constant (\textit{z} for $T \lesssim P_{b}/10$) or linearly evolving ``jerk'' (\textit{w} for $T \lesssim P_{b}/15$) line-of-sight pulsar acceleration, where $z$ and $w$ correspond to the number of Fourier frequency (and frequency derivative, respectively) bins that the signal drifts through in an observation \citep{2001PhDT.......123R, Andersen_2018}. The parameters ${z}_{\rm max}$ and ${w}_{\rm max}$ define the maximum number of bins over which a signal can still be recovered in the search. This makes our periodicity search particularly sensitive to highly accelerated pulsars that may be directly orbiting Sgr A* and/or stellar-mass BHs, as well as other standard known compact binary configurations. 

\subsection{Acceleration Searches}
\begin{figure}
\centering
    \includegraphics[width=\linewidth]{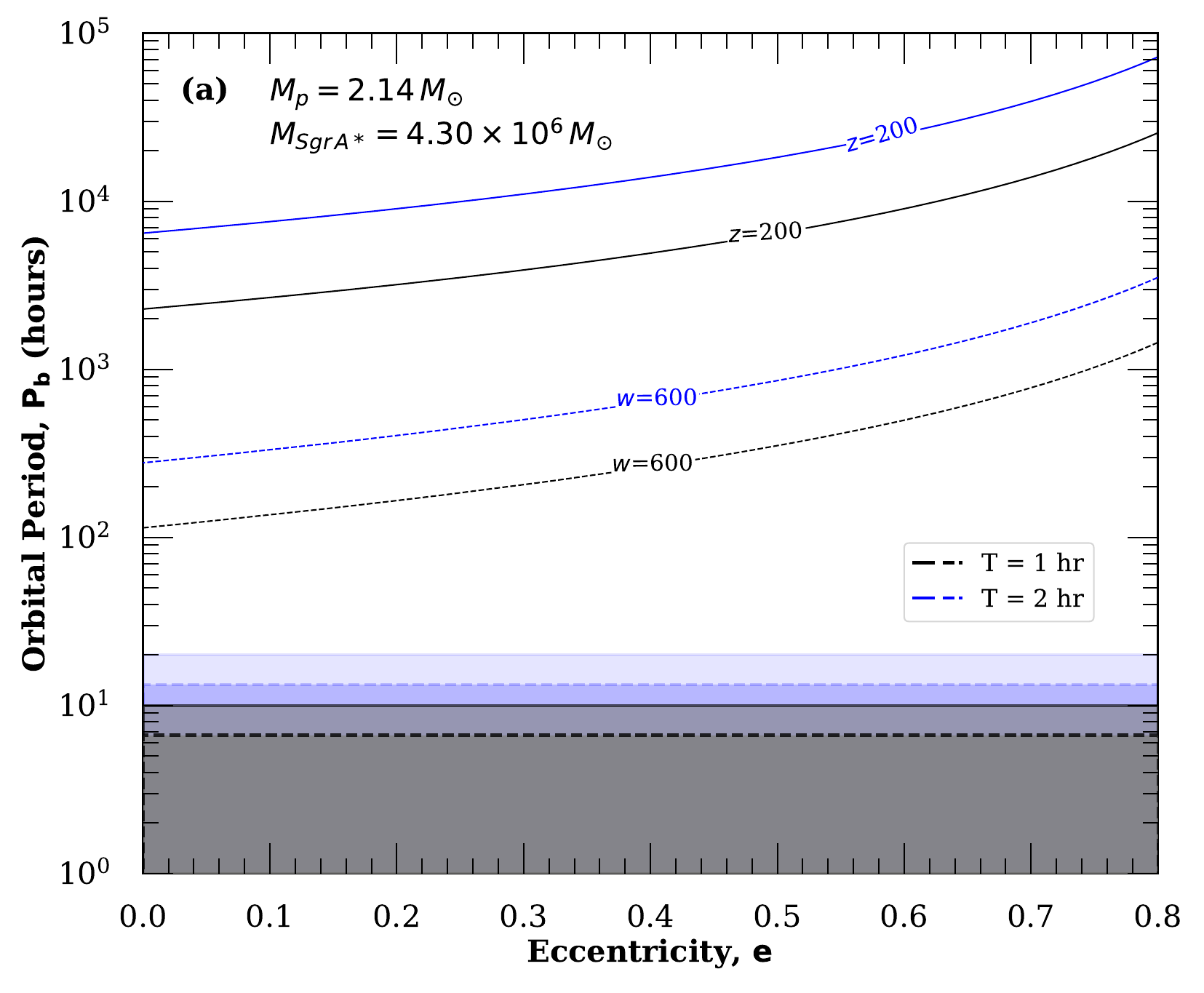}\\[-0.6em]
    \includegraphics[width=\linewidth]{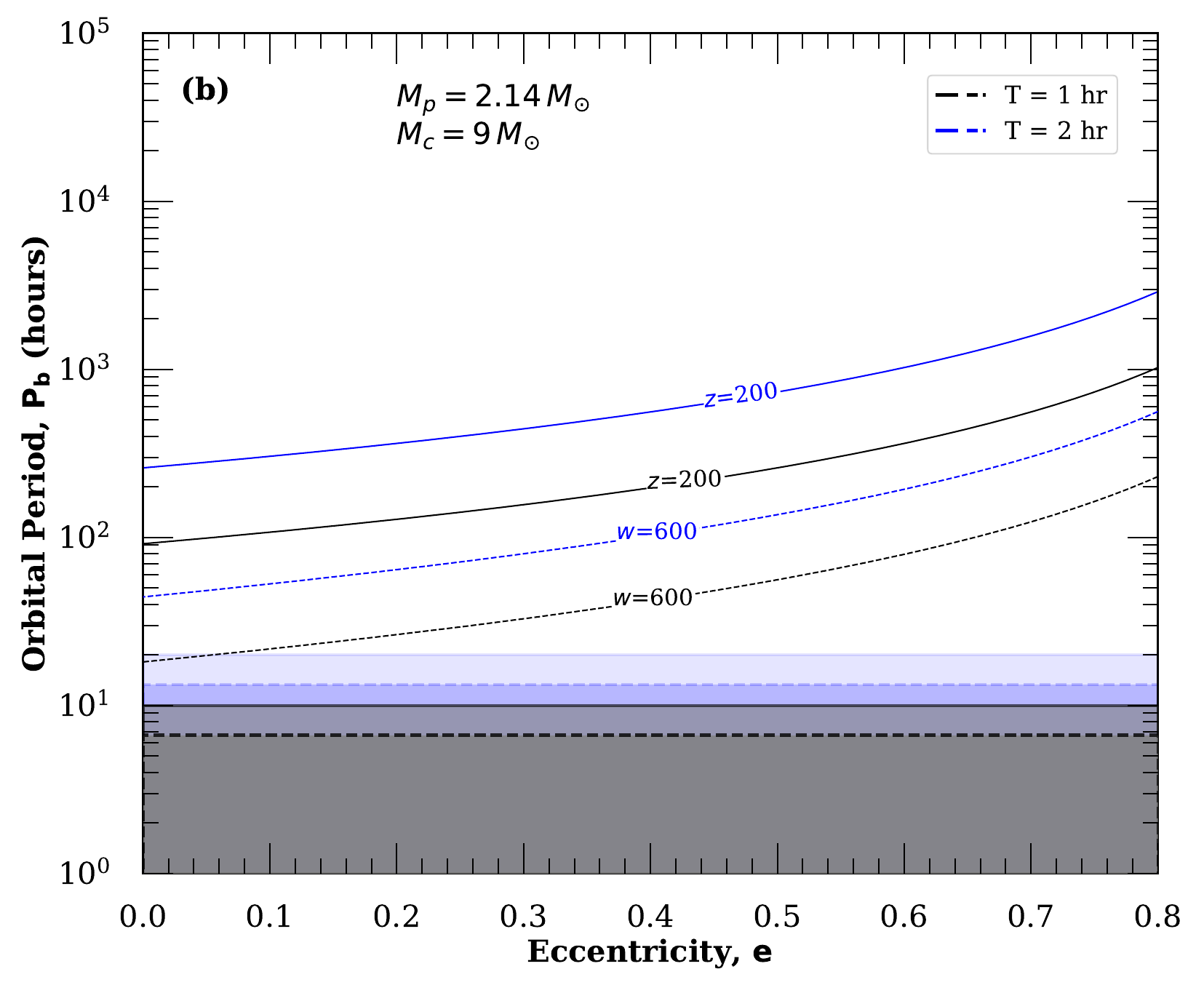}\\[-0.6em]
    \includegraphics[width=\linewidth]{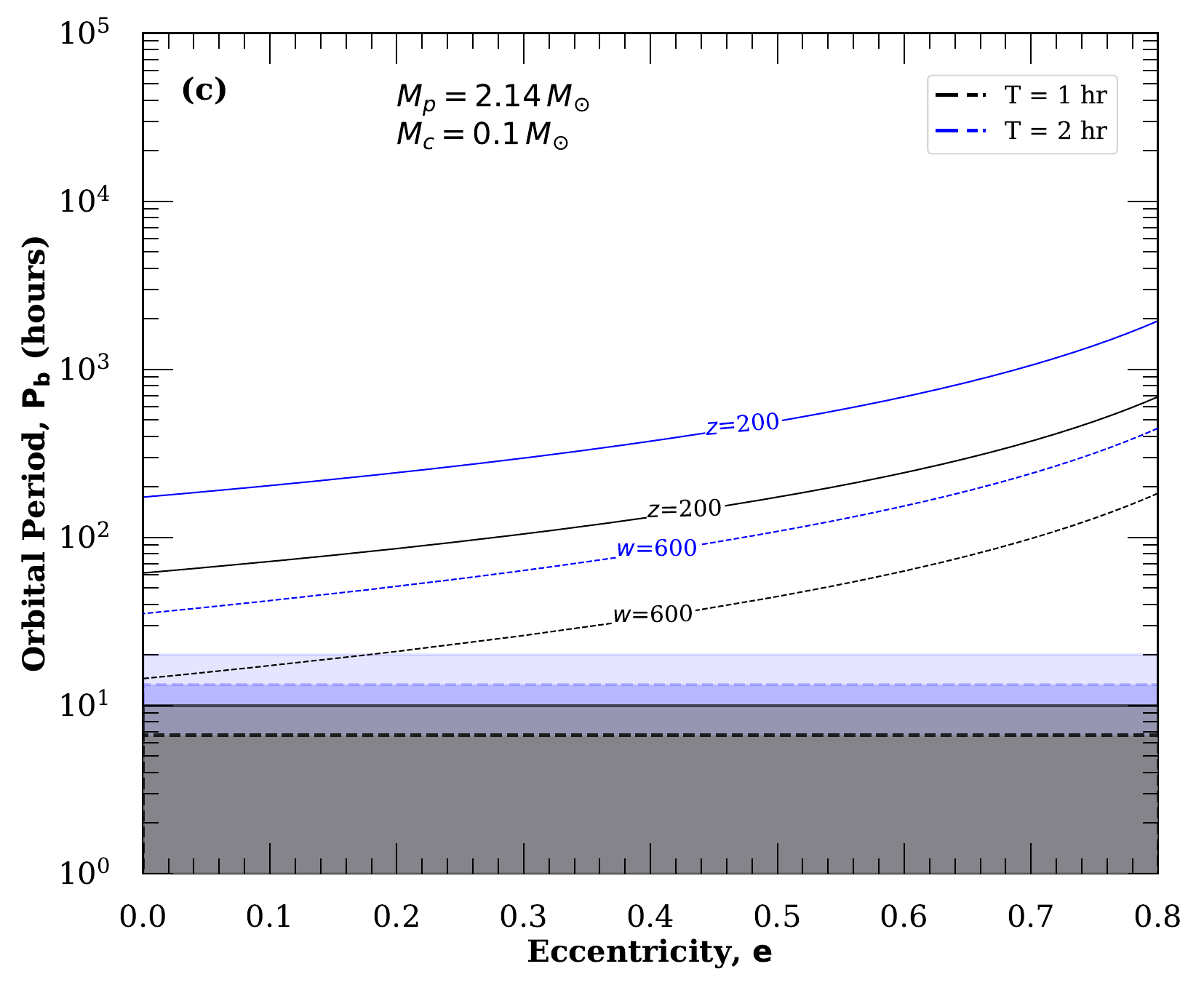}\\[-1em]
    \caption{Contours of maximum Fourier-domain drift for a 10\,ms pulsar with $h$=8 harmonics, sampled over orbital period and eccentricity. Panel shows a pulsar orbiting: \textbf{a)} the GC SMBH; \textbf{b)} a $9 M_{\odot}$ black hole; and \textbf{c)} a $0.1 M_{\odot}$ companion. Solid and dashed lines correspond to constant ($z$) and linearly-changing ($w$) acceleration thresholds, respectively. Black and blue lines show results for 1- and 2-hr integrations, respectively. Shaded regions indicate the absolute minimum $P_b$ detectable given our integration times.}
\label{fig:contour_lines}
\end{figure}

We conducted blind searches for isolated and binary pulsars in the frequency range 0.01--10,000 Hz, sampling a five-dimensional parameter space DM--$f$\,(spin frequency)--${{N}_{h}}$\,(number of harmonics)--${z}_{\rm max}$--${w}_{\rm max}$. Our search encompassed both CPs and MSPs, as our detection threshold is sensitive to a wide range of spin periods, including hundreds of slow pulsars and at least a few MSPs based on our sensitivity calculations for the known pulsar population (see Section \ref{sec:survey_sensitivity} for details) We adopt the recommended values of $z_{max}$=200 and $w_{max}$ = 600 from \citet{Andersen_2018}.

To determine the detectability of pulsars in compact orbits in our survey, we used the line-of-sight pulsar acceleration ($a_{l}$) and jerk ($j_{l}$) equations \citep{Bagchi_2013, Liu_2021}
\begin{equation}
    a_l= -\,\left( \frac{2\,\pi}{P_b} \right)^{2} \, \frac{a_p\,{\rm sin}\,i}{(1-e^{2})^{2}} \, (1+e\,\text{cos}\,A_T)^{2}\,\text{sin}\,(A_{T} + \omega)
    \label{eqn:bacghi_al},
\end{equation}
and 
\begin{align}
    j_l = & -\left( \frac{2\pi}{P_b} \right)^3 \frac{a_p\,{\rm sin}\,i}{(1 - e^2)^{7/2}} (1 + e \cos A_T)^3 \nonumber \\
    & \times \left[ e \cos \omega + \cos(A_T + \omega) - 3e \sin(A_T + \omega) \sin A_T \right]
    \label{eqn:bagchi_jl},
\end{align}
where $e$ is the orbital eccentricity, $P_b$ is the orbital period, $a_p$ is the semi-major axis, $i$ is the orbital inclination, $\omega$ is the argument of periapsis, and $A_T$ is the true anomaly. We also use the corresponding Fourier-domain drift equations for acceleration and jerk \citep{2001PhDT.......123R, Andersen_2018}:

\begin{equation}
    z= h\,\left( \frac{a_{l}\,f\,T^{2}}{c} \right)
    \label{eqn:zcalc}
\end{equation}
and 
\begin{equation}
    w= h\,\left( \frac{j_{l}\,f\,T^{3}}{c} \right),
    \label{eqn:wcalc}
\end{equation}
where $h$ is the harmonic number (with $h = 1$ being the fundamental frequency), $f$ is the pulsar spin frequency, $c$ is the speed of light in vacuum, and $T$ the length of the observation. 

We considered a $M_{p} = 2.14\, M_{{\odot}}$ pulsar --- the highest measured neutron star mass to date \citep{Cromartie_2020} --- in an edge-on orbit (inclination $i=90$\textdegree). To estimate the maximum Fourier-domain drift, we evaluated $a_{l}$ at  $A_{T}=0$\textdegree\ and $\omega=\pi/2$, which together maximize the projected acceleration. For jerk, we kept $A_{T}=0$\textdegree\ and instead set $\omega=0$\textdegree\, which maximizes the projected jerk. This 90\textdegree\ phase shift arises from the analytic dependence on the \text{sin}\,($A_{T}$ + $\omega$) and \text{cos}\,($A_{T}$ + $\omega$) terms in Equations \ref{eqn:bacghi_al} and \ref{eqn:bagchi_jl}, and corresponds to a shift of 90\textdegree\ in the projected orbital phase. 
We used our 1- and 2--hr integration times and applied Kepler's third law to solve for $a_p$, then substituted the resulting expressions into Equations~\ref{eqn:zcalc} and \ref{eqn:wcalc}, yielding $z$($A_{T}$, $\omega$) and $w$($A_{T}$, $\omega$). Applying $A_{T}=0$\textdegree\ and $\omega=\pi/2$ for $z$ and $\omega$=0\textdegree\ for $w$ yields:

\begin{equation}
\begin{aligned}
z(0^{\circ},\pi/2) = h\frac{G^{1/3}}{c\,(4\pi^2)^{2/3}} 
\frac{(M_{\rm c} + M_{\rm p})^{1/3} (1 + e)^2\, T^2}
{P_0\, P_b^{4/3} (1 - e^2)^2}
\end{aligned}
\label{eqn:z_derived}
\end{equation}
and 
\begin{equation}
w(0^{\circ},0^{\circ}) = h\frac{8\,\pi^3\, G^{1/3}}{c\, (4\pi^2)^{1/3}}
\frac{(M_{\rm c} + M_{\rm p})^{1/3} (1 + e)^4\, T^3}
{P_0\, P_b^{7/3} (1 - e^2)^{7/2}}
\label{eqn:w_derived},
\end{equation}
where $M_{\rm c}$ is the mass of the binary companion and $P_0$ is the intrinsic pulsar spin period.  
We sampled parameter grids of $P_{b}$ $\in$ [1, 1$\times$$10^{5}$]\,hr and $e$ $\in$ [0.0, 0.8], assuming a $P_{0}$\,=\,10\,ms pulsar in orbit around either the SMBH Sgr A* ($M_{\rm Sgr\,A*} = 4.30 \times 10^{6}\, M_{{\odot}}$), a stellar-mass BH ($M_{\rm BH} = 9\, M_{{\odot}}$), or a $0.1 M_{\odot}$ companion---the latter being most representative of typical MSP companions. To reflect the conditions of our search pipeline, we used $h = 8$, consistent with the harmonic summing used in our Fourier-domain searches. Higher harmonic summing improves sensitivity to pulsars with narrower duty cycles \citep{Ransom_2002} and those in larger orbits with small accelerations. For each ($e$, $M_c$) combination, we determined the shortest $P_{b}$ such that the projected acceleration and jerk remain at or below our thresholds, satisfying $z(0^{\circ},\pi/2)$=200 and $w(0^{\circ},0^{\circ})$=600. This $P_{b, \rm{min}}$ defines the shortest orbital period for which a system would remain detectable at all orbital phases within our search limits. More compact binaries (with shorter $P_b$) can exceed $z$=200 and/or $w$=600 at other orbital phases, but may still be detectable by our search if observed near favorable phases. However, because our search aims to be phase-independent, we define our recoverability conservatively, requiring detectability at all orbital phases, and adopt the maximum projected acceleration and jerk values in Equations \ref{eqn:z_derived} and \ref{eqn:w_derived} to set $P_{b, \rm{min}}$ limits.

Figure \ref{fig:contour_lines} shows the resulting contours of constant absolute $z_{\rm max}$ (solid lines) and $w_{\rm max}$ (dashed lines) for both integration times (1- (black) and 2-hr (blue)) under the three binary scenarios: a pulsar orbiting  Sgr A* (Figure \ref{fig:contour_lines}a), a stellar-mass BH (Figure \ref{fig:contour_lines}b), and a $0.1 M_{\odot}$ companion (Figure \ref{fig:contour_lines}c). These contours, plotted in the ($e$, $P_b$) plane, represent the $P_{b, \rm{min}}$ accessible to our search at each $e$, assuming full orbital phase coverage, and search limits of $z_{max}=200$ and $w_{max}=600$ (see Section \ref{sec:accel_long_pointings}, where $w_{max}=600$ includes $z_{max}=200$). Systems lying above the contours (i.e., wider orbits at the same $e$) produce lower maximum projected acceleration and jerk ($z$ $<$ 200; $w$ $<$ 600) and are therefore fully detectable by our search at all orbital phases. In contrast, systems below the contours (i.e., more compact orbits at the same $e$), would either (i) require a more sensitive search with higher acceleration and/or jerk limits, or (ii) be detectable only when observed near favorable orbital phases. This behavior holds for all eccentricities. As $e$ increases, $z$ and $w$ become more extreme, shifting the $P_{b, \rm{min}}$ contours upward. That is, more eccentric systems require longer orbital periods to remain detectable within our limits.

In the Sgr A* case, our search of $z_{max}=200$ allows detection of orbital periods down to $P_{b, \rm{min}}$ $\sim$0.7\,yr for $e\sim0.5$, and $\sim$0.3\,yr for $e\sim0.1$, assuming detectability at all orbital phases. Jerk searches with $w_{max}=600$ extend this coverage down to $\sim$ 15\,days and $\sim6$\,days, respectively. For a 9\,$M_{{\odot}}$ BH companion, the $z_{max}=200$ reaches down to $P_{b, \rm{min}}$ $\sim 11$\,days for $e\sim0.5$, and $\sim$4\,days for $e\sim0.1$. Jerk searches extend this to $P_{b, \rm{min}}$ $\sim$ 2\,days and $\sim$22\,hrs, respectively. Finally, for the $0.1 M_{\odot}$ companion case, $z_{max}=200$ allows detection down to $P_{b, \rm{min}}$ $\sim 7$\,days for $e\sim0.5$, and $\sim$3\,days for $e\sim0.1$. Jerk $w_{max}=600$ searches extend these limits to $P_{b, \rm{min}}$ $\sim$ 1.9\,days and $\sim$17\,hrs, respectively. These estimates are consistent with those reported in \cite{Eatough2021} and \cite{Liu_2021}. Although more computationally expensive, jerk searches significantly extend orbital coverage to pulsars in highly compact orbits, such as those orbiting Sgr A*, relativistic double neutron star systems, or systems with high eccentricities, that would otherwise require values much larger than $z_{max}=200$ to detect. This is evident from the 1--2 order of magnitude improvement in the $P_{b, \rm{min}}$ shown in Figure \ref{fig:contour_lines}. 

It is also evident from Figure \ref{fig:contour_lines} that, as the assumed companion mass decreases, the search becomes limited by the absolute minimum $P_b$ detectable given our integration times. This is indicated by the onset of the shaded black and blue exclusion regions. These detection constraints arise from the condition of $T$ $\lesssim$ 0.10\,$P_{b}$ for constant acceleration searches, which limits our parameter range to orbits with $P_{b}$ $\gtrsim$ 10 hours for the 1-hr integrations and $P_{b}$ $\gtrsim$ 20 hours for the 2-hr integrations \citep{Andersen_2018}. Likewise, for jerk searches, the condition $T$ $\lesssim$ 0.15\,$P_{b}$ restricts detections to $P_{b}$ $\gtrsim$ 6.67\,hr and $P_{b}$ $\gtrsim$ 13.33\,hr for each integration time, respectively \citep{Andersen_2018}. We note that these threshold rules are based on the assumption of a solar-mass companion. In systems with more massive companions---such as a stellar-mass or supermassive black hole---the orbital phases over which a signal is detectable is reduced due to the rapid variations in line-of-sight accelerations and its jerk. This effect has been demonstrated in \cite{Liu_2014}. While shorter integrations can mitigate this partially, high eccentricities may still prevent detection, particularly near periastron where acceleration changes the most rapidly.

\subsubsection{Short Pointings}
\label{sec:short_pointings}
Given the short 5-minute duration of these pointings, we performed a blind, zero-acceleration search ($z_{\rm max}=0$) using \texttt{accelsearch} on all 37 short pointings to target isolated NSs. Given the short integrations and zero acceleration, we conducted the searches using the local workstation, with each scan taking $\sim$1 minute. With a cadence of six per pair, we analyzed a total of 114 mid-time resolution \texttt{filterbank} files, applying a 2$\sigma$ detection threshold on the harmonic-summed power spectra. Note that $\sigma$ here includes a correction for the number of independent frequencies and derivatives searched in each run \citep{2001PhDT.......123R}. Periods were searched up to eight harmonics, after which the \texttt{sifting} algorithm was used to group hits in adjacent trial DMs and frequencies $f$ into distinct candidates, while removing any duplicate candidates across DM, P, \(\dot{\rm P}\), and harmonics. \texttt{sifting} yields single-trial significance values, $\sigma_{ps}$, which are interpreted as the equivalent Gaussian significance of a frequency $f$ in the $\chi^{2}$-distributed power spectrum (assuming a de-reddened, Gaussian white-noise background). This allows for a direct comparison of candidates across different searches. Final candidates were folded on the raw \texttt{filterbank} files for visual inspection using \texttt{prepfold}. Folding was performed on the local workstation utilizing parallelization, with each fold requiring $\sim$10\,mins. 

\subsubsection{Long Pointings}
\label{sec:accel_long_pointings}
We conducted two constant acceleration searches ($z_{\rm max}=0$ and $z_{\rm max}=200$) and one linearly-changing acceleration (jerk) search ($z_{\rm max}=200$, $w_{\rm max}=600$) on Epoch 1 and 2, which comprised five 1-hr and three 2-hr high-time resolution \texttt{filterbank} files. Due to the computational intensity involved, we executed \texttt{accelsearch} on Columbia University's High-Performance Computing (HPC) cluster, equipped with Dual Intel Xeon Gold 6226R processors across 286 nodes (32 cores/node). We modified our analysis pipeline to run these searches in parallel --- 50 de-dispersed time series simultaneously --- to optimize runtime. Memory allocation was set at 8\,GB per core for constant acceleration searches and 16\,GB per core for the jerk searches. Typical run times were approximately 10-minutes per search for $z_{\rm max}=0$, 30-minutes for $z_{\rm max}=200$, and $\approx 1$~day for the jerk search ($z_{\rm max}=200$, $w_{\rm max}=600$). Given our large number of de-dispersed time series (500), utilizing parallelization was essential. We again applied a 2$\sigma$ threshold to ensure that no potential candidates were overlooked. Additionally, since shorter integration times improve sensitivity to a broader range of orbital periods (see Figure \ref{fig:contour_lines}), we also divided the 2-hr integrations into 1-hr segments to maximize our detection capability. Such a comprehensive approach is especially important given the poorly understood electron density environment in the GC and the uncertainties in the location, uniformity, and/or patchiness of the scattering screen --- all of which can contribute to signal intermittency. Additional factors such as binary eclipses may also obscure pulsar signals at certain orbital phases \citep{Perez_2023}. 

Like the short pointings, periods were searched up to 8 harmonics, and candidates were grouped using the \texttt{sifting} routine, resulting in $\sigma_{ps, {\rm min}}=6$ as the final detection threshold for comparison. The parameters from the blind searches were then used to fold the remaining candidates on the raw \texttt{filterbank} files for visual inspection using \texttt{prepfold}. Folding was performed on the HPC cluster, as each 1-hr observation required one day of folding per candidate. To expedite the analysis for all candidates, we also parallelized the folding tasks across multiple compute nodes. 

\section{Results}
\label{sec:results}

\subsection{Short Pointings}
\label{sec:shortpointings_results}
Across all 37 short, 5-min pointings, we detected a total of 455 signal candidates, including some of which were duplicated across multiple pointings. Table \ref{tab:shortpointings_log} summarizes the number of candidates detected per pointing. To mitigate false positives, we used a position-switching strategy (see Section \ref{sec:observing_strategy}). This allowed us to reject 11 of these 455 candidates, as these signals were detected in paired pointings separated by more than two beam-widths --- behavior indicative of terrestrial RFI. Some of these signals also exhibited low DMs, high duty cycles, and/or were only present in a few time samples.

To further vet our candidates, we implemented a second filter to determine potential sidelobe contamination. Although sidelobe effects are generally minimal with the GBT, we adopted a conservative approach by cross-checking for each of the remaining 444 candidates beyond their adjacent pointings. Based on GBT measurements at X-band --- where the first sidelobe is $\sim -27$\,dB below the main lobe \citep{Frayer_2017} --- we estimate that the first sidelobe occurs at an angle of $\approx 1\farcm86$ away from the main lobe (over a beam-width away), meaning that any candidate detected in directly adjacent pointings is likely a sidelobe artifact. Similarly, the second sidelobe, with an estimated power level of $\sim -29$\,dB, is expected at an angle of $\sim 2\farcm76$ away from the main lobe (two beam-widths away). 

A candidate was flagged for sidelobe contamination if it appeared in multiple pointings. To distinguish a true detection from RFI, we utilized a beam adjacency mapping, assigning each beam to its immediate neighbors (e.g., beam A00 maps to beams B01--B06). Candidates appearing in non-adjacent beams were classified as RFI and removed from further analysis, since signals detected two beam-widths away are almost certainly spurious. Ultimately, we rejected all 444 candidates, i.e., no pulsars were detected.  Additionally, a visual inspection of all 455 candidates revealed time and frequency features not characteristic of a typical pulsar (large duty cycles, non-persistent signal, reduced $\chi^2$ decreasing or staying constant), further confirming their classification as RFI. This RFI behavior is consistent with that observed in the GC survey at C-band (4--8\,GHz) for the short pointings surrounding A00 \citep{Suresh2022}, where the lower frequencies led to a higher false-positive rate, highlighting the advantages of higher-frequency surveys in the GC.

\subsection{Long Pointings}
\label{sec:long_pointings}
We detected a total of 4,827 periodicity candidates across both epochs. Table \ref{tab:A00_obs_log} shows the number of candidates detected in each acceleration search per scan. For Epoch 2, labels ``2.n.1" and ``2.n.2" indicate the first and second halves of the 2-hr integrations, respectively, with each treated as a separate 1-hr segment. Duplicate candidates within a single scan (e.g., those detected in both $z_{\rm max}=0$ and $z_{\rm max}=200$ searches) were counted only once; however, some candidates in Epoch 2 may have been recorded twice if they were detected in both the full 2-hr integration and in the split 1-hr segments. Using the full 2-hr integrations is especially advantageous for detecting extremely dim pulsars (see Section \ref{sec:survey_sensitivity} for more details). As expected, the number of detections increases with integration time: from Equations \ref{eqn:zcalc} and \ref{eqn:wcalc}, both the $z$ and $w$ bins scale with the square or cube of the scan integration time, so longer scans and higher bin values yield more candidates.

\begin{figure*}
    \centering
    \includegraphics[width=\textwidth]{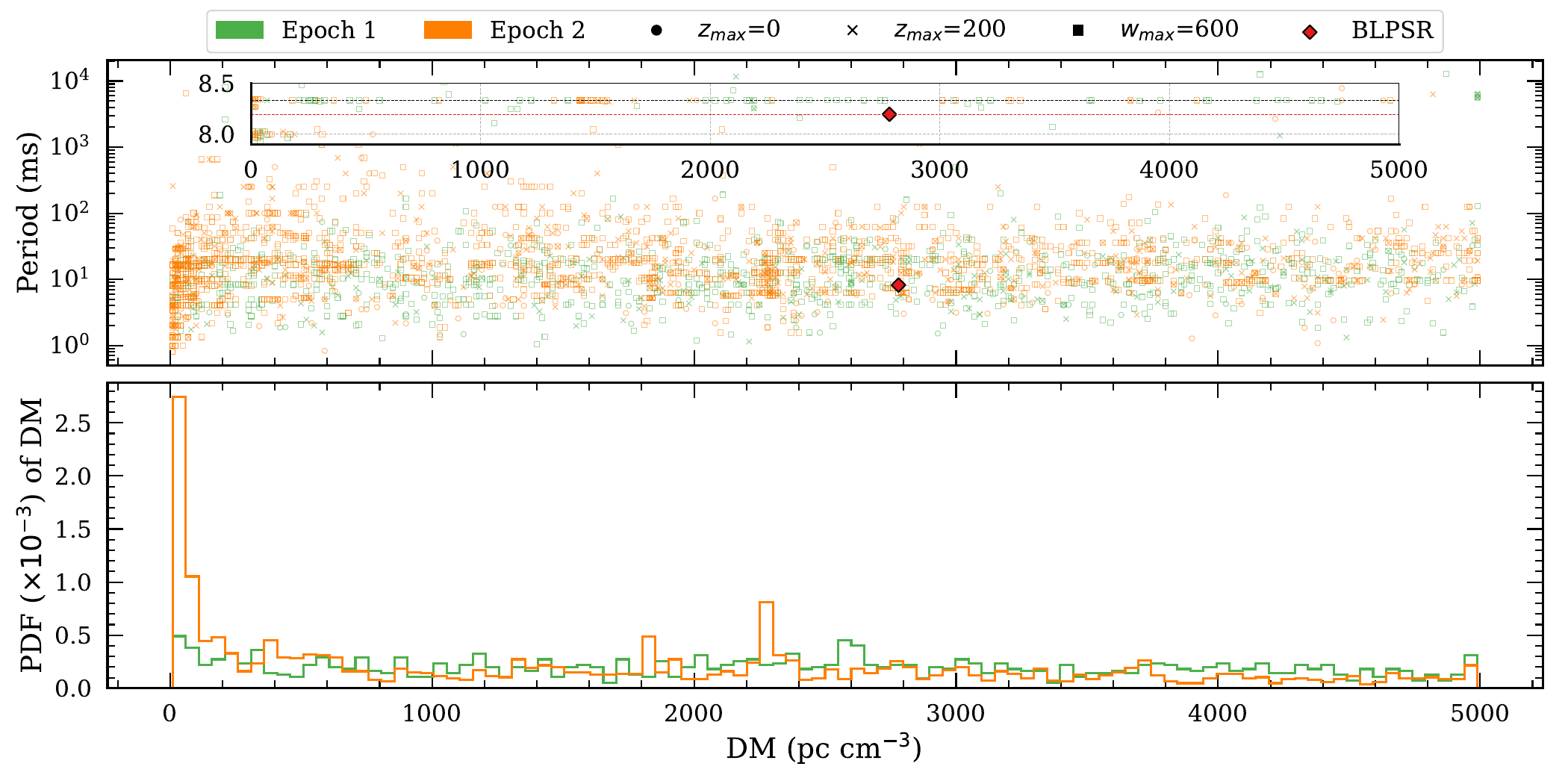}
    \caption{Top panel: Period vs. DM for all candidates detected blindly at all three of our accelerated searches $z_{\rm max}=0$ (circles), $z_{\rm max}=200$ (crosses), and $z_{\rm max}=200$, $w_{\rm max}=600$ (squares) across both epochs, using 4096 sub-bands. The BLPSR candidate is marked by the red diamond, and each epoch is color-coded accordingly. The inset plot provides a zoomed-in view of candidates near BLPSR. The red dashed line indicates BLPSR’s period (8.19\,ms), with no other candidate matching this value. The black dashed line marks a recurring 8.33\,ms candidate, corresponding to twice the frequency of 60\,Hz power lines. Bottom panel: Probability distribution function (PDF) of candidate DMs binned uniformly to 50~pc\,cm$^{-3}$ resolution, per epoch.}
  \label{fig:DM_dist_plots}
\end{figure*}

\begin{figure*}
    \centering  \includegraphics[width=\textwidth]{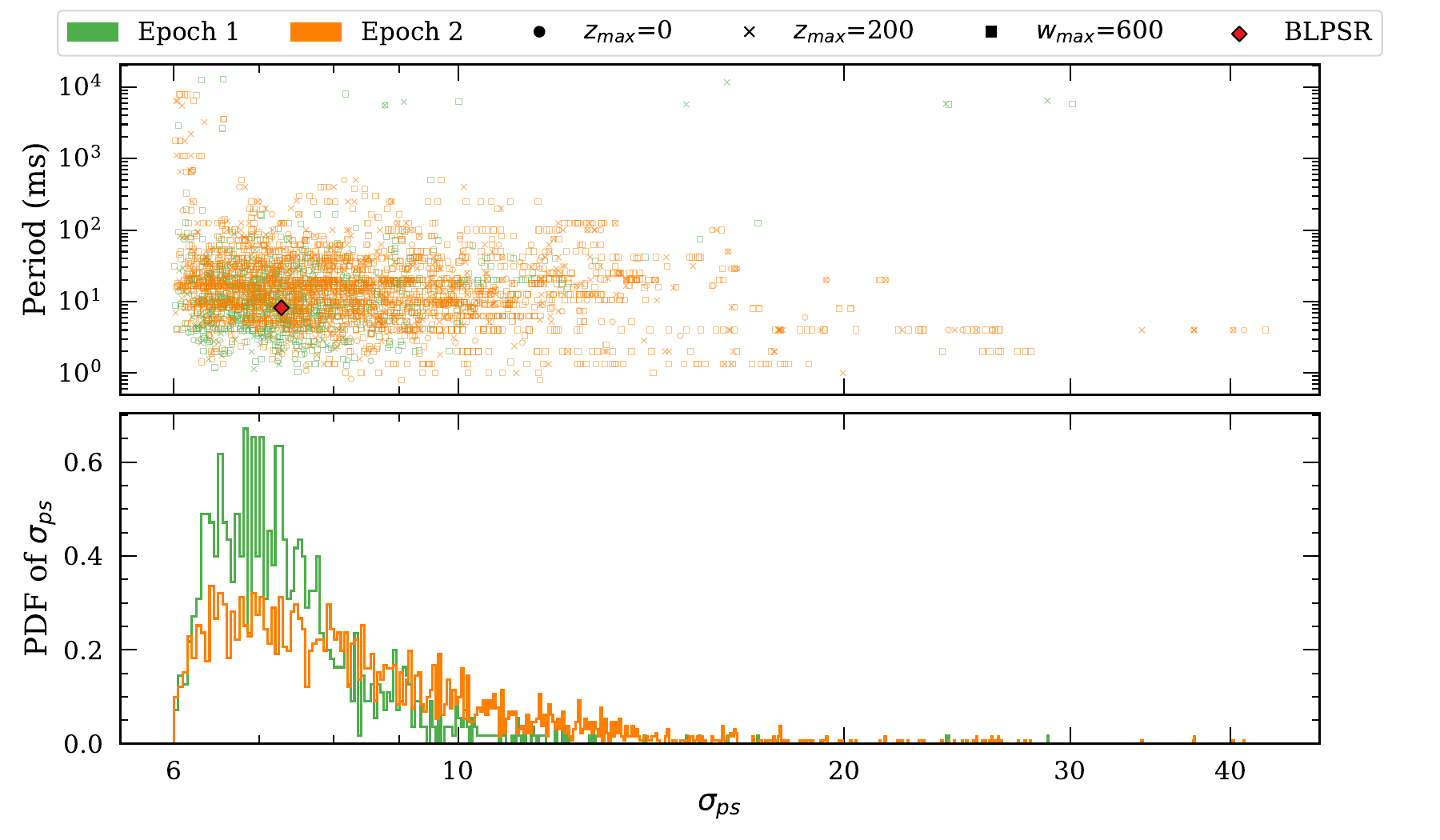}
    \caption{Top panel: Period vs. $\sigma_{ps}$ for all candidates detected blindly at all three of our accelerated searches $z_{\rm max}=0$ (circles), $z_{\rm max}=200$ (crosses), and $z_{\rm max}=200$, $w_{\rm max}=600$ (squares) across both epochs, using 4096 sub-bands. The BLPSR candidate is marked by the red diamond, and each epoch is color-coded accordingly. Bottom panel: Probability distribution function (PDF) of equivalent Gaussian significance ($\sigma_{ps}$) of candidates in power spectra binned uniformly to ln $\sigma_{ps}$ with width $\approx$ 0.05, per epoch.}
  \label{fig:sigma_dist_plots}
\end{figure*}

Figures \ref{fig:DM_dist_plots} and \ref{fig:sigma_dist_plots} present the statistical distributions of DM, period, and $\sigma_{ps}$ for the 3,289 candidates detected in the 1-hr integrations, with a median of $\tilde x_{\rm DM} = 1830$~pc~cm$^{-3}$, $\tilde x_{P} = 13.4$~ms, and $\tilde x_{\sigma_{ps}} = 7.7$. We see candidates span a wide range of periods, from millisecond to sub-second scales, revealing a mix of potential pulsars and interference. However, the majority fall between $P$ $\in$ [1, 100] ms, indicating a statistical sampling effect and periodic RFI. To maintain consistency across epochs, candidates from the full 2-hr integrations are excluded from these plots (though they are reported in Table \ref{tab:A00_obs_log} for completeness). Candidates are distinguished by acceleration search parameters: dots represent $z_{\rm max}=0$, crosses denote $z_{\rm max}=200$, and squares indicate $z_{\rm max}=200$ $w_{\rm max}=600$; candidates from Epoch 1 are shown in green and those from Epoch 2 in orange. The red diamond, denoted as the Breakthrough Listen Pulsar (BLPSR), marks an interesting 8.19\,ms MSP candidate found in our searches and will be discussed in detail in the following subsection.  

Figure \ref{fig:DM_dist_plots} displays the DM--period distribution for our candidate detections, with the inset zooming into the 8--8.5\,ms period range to highlight the location of BLPSR relative to other candidates, including the signal associated with the 60\,Hz power line (dashed, black line). We observe a cluster of candidates in Epoch 2 with DMs between 0--200\,pc\,cm$^{-3}$. This epoch yielded more candidates overall, likely attributable to the wider bandwidth and hence increased sensitivity to RFI. There was also an increased incidence of detections toward the end of the session (second half of the last scan) as the GC approached the horizon, leading to increased interference at lower DMs. A spike in candidates at DMs\,$\approx$\,2250--2300 was also identified, driven by two strong narrowband RFI signals and their harmonics; these alone accounted for $\approx 80$\% of candidates in that DM range across all acceleration searches in the second half of the final scan.

Figure \ref{fig:sigma_dist_plots} shows the $\sigma_{ps}$--period distribution, where the majority of candidates have $\sigma_{ps} < 10$. Although pulsars near the GC are expected to be faint (and hence a low $\sigma_{ps}$), RFI is also a culprit, with very strong RFI producing higher $\sigma_{ps}$. Similar to the short pointings, many more false positives were found for the long 6-hr A00 pointings at C-band \citep{Suresh2022} than at X-band, again showing the importance of conducting higher-frequency pulsar surveys in the GC. Following visual inspection of all 4,827 candidates, all were classified as RFI. Although several pulsar-like profiles were detected at different periods, they all showed a reduced $\chi^{2}$ peak near DM=0\,pc\,cm\(^{-3}\) and exhibited no broadband properties. A single outlier --- the BLPSR candidate --- stood out and warranted closer examination.

\begin{deluxetable*}{cccccc}[!t]
\tablecolumns{6}
\tablewidth{0pt}
\tablecaption{Summary of GBT X-band Long Observations
\label{tab:A00_obs_log}}
\tablehead{
    \colhead{Epoch} & \colhead{Scan\tablenotemark{a}} & \colhead{Date} & & \colhead{\# of candidates} & \\
    & & \colhead{(MJD)} & 
    \colhead{$z_{\rm max} = 0$} &
    \colhead{$z_{\rm max} = 200$} &
    \colhead{$z_{\rm max} = 200, w_{\rm max} = 600$}
      }
    \startdata
    \hline
    1 & 1.1 & 59740.146 & 13 & 24 & 97 \\
    & 1.2 & 59740.188 & 18 & 30 & 81 \\
    & 1.3 & 59740.230  & 11 & 35 & 79 \\
    & 1.4 & 59740.272 & 14 & 26 & 75 \\
    & 1.5 & 59740.313 & 14 & 38 & 111 \\
    \hline
    2 & 2.1 & 60287.632 & 44 & 106 & 324 \\
    & 2.1.1 & -- & 59 & 54 & 117 \\
    & 2.1.2 & -- & 20 & 126 & 207\\
    & 2.2 & 60287.715 & 41 & 69 & 257 \\
    & 2.2.1 & -- & 38 & 56 & 120 \\
    & 2.2.2 & -- & 28 & 67 & 209 \\
    & 2.3 & 60287.799 & 68 & 112 & 517 \\
    & 2.3.1 & -- & 68 & 107 & 386 \\
    & 2.3.2 & -- & 46 & 200 & 715
    \enddata
    \tablenotetext{a}{For Epoch 2, labels ``2.n.1" and ``2.n.2" indicate the first and second halves of the 2-hr integrations, respectively, with each treated as a separate 1-hr segment}
\end{deluxetable*}

\subsection{BLPSR Candidate}
\label{sec:blpsr_cand}
The MSP candidate, BLPSR, was detected during our $z_{\rm max}$\,=\,200 acceleration searches with a period of $P$= 8.19\,ms in the first 1\,hr scan of Epoch 1 (Scan 1.1) during the long-pointing observations of A00. It was detected at a DM of 2780~pc\,cm$^{-3}$, $\sigma_{ps} = 7.28$, $z = 2.75$ and $w$ = 0, with a $\dot{P}$ =  $-1.23\times10^{-11}$ s\,s$^{-1}$ corresponding to a line-of-sight acceleration of $a_l = -0.45$ m\,s$^{-2}$, where a negative $a_l$ is defined as accelerating towards the observer. Although its DM is substantially higher than that of the nearest known neutron star to the GC --- magnetar PSR J1745--2900 (DM = 1778~pc\,cm$^{-3}$) --- it is consistent with expectations for sources located in the inner regions of the GC, given the distance uncertainty of the patchy scattering screen (see Section \ref{sec:orbital_params} for further discussion). Applying the NE2001 Galactic electron density model using \texttt{PyGEDM} predicts a distance of $d_{\rm NE2001} = 8.59$\,kpc \citep{Cordes_2003_NE2001model, Price_2021}, supporting the candidate's location near Sgr A*. 

\begin{figure*}[t]
\centering
    \includegraphics[width=0.7\textwidth]{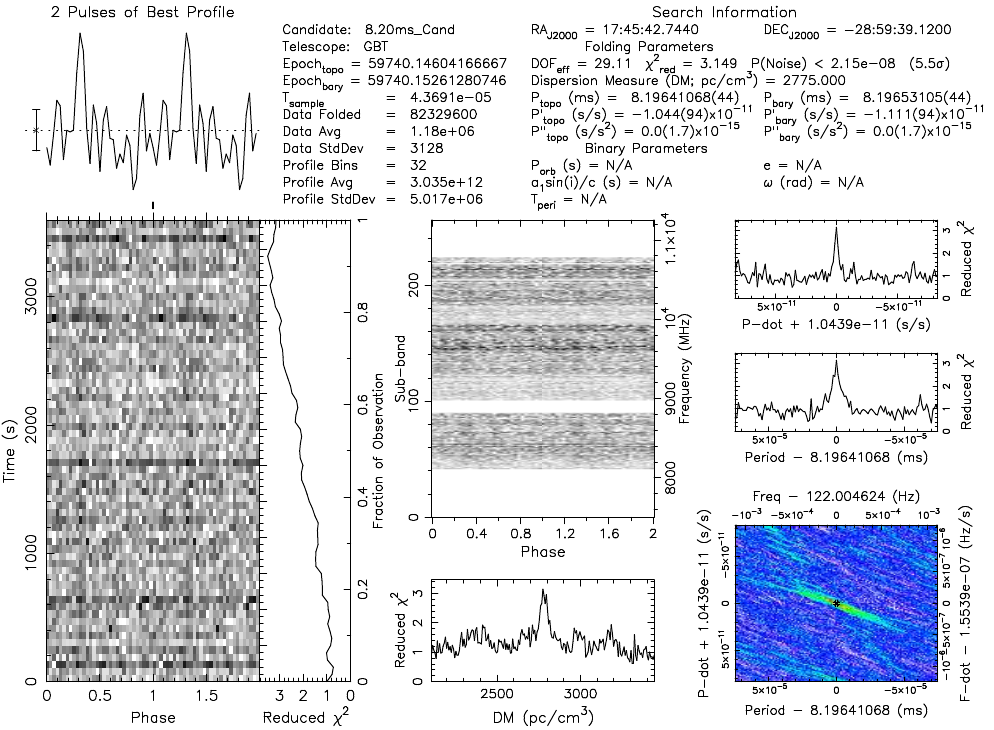}	
    \caption{PRESTO \citep{Presto_Ransom_2011} plot of the BLPSR candidate, folded using parameters from the .par file derived from TOA measurements (see Section \ref{sec:confirm_blpsr}). Data was obtained with the GBT at a center frequency of $\nu_{c}=9.37$\,GHz on 2022 June 10, with a total integration time of 1\,hr. The folded profile was generated using 32 phase bins and 256 sub-bands (each 14.65~MHz wide), with $\sim$\,1.2\,GHz removed for signal clarity.}
    \label{fig:MSP_Cand_origdetect}
\end{figure*}

BLPSR was initially selected based on its detection significance ($>6$$\sigma_{ps}$), coherent power, and its temporal and spectral characteristics after folding. Its centroid --- the signal's approximate location in the normalized time series --- is 0.476 at the fundamental frequency. Similarly, its purity --- a measure of the root-mean-squared dispersion of the pulsations in time with respect to the centroid --- is 0.899. This indicates that the signal is present throughout nearly the entire scan, supporting a real, temporally stable signal. As seen in Figure \ref{fig:MSP_Cand_origdetect}, the signal persists for a large fraction of the entire 1-hr integration and appears present across most of the band in the dynamic spectrum. The integrated profile's reduced $\chi^{2}$ increases over time, as does the reduced $\chi^{2}$ as a function of $P$ and $\dot{P}$. However, we also observe variations in signal intensity across both time and frequency, which may result from a lower S/N and/or complex propagation effects in the interstellar medium. While the NE2001 model \citep{Cordes_2003_NE2001model} predicts strong scattering ($\tau_{ss} = {2000}\,{\nu^{-4}}$\,s), corresponding to pulse broadening of $\sim 0.2$\,s at X-band ($\nu_{c}=10$\,GHz), the weak scattering scenario observed for the GC magnetar PSR\,J1745$-$2900 \citep{Spitler_2014}, with $\tau_{ws} = {1.3}\,{\nu^{-4}}$\,s, would produce negligible pulse broadening of $\approx$\,130\,$\mu$s for our putative BLPSR candidate. We attempted to measure the scintillation timescale using the autocorrelation function of the on-pulse intensity over time, and found a pulse intensity scintillation time-scale of $\sim$ 1800\,s. This timescale is inconsistent with the expected longer timescales of diffractive scintillation at this frequency. Thus, neither pulse broadening nor scintillation appears to account for the observed short-timescale variability.

\begin{figure*}
\centering
    \includegraphics[width=0.45\linewidth]
  {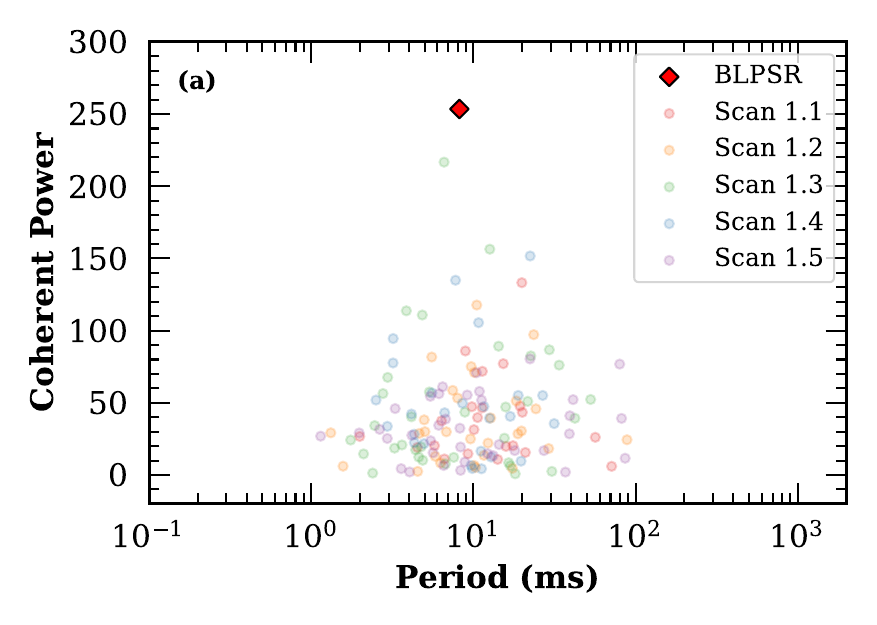}
    \hspace*{0.5cm}
    \includegraphics[width=0.45\linewidth]{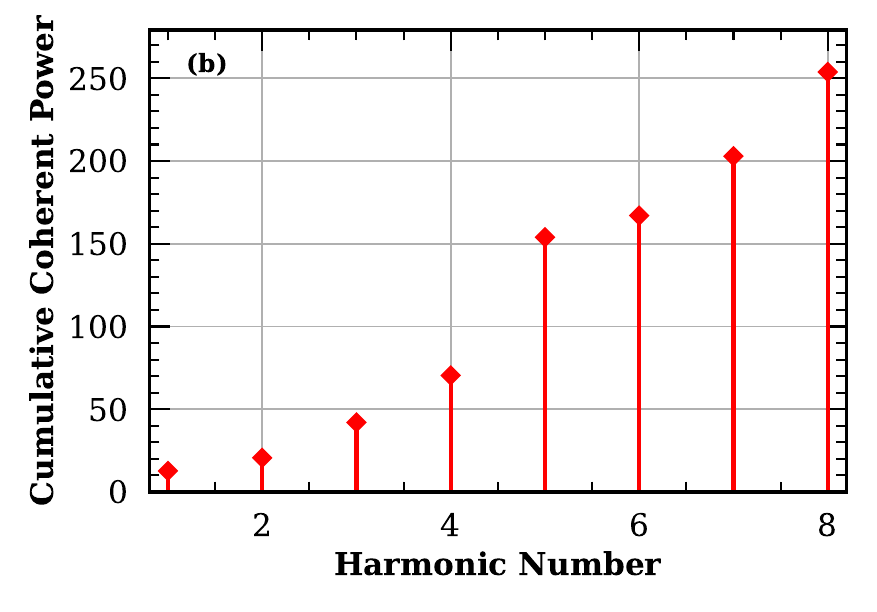}
\caption{a) Coherent power as a function of period (in ms) for all candidates detected with $z_{\rm max}=200$ during Epoch 1 using 4096 sub-bands. The BLPSR candidate, depicted by the red diamond, exhibits the highest coherent power among all detected candidates, strongly suggesting its astrophysical nature. b) Cumulative coherent power as a function of harmonic number ($n$) for BLPSR. Each stem line represents the aggregated power up to $n = 8$, illustrating how the coherent power accumulates with each harmonic step as that expected for a real pulsar. 
\label{fig:coherent_power}}
\end{figure*}

In the broader candidate population, BLPSR was notable despite being centrally located in the $\sigma_{ps}$--period distribution cluster (see Figure \ref{fig:sigma_dist_plots}). Specifically, it is the only candidate detected at its specific period (see Figure \ref{fig:DM_dist_plots}). Modified $z$ scores (standard deviations from the median) for DM, $P$, and $\sigma_{ps}$ based on the long-pointing candidate distributions in Figures \ref{fig:DM_dist_plots} and \ref{fig:sigma_dist_plots} yield $z_{\rm DM} = 0.48$, $z_{P} = -0.49$, and $z_{\sigma_{ps}} = -0.3$. Although this indicates that BLPSR is not an outlier among the 4,827 detected candidates, visual inspection revealed it to be the only one with pulsar-like properties, requiring further examination. BLPSR also exhibited the highest coherent power (253.5; marked by a red diamond in Figure \ref{fig:coherent_power}a) among all candidates detected at $z_{\rm max} = 200$ during Epoch 1, with a $z$ score of $z_{\rm power} = 7.84$, showing that BLPSR is an outlier when considering this parameter. We also computed the total coherent power of BLPSR's eight harmonics by iteratively summing their respective complex Fourier amplitudes and phases, demonstrating how the signal power amplifies with each harmonic step, as expected for a true pulsar signal (see Figure \ref{fig:coherent_power}b)\footnote{However, it should be noted that Gaussian noise can also produce peaks in the power spectrum which, when incrementally summed over harmonics, may lead to an apparent increase in total coherent power at each step.}.

Although the detection Gaussian significance of BLPSR is not particularly high, we chose to examine BLPSR closely due to the importance of finding pulsars near the GC and minimizing the false-negative rate. Specifically, we deliberately examined candidates with lower folded profile reduced $\chi^{2}$ values, among which BLPSR emerged. If we had imposed a stricter detection threshold (e.g., S/N $>$ 5), BLPSR would not have been flagged for visual inspection. Although BLPSR did cross our detection threshold, it is still near the borderline and could not be categorically ruled out as a signal originating from either RFI or noise fluctuations. The following section aims to qualitatively examine BLPSR--representing one of the most detailed inspections carried out on any potential pulsar candidate from a single-dish telescope to date.

\begin{deluxetable}{ll}
\tablecolumns{2}
\tablewidth{0pt}
\tablecaption{
Parameters of the Unconfirmed BLPSR Candidate 
\label{tab:blpsr_params}}
\tablehead{
    \colhead{Parameter} & \colhead{Value}} 
    \startdata
    \hline 
    R.A. (J2000)\tablenotemark{a} & $17^{\rm h}45^{\rm m}40^{\rm s}\!.04 \pm  5^{\rm s}\!.6$ \\
    Decl. (J2000)\tablenotemark{a} & $-29\degr 00\arcmin 28\farcs10 \pm 1\farcm4$ \\
    $P_{s,0}$ (ms) & 8.196531(02) \\
    $\dot{P}$ (s\,s$^{-1}$) & $-1.111(12)\times10^{-11}$ \\
    DM (pc\,cm$^{-3}$) & $2775 \pm 24\,(1\sigma)$ \\
    $d_{\rm{NE2001}}$ (kpc) & 8.59 \\
    $d_{\rm{YMW16}}$ (kpc) & 11.33 \\
    10\,GHz min. flux density, $S_{\rm min, 10\,GHz}$ (mJy)\tablenotemark{b} & 0.009 \\
    $L_{\rm 10\,GHz} = S_{\rm 10\,GHz}\,d^{2}$ \,(mJy\,kpc$^{2}$) & 0.61 \\
    Pulse width at 10\,GHz, $W_{50}$  (ms) & $<0.7$
    \enddata   
    \tablenotetext{a}{The position error corresponds to our $1\farcm4$ beam width.}
    \tablenotetext{b}{Assuming BLPSR was detected at the A00 pointing beam center}
\end{deluxetable}

\section{Examination of BLPSR detection properties}
\label{sec:confirm_blpsr}
We initially folded BLPSR using the detection parameters obtained from \texttt{accelsearch}. To refine these parameters, we re-folded the \texttt{filterbank} data using \texttt{prepfold}, and measured the DM that maximized the S/N (see Section \ref{sec:peak_flux_with_dm}). While a timing-based DM fit was attempted, it produced unrealistically small formal uncertainties. Instead, we adopted the DM and associated error estimated from the analysis in Section \ref{sec:peak_flux_with_dm} and Figure \ref{fig:SNR_vs_DM}, which we consider more robust. Pulse times of arrival (TOAs) were then derived from the prepfold file using the \texttt{pat} utility (part of the \texttt{PSRCHIVE} suite) employing Fourier-Domain Monte Carlo (FDM) techniques. The template profile for TOA generation was created by fitting a single Gaussian component to the full-average prepfold profile, using the \texttt{paas} routine in \texttt{PSRCHIVE}. Splitting the full 1-hr integration of Scan 1.1 into eight 7.5-minute sub-integrations resulted in a median TOA uncertainty of $\sim 1$\,ms, or about 0.12 rotations.  We then used \texttt{tempo2} with these TOAs to further refine the fundamental frequency and its first derivative during this observation. Figure \ref{fig:MSP_Cand_origdetect} shows the BLPSR candidate folded using \texttt{prepfold} with this refined ephemeris. Although BLPSR is speculative, we summarize its final parameters in Table \ref{tab:blpsr_params} to reference our discussion. We used the newly derived DM of 2775 pc\,cm$^{-3}$ for the electron density models\footnote{Distance discrepancies between electron density models (NE2001 and YMW16) arise from different assumptions, including the distance of the GC from the Sun, the modeled Galactic components, and the use of interstellar scattering measurements \citep{ Cordes_2003_NE2001model, Yao_2017_ymw16, Price_2021}.}. We noticed that after folding with the derived ephemeris, the folded profile exhibited a peak close to \(\rm{S/N}\approx3\). For additional verification, we dedispersed and folded the data using \texttt{dspsr} \citep{DSPSR_vanStraten:2010hy} to generate a \texttt{PSRCHIVE}-formatted \texttt{archive} file \citep{PSRCHIVE_Hotan_2004}. We were able to see a similar, although relatively weaker, pulse profile (not shown here) in these data, confirming that the original detection shown in Figure \ref{fig:MSP_Cand_origdetect} is not an artifact of our \texttt{presto}-based pipeline. 

\subsection{Data Quality test}
\label{sec:data_quality_test}
\begin{figure}
\centering
\includegraphics[width=0.9\linewidth]{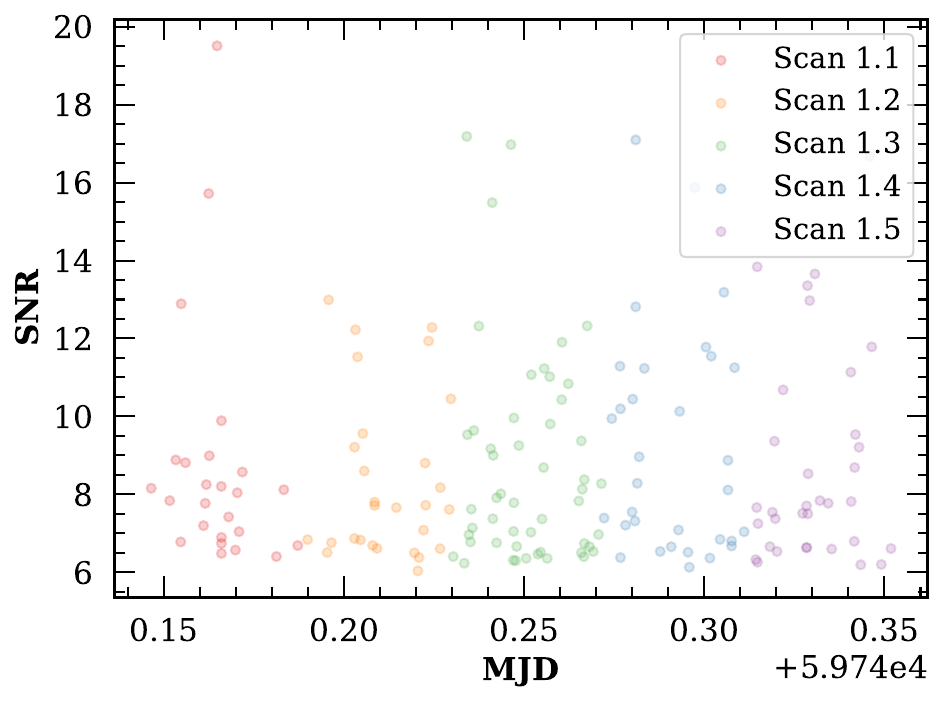}
\caption{Single-pulse detections of the GC magnetar SGR~J1745$-$2900 as a function of observation MJD. Each color corresponds to a different 1-hr scan from Epoch 1. The BLPSR candidate was only detected in the first scan, for which the magnetar single pulses are shown in red. The consistency of magnetar pulses across all scans indicates no apparent data quality issues during our observations.}
\label{fig:magnetar_snr_mjd}
\end{figure}

As noted in Section \ref{sec:blpsr_cand}, BLPSR was only detected in the first 1-hr scan of Epoch 1 at a relatively low significance of the folded profile (around 3\(\sigma\) as seen in Figure \ref{fig:MSP_Cand_origdetect}). To verify this detection was not due to data quality issues, we attempted to find single pulses from the bright GC magnetar, SGR J1745$-$2900 (DM $\approx$1778\,pc\,cm$^{-3}$), which is located well within our beam. This magnetar is known to regularly produce strong single pulses, which serves as an excellent benchmark for data quality. We used \texttt{Heimdall} \citep{Heimdall_Barsdell_2012} with our specialized machine learning candidate-sorting algorithm \texttt{SPANDAK} \citep{Gajjar_2022} to detect single pulses from all Epoch 1 scans (see \citealt{Gajjar_2021} for more details). Figure \ref{fig:magnetar_snr_mjd} shows the magnetar's single-pulse S/N as a function of MJD across the full five 1-hr scans in Epoch 1. It is consistently detected at a stable S/N level with no noticeable drop-outs or systematic trends, indicating that our data quality remained consistent throughout the full epoch. A one-way Analysis of Variance across the five scans showed no statistically significant difference ($F = 0.3548$; p-value\,=\,0.84) in the magnetar’s mean S/N, including the first segment (Tukey's Honestly Significant Difference test also confirmed this; p-value\,=\,0.97). Hence, the non-detections (see Section \ref{sec:blpsr_followup} for further discussion) of BLPSR in scans beyond the first hour cannot be solely attributed to any sensitivity loss or other data quality issues. 

\subsection{Segmentation test}
\label{sec:segmentation_test}
\begin{figure}
    \centering
    \includegraphics[scale=0.5]
    {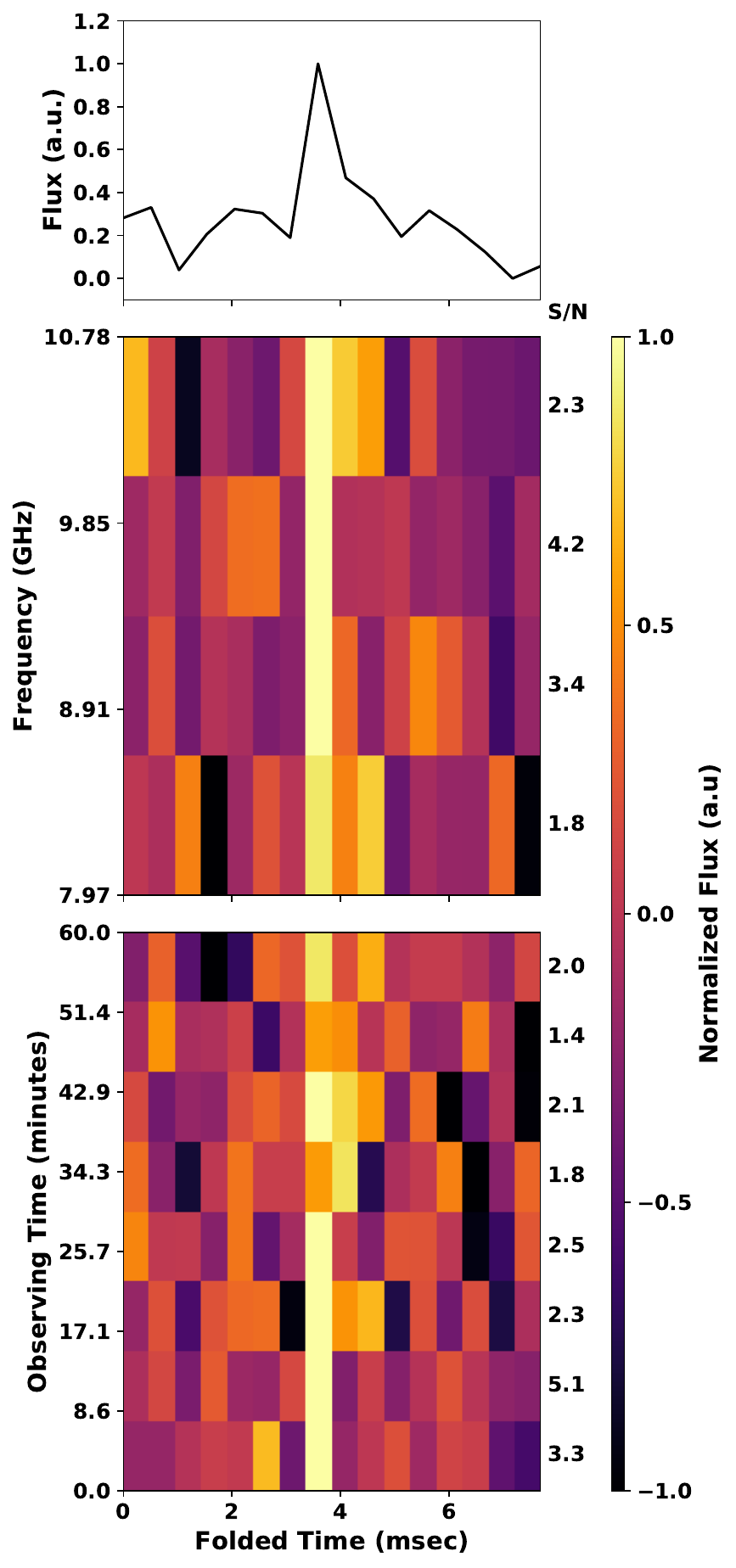}
    \caption{The folded profile of the candidate pulsar, BLPSR, detected from our survey is shown across 4 sub-bands and 8 sub-integrations, demonstrating consistent peaks across most segments. The corresponding peak S/N values are also shown separately for each sub-integration and sub-band, highlighting the persistence of the signal across time and frequency. The overall average S/N of the folded profile is $\approx3\sigma$, and the signal appears to gradually fade toward the end of the observation.}
    \label{fig:BLPSR_time_frequency}
\end{figure}

A key test to validate a pulsar detection is to verify that the pulse profile persists across both time and frequency, as spurious interference or RFI seldom display such consistent behavior. Figure \ref{fig:BLPSR_time_frequency} shows the pulse profile divided into 8 sub-integrations and 4 frequency sub-bands. A relatively weaker, but statistically significant, integrated profile with \(\sim 3\sigma\) significance can be seen. The profile remains visible in most sub-integrations --- except near the end of the observation --- and is visible across most sub-bands, except at the lower frequencies. It is also evident that beyond $\sim 30$\,minutes in observing time, the per-subintegration S/N values drop to close to unity, indicating that the on-pulse bin is statistically indistinguishable from the off-pulse baseline; thus, the pulsar is not detected in the latter part of the observation. To quantify signal persistence, we performed a Kolmogorov–Smirnov (KS) test on each profile bin, comparing the on-pulse power distribution to the off-pulse distribution. For a real pulsar, we expect the power distribution for on-pulse bins to show significant deviations from the off-pulse bins \citep{Keith_2009_MNRAS}. We refer to the measured KS statistics (and corresponding p-values) from sub-band comparisons across profile bins as the parameter for broadbandness and those from sub-integration comparisons across profile bins as the parameter for time persistence as, \(D_{\rm broad}^{\rm KS}\) and \(D_{\rm time}^{\rm KS}\), respectively (see Figure \ref{fig:BLPSR_time_frequency}).

We found \(D_{\rm broad}^{\rm KS} = 0.94\) (p-value \(< 4\times10^{-4}\)) for broadbandness across the full bandwidth and \(D_{\rm time}^{\rm KS} = 0.62\) (p-value \(< 2\times10^{-3}\)) for time persistence across the full Scan 1.1, indicating that we can reject the null hypothesis that on-pulse and off-pulse distributions are drawn from the same population. The relatively lower p-value associated with time persistency is likely due to BLPSR exhibiting an almost negligible profile roughly after the first 45 minutes (also seen in Figure \ref{fig:MSP_Cand_origdetect}); by ignoring the last 15 minutes, we found an expected improved p-value  of \(< 2\times10^{-4}\). It should be noted, however, that although these probabilities and Figure \ref{fig:BLPSR_time_frequency} suggest the presence of a broadband signal persisting across a large fraction of the observing time --- making spurious RFI unlikely --- the observed \(S/N\) in each sub-band and sub-integration is low enough that noise fluctuations could still mimic such behavior. Furthermore, the derived KS statistics are based on a small number of samples (4 sub-bands and 8 sub-integrations across 16 profile bins). In other words, we cannot rule out the possibility that the detection is due to spurious noise or low-level RFI, and further tests were carried out in Section \ref{sec:randomtests}.

\subsection{Peak flux with DM}
\label{sec:peak_flux_with_dm}
\begin{figure}
    \centering
    \includegraphics[width=1\linewidth]{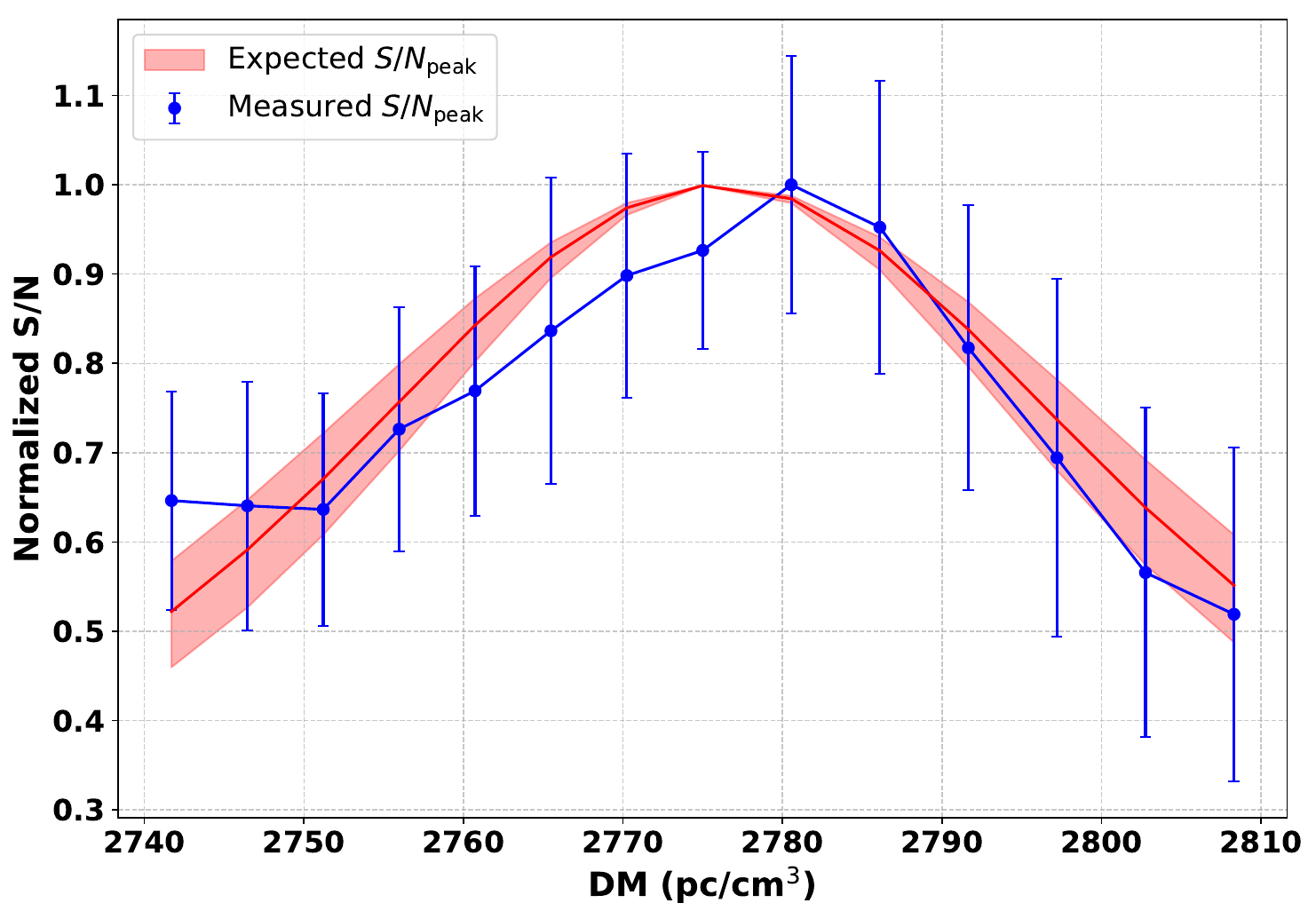}
    \caption{Measured peak S/N (blue) vs. DM for BLPSR, normalized to highest measured S/N. The red curve is the best‐fit model from \cite{Cordes_McLaughlin_2003}, treating the pulse width as a free parameter. The pink shaded region indicates the range of theoretical S/N vs. DM curves spanned by the 1‐sigma uncertainty in the fitted width.}
    \label{fig:SNR_vs_DM}
\end{figure}

An important diagnostic for evaluating the astrophysical nature of BLPSR is a comparison of its folded profile peak amplitude \(S/N\) as a function of DM. Following \cite{Cordes_McLaughlin_2003}, the reduction in \(S/N\) due to an incorrect DM (\(\delta \mathrm{DM}\)) can be expressed by
\begin{equation}
\frac{S/N(\delta \mathrm{DM})}{S/N^{\rm max}} \;=\; \frac{\sqrt{\pi}}{2}\,\zeta^{-1}\,\mathrm{erf}(\zeta),
\label{eq:fluxratio}
\end{equation}
where 
\begin{equation}
\zeta \;=\; 6.91\times10^{-3}\,\bigl(\delta \mathrm{DM}\bigr)\,
\frac{\Delta{\mathrm{B}}}{W_{\mathrm{ms}}\,\nu_{\mathrm{c}}^3}.
\end{equation}
Here, \(\Delta{\mathrm{B}}\) is the total bandwidth in MHz, \(\nu_{\mathrm{c}}\) is the center observing frequency in GHz, \(W_{\mathrm{ms}}\) is the pulse width in milliseconds, and \(\mathrm{erf}\) is the error function.

To test BLPSR, we began with the original \texttt{filterbank} file, dedispersed it at a DM of 2775\,pc\,cm\(^{-3}\), and used \texttt{prepfold} to fold the data into 4096 sub-bands with 64 profile bins each, using the ephemeris listed in Table \ref{tab:blpsr_params}. We then dedispersed the sub-banded folded data (in \texttt{pfd} format) across a range of trial DMs, where, for each DM, a mean-subtracted folded profile was obtained. Visual inspection confirmed the expected profile broadening towards the trailing and leading edges when it was under- and over-dispersed, respectively. To measure the peak \(S/N\), each folded profile was multiplied with a top-hat (boxcar) function of varying widths \citep{Presto_Ransom_2011,Heimdall_Barsdell_2012}. We then visually inspected and selected the width that maximized \(S/N\)  within the on-pulse region. The off-pulse region was used to calculate the standard deviation and calculate the error in the measured \(S/N\). We calculated these for each DM and normalized them by the maximum \(S/N^{\rm max}\) across all DMs to compare the measured ratio with the ratio given in Equation \ref{eq:fluxratio}.

The measured \( \frac{S/N(\delta \mathrm{DM})}{S/N^{\rm max}} \) as a function of DM, along with their errors, are plotted in Figure \ref{fig:SNR_vs_DM}. We then fitted this data with the expected distribution from Equation \ref{eq:fluxratio}, treating both the pulse width and the DM offset as free parameters. The fitted model loosely matches the data with some degree of confidence (\( \chi^2_{reduced} \approx 0.23; ~p\text{-value}>99\%\)). The derived DM that provides the maximum \(S/N\) is \(2775\pm24\,(1\sigma)\) pc\,cm\(^{-3}\). Although the peak \(S/N\) follows the expected distribution, the relatively large error bars (as indicated by the low \( \chi^2_{reduced}\)) imply that the variations in the peak \(S/N\) across DM are not statistically significant. As a result, we cannot conclusively determine that our candidate follows the expected distribution, so this test is inconclusive given the marginal quality of the detection.

\subsection{Randomization tests}
\label{sec:randomtests}
Figure~\ref{fig:BLPSR_time_frequency} offers a limited but suggestive indication that BLPSR could be a genuine astrophysical signal, and as discussed in Section \ref{sec:blpsr_cand}, its detection properties warrant a deeper examination. However, BLPSR's relatively weak folded profile (\(\chi^2_{\mathrm{reduced}} \sim 3\)) and inconclusive peak \(S/N\)--DM variation (see Section \ref{sec:peak_flux_with_dm}) motivate a statistical probability estimate of two additional possibilities: that its detection could arise from very low-level RFI or random noise fluctuations within Scan 1.1. 

To investigate this, we designed a novel randomization test to jointly evaluate the significance of BLPSR's observed properties. As a first step, and to reduce computational load, we downsampled the original \texttt{filterbank} data from 40,960 to 1,024 frequency channels, preserving the $43.69\,\mu$s sampling time and omitting any dispersion correction. We also used \texttt{prepfold} to verify that BLPSR remained detectable with the previously derived ephemeris. With this scrunched data, we found \(D_{\rm broad}^{\rm KS} = 0.96\) (p-value \(< 4\times10^{-4}\)) for broadbandness across the full bandwidth and \(D_{\rm time}^{\rm KS} = 0.91\) (p-value \(< 2\times10^{-6}\)) for time persistence. The slight increase in KS statistics compared to the original data is expected since no frequency channels were flagged during the collapse. A blind $z_{\rm max}$\,=\,200 search on this scrunched \texttt{filterbank} file again recovered BLPSR as the top candidate in both $\sigma_{ps}$ and coherent power, out of 727 raw candidates. After applying the \texttt{sifting} algorithm to remove duplicates, only 36 unique candidates remained, all of which --- except for BLPSR --- were ultimately identified as spurious noise. 

We then generated a fully randomized version of the aforementioned downsampled \texttt{filterbank} file by shuffling all time-frequency samples independently, thereby randomizing the entire dynamic spectrum while preserving the overall structure of 1024 frequency channels and $43.69\,\mu$s sampling time. This procedure destroyed any coherent, pulsar-like signal or low-level RFI, allowing us to assess the probability of recovering a BLPSR-like candidate from noise alone. We processed this randomized data through our standard pulsar search pipeline (see Figure~\ref{fig:presto_flowchart}) using the same parameters as in the original search. This yielded 533 raw candidates, which were subsequently reduced to 27 unique candidates after applying the \texttt{sifting} algorithm. However, because the data is entirely randomized, all candidates can be considered statistically independent noise realizations across all detected DMs and periods. Thus, to better characterize the noise distribution, we used all 533 raw candidates in the subsequent analysis. 

Each candidate was folded using \texttt{prepfold} into 256 sub-bands, 64 sub-integrations, and 16 profile bins, producing corresponding folded \texttt{pfd} files. These files were then collapsed into 4 sub-bands and 8 sub-integrations (similar to Figure \ref{fig:BLPSR_time_frequency}) to evaluate their broadbandess and time persistence KS-statistics, \(D_{\rm broad}^{\rm KS}\) and \(D_{\rm time}^{\rm KS}\), respectively, as described in Section \ref{sec:segmentation_test}. While we did calculate p-values for each candidate, a more robust significance estimate comes from the randomization test described below, which benefits from a larger number of samples than the individual candidate KS-derived p-values. 

\begin{figure*}
    \centering
    \includegraphics[width=0.75\linewidth]{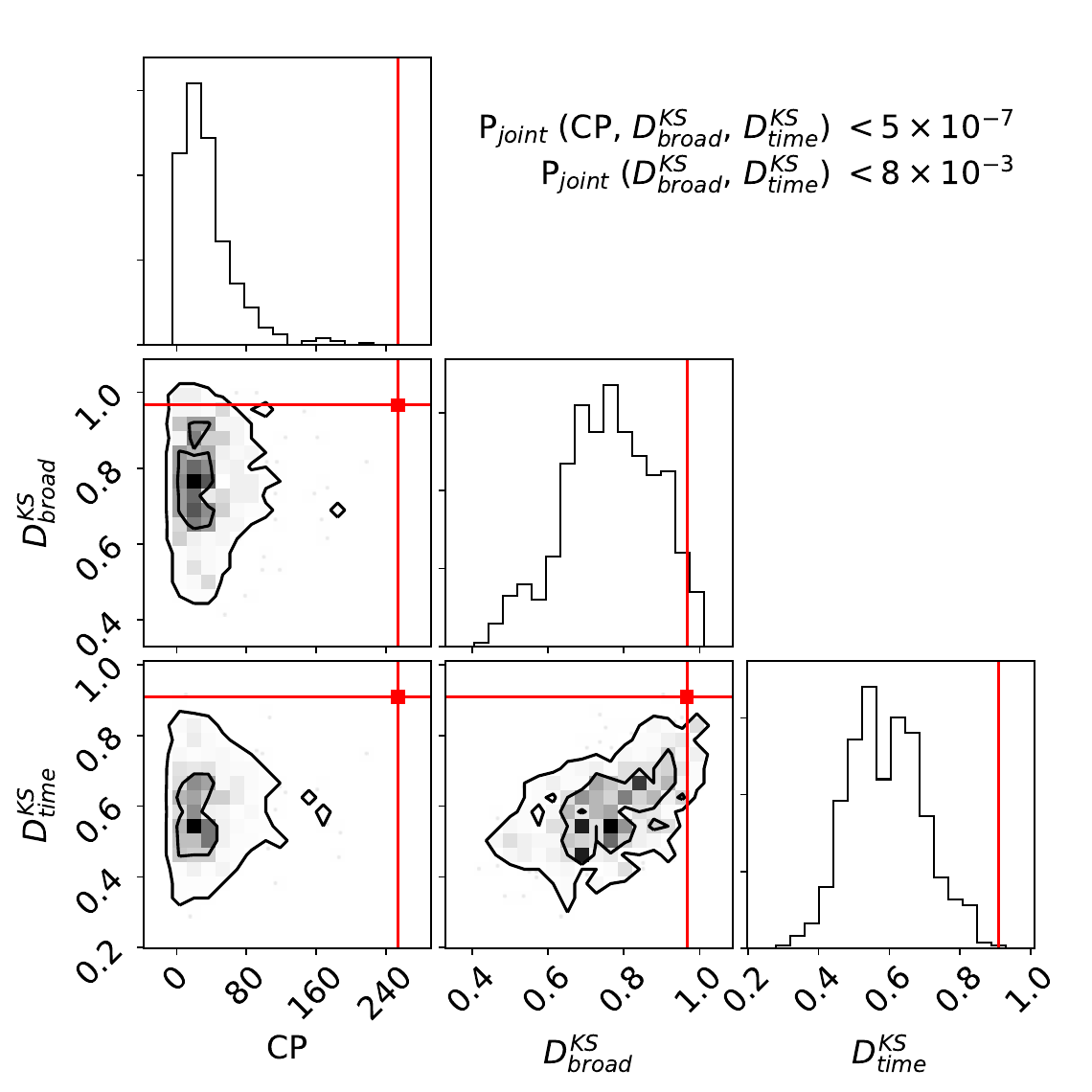}
    \caption{Corner plot showing the joint distribution of coherent power, broadband KS-statistic (\(D^{\rm KS}_{\mathrm{broad}}\)), and time-persistency KS-statistic (\(D^{\rm KS}_{\mathrm{time}}\)) for candidates from the randomized \texttt{filterbank} search. Contour lines enclose 50\% and 99\% of the total probability density under the KDE. Red lines mark the BLPSR candidate values, which lie at the tail end of all three distributions. The computed joint probability of BLPSR under the 3D KDE model is \(P_{\mathrm{joint}} < 5 \times 10^{-7}\), indicating a highly significant deviation from noise (primarily due to its high coherent power). For comparison, the joint p-value using only the two KS-statistics is less significant at \(P_{\mathrm{joint}} < 8 \times 10^{-3}\).}
    \label{fig:Corner_plot_filterbank_randomization}
\end{figure*}

We computed a joint probability distribution for BLPSR based on three independent parameters of this randomized candidate set: (i) the coherent power measured, and the KS-test statistics for (ii) \(D_{\rm broad}^{\rm KS}\) and (iii) \(D_{\rm time}^{\rm KS}\) of all 533 candidates. Although coherent power and \(D_{\rm time}^{\rm KS}\) may appear correlated, they are not strictly dependent. Coherent power is computed from the sum of complex Fourier amplitudes, and can produce a significant peak with harmonics in the Fourier transform even if a strong periodic signal is present for only a brief portion of the observation. Thus, coherent power does not directly test for time persistence and can be treated as an independent property of BLPSR. Figure \ref{fig:Corner_plot_filterbank_randomization} shows the 3D distribution and correlation of these three critical parameters among the randomized candidates. To quantify the joint probability of our BLPSR candidate's parameters, we constructed a joint probability density using a Gaussian kernel density estimator (KDE), \texttt{scipy.stats.gaussian\_kde}. The KDE is given by
\begin{equation}
\hat{f}(\vec{x}) = \frac{1}{n h^d} \sum_{i=1}^{n} K\!\left(\frac{\vec{x}-\vec{x}_i}{h}\right),
\end{equation}
\begin{equation}
\quad \text{with } K(\vec{z}) = \frac{1}{(2\pi)^{d/2}} \exp\!\left(-\frac{1}{2}\|\vec{z}\|^2\right),
\end{equation}
where \(n\) is the number of candidate samples, \(d\) is the number of parameters, and \(h\) is the smoothing bandwidth. 
We follow Scott’s normal–reference rule \citep{Scott1992Multivariate}, which gives the asymptotic mean integrated squared error–optimal bandwidth for a $d$-variate KDE with a Gaussian kernel: 
\(h_j = c_d\,\hat{\sigma}_j\,n^{-1/(d+4)}\), where \(\hat{\sigma}_j\) is the sample standard deviation of parameter \(j\) and \(c_d=(4/(d+2))^{1/(d+4)}\). 
In our implementation using \texttt{scipy.stats.gaussian\_kde}, this scalar bandwidth factor \(n^{-1/(d+4)}\) is applied in combination with the full sample covariance matrix of the parameters, which automatically accounts for differences in scale and correlations between dimensions \citep[see][]{Virtanen2020SciPy}. This approach ensures that the KDE reflects the underlying data density. The resulting KDE thus provides a continuous, smooth estimate of the probability density function in the joint parameter space spanned by coherent power, \(D_{\rm broad}^{\rm KS}\), and \(D_{\rm time}^{\rm KS}\).

As is evident from Figure \ref{fig:Corner_plot_filterbank_randomization}, BLPSR lies at the extreme tail of all three parameter distributions. To thoroughly explore the tail of this multidimensional distribution, we implemented a simple random-walk Markov Chain Monte Carlo (MCMC) sampler based on the Metropolis--Hastings algorithm. Starting from the BLPSR candidate's parameter vector, \(\vec{x}_{\mathrm{det}}\), the MCMC sampler generated \(10^8\) samples, \(\{\vec{x}_j\}_{j=1}^{N_{\mathrm{MCMC}}}\), from the KDE. For each sample, the KDE density \(\hat{f}(\vec{x}_j)\) was computed, and the joint p-value was then estimated as
\begin{equation}
  P_{\mathrm{joint}} = \frac{1}{N_{\mathrm{MCMC}}} \sum_{j=1}^{N_{\mathrm{MCMC}}} \mathbf{1}\Bigl(\hat{f}(\vec{x}_j) \le \hat{f}(\vec{x}_{\mathrm{det}})\Bigr),
\end{equation}
where \(\mathbf{1}(\cdot)\) denotes the indicator function, which returns 1 if the condition is satisfied and 0 otherwise. In this case, it counts the fraction of MCMC samples whose estimated KDE density is less than or equal to that of BLPSR \(\vec{x}_{\mathrm{det}}\), thereby providing a quantifiable estimate of the joint tail probability of observing a candidate as rare or rarer than BLPSR among noise-generated candidates.  

This MCMC-based resampling method allowed for a high-resolution estimation of the tail probability within the joint parameter space of coherent power, \(D_{\rm broad}^{\rm KS}\), and \(D_{\rm time}^{\rm KS}\). We find a resulting joint p-value of \(P_{\mathrm{joint}} < 5\times 10^{-7}\), indicating that BLPSR's overall properties are extremely rare in the multidimensional space. By comparison, a joint p-value computed using only the KS-statistics yielded a less significant \( P_{\mathrm{joint}} < 8\times10^{-3}\). This outcome is expected at the very low folded profile significance of BLPSR \(\sim 3\sigma\), which gives rise to weaker sub-band and sub-integration profiles. For completeness, we also applied the same analysis to the 727 raw candidates detected in the original downsampled 1,024-channel frequency-scrunched \texttt{filterbank} file. These candidates showed similarly extremely low \(P_{\mathrm{joint}}\) probabilities; however, because the raw candidates are not completely independent as stated earlier, we do not consider them for our statistical interpretation. 

From this randomization test, there remains approximately a one-in-a-million probability of finding a BLPSR-like candidate with similar properties purely from noise. It should be noted that, in order to maximize the number of candidates in our analysis, no sifting was applied during the randomization test. However, had we restricted the analysis to only candidates that did pass our duplicate rejection criteria--as BLPSR did--the likelihood of finding such a candidate from noise alone would be even lower. While these odds are noteworthy, they must be interpreted within the broader context of the
large number of trials performed across five dimensions --- DM--f--N$_{h}$--z$_{\rm max}$--w$_{\rm max}$ --- as mentioned in Section \ref{sec:periodicity_searches}. In other words, although BLPSR may indeed be a real pulsar, current evidence remains insufficient to support that claim with high statistical significance.  

\begin{deluxetable*}{lccccccc}[t]
\tablecolumns{10}
\tablewidth{0pt}
    \tablecaption{Comparison of This Survey with Previous Deep Pulsar Searches in the GC 
    \label{tab:past_surveys}}
\tablehead{
    \colhead{Survey} & \colhead{Telescope} & \colhead{$\nu_{c}$} &  \colhead{$\Delta\nu$} & \colhead{$t_{\rm int}$} & \colhead{$L_{\rm min}(\nu_c)$\tablenotemark{a}} & \colhead{$L_{\rm min}(10\,\mathrm{GHz})$\tablenotemark{a,b}} & \colhead{\# detected} \\ 
    & & (GHz) & (GHz) & (hr) & (mJy\,kpc$^{2}$) & (mJy\,kpc$^{2}$)& }
    \startdata
    \hline
    \citet{Johnston2006} & Parkes & 3.1 & 0.576 & 1.2 & 66.9 & $\approx$\,7 & 2 \\
    \citet{Johnston2006} & Parkes & 8.4 & 0.864 & 1.2 & 13.4 & $\approx$\,10 & 0 \\
    \citet{Deneva2009} & GBT & 1.95 & 0.600 & 1 & 434.9 & $\approx$\,18 & 3 \\
    \citet{Macquart2010} & GBT & 14.6 & 0.800 & $<$10\tablenotemark{c} & 40 & $\approx$\,84 & 0 \\
    \citet{Eatough2021} & Effelsberg & 8.35 & 0.5 & 2.4\tablenotemark{d} & 4.68 & $\approx$\,3 & 0 \\
    \citet{Eatough2021} & Effelsberg & 18.95 & 2 & 2.4\tablenotemark{d} & 3.35 &  $\approx$\,12 & 0 \\
    \cite{Liu_2021} & ALMA & 86.3 & 2 & 5.2\tablenotemark{e} & 2.01 & $\approx$\,134 & 0 \\
    \citet{Torne2021} & IRAM 30\,m & 120 & 32 & 2.2\tablenotemark{f} & 3.95 & $\approx$\,502 & 0 \\
    \citet{Suresh2022}\tablenotemark{g} & GBT & 6.1 & 3.0 & 0.5 & 2.51 & $\approx$\,1 & 0 \\
    \hline
    This survey: & & & & & & \\
    Short pointings & GBT & 9.37 & 3.69 & 5-min & 1.60 & 1.60 & 0 \\
    Long pointings E1 & GBT & 9.37 & 3.75 & 1 & 0.44 & 0.44 & 1 cand. \\
    Long pointings E2 & GBT & 9.94 & 3.75 & 1 & 0.37 & 0.37 & 0 \\
    Long pointings E2 & GBT & 9.94 & 4.88 & 2 & 0.26 & 0.26 & 0 
    \enddata
    \tablenotemark{a} Using $d_{\rm GC}=8.18\,\mathrm{kpc}$ \citep{Gravity_Collab_2019}.
    \tablenotetext{b}{\(L_{\rm min} \propto \nu^\alpha\) is invoked, where \(\alpha = -1.95\) is the mean spectral index of MSPs \citep{Torne2021}. These estimates do not account for scattering differences at 10\,GHz.}
    \tablenotetext{c}{Maximum integration of a combined scan was 10\,hr.}
    \tablenotetext{d}{Maximum scan duration \citep{Eatough2021}.}
    \tablenotetext{e}{Composed of individual scans of $\approx$\,3--6\,min each that were coherently connected.}
    \tablenotetext{f}{Mean \(t_{\rm int}\) calculated from the 28 observations in \citet{Torne2021}.}
    \tablenotetext{g}{Recalculated luminosity from that derived in \cite{Suresh_2021}. We apply our median values of $\alpha$ and $\delta$ for MSPs and CPs for greater consistency in comparing it to this survey, and use their \((S/N)_{\rm min}=6\).}
\end{deluxetable*}

\section{Discussion}
\label{sec:discussion}
\subsection{Known GC Region Pulsars}
\label{sec:known_pulsars}
Currently, six pulsars --- excluding the GC magnetar --- are known within 1$^{\circ}$ of Sgr A*. Of these, three (PSR\,J1746$-$2850I, PSR\,J1746$-$2850II, and PSR\,J1745$-$2910) lie within 12\,\arcmin\, ($\sim$\,30\,pc in projection) of the SMBH \citep{Deneva2009}. Three additional pulsars are located within 18\,\arcmin\, ($\sim$\,40\,pc in projection): PSR\,J1745$-$2912 and PSR\,J1746$-$2856 \citep{Johnston2006}, and PSR\,J1744$-$2946 \citep{Lower_2024}. The latter is the fastest ($P = 8.4$\,ms) MSP detected near the GC to date. However, its low DM ($\sim 674$\,pc\,cm\(^{-3}\)) implies that it resides in the foreground along the line-of-sight and is not physically close to the GC. While these discoveries support the presence of a large pulsar population near the GC, none lie close enough (within a parsec) to the SMBH to probe its gravitational field \citep{Liu_2012}. Furthermore, all known pulsars in the region fall outside the primary beam of our survey and were therefore not detected in our observations. In the following subsection, we detail our sensitivity limits and compare them to other previous surveys conducted toward the GC.  

\subsection{Highest Sensitivity to GC Pulsars}
\label{sec:survey_sensitivity}
\begin{figure*}
\centering
    \includegraphics[width=0.8\textwidth,]{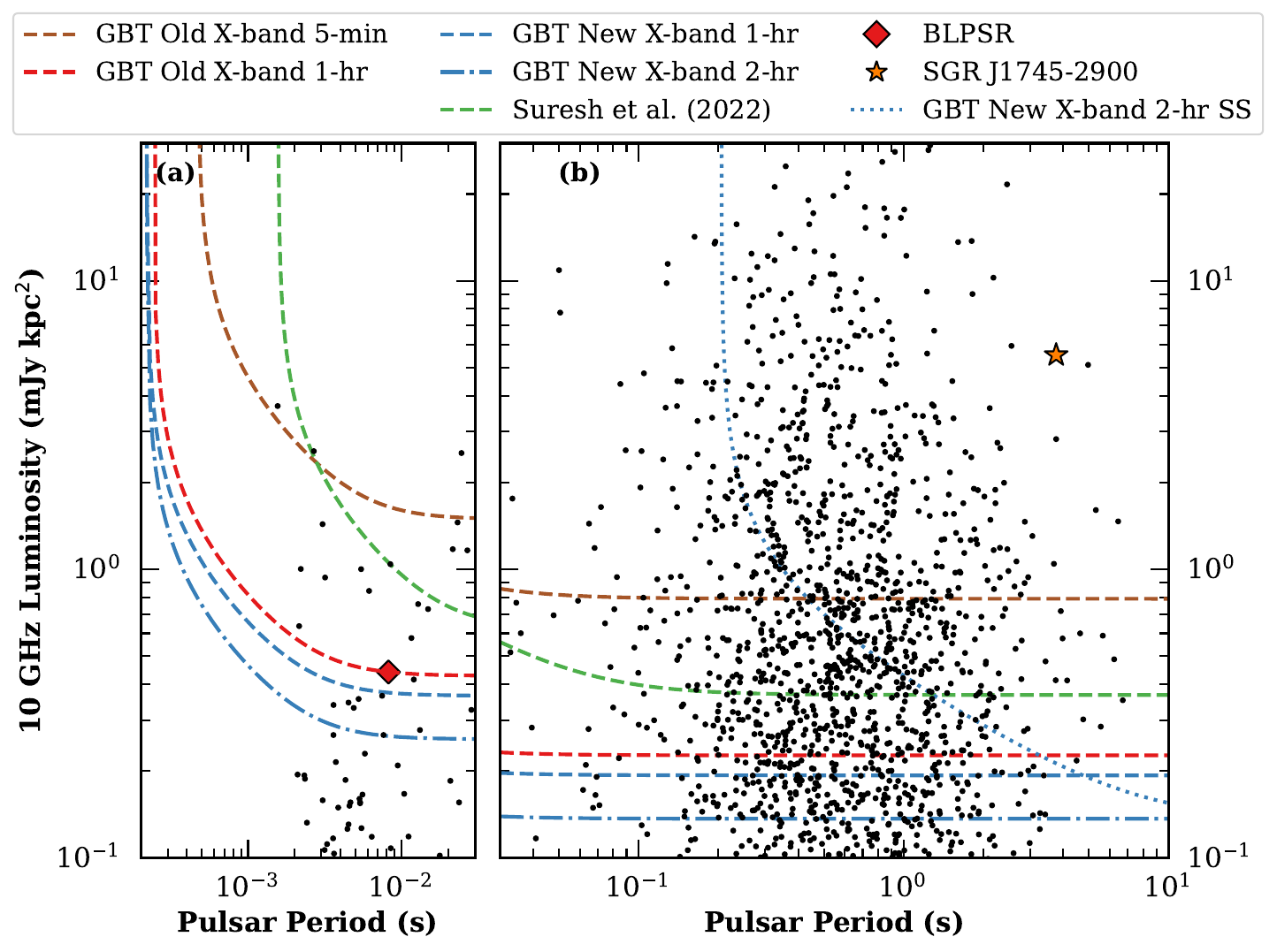}	
    \caption{GBT 3$\sigma$ sensitivity curves at X-band for our 5-min, 1-hr, and 2-hr integrations. Black dots indicate known field pulsars from the ATNF catalogue, plotted in period--pseudo-luminosity space. Their flux densities have been scaled to 10\,GHz assuming a power-law spectra with spectral indices of $\alpha$ = -1.76 for CPs and $\alpha$ = -1.95 for MSPs. The figure is split into two panels at $P_0$ = 30\,ms to separate (a) MSPs ($P_0 < 30$\,ms) and (b) CPs ($P_0 > 30$\,ms), with different duty cycles and spectral indices used for each. The brown dashed line corresponds to the 5-min integrations used for all 37 short pointings. The red dashed line represents the 1-hr integration used for Epoch 1 with the old X-band receiver (7.50--11.25\, GHz). The blue dashed-dotted line denotes the 2-hr integration used for Epoch 2, and the blue dashed line shows their 1-hr segments with the new X-band receiver (7.50--12.38\, GHz). For comparison, the green dashed line shows the predecessor BL-GC survey at C-band (4--8\,GHz) conducted with the GBT \citep{Suresh2022}. All curves assume weak scattering toward the GC. However, under strong scattering conditions, even our most sensitive observation (Epoch 2; 2-hr integrations) would only detect pulsars with $P_0$ $\gtrsim$ 0.2\,s (dotted blue line). The orange star denotes the lowest flux density recorded for the GC magnetar J1745$-$2900, although it has been observed to fluctuate over the years, detected as high as $\sim 440$\,mJy \citep{Eatough_2013b,Yan_2015, Pearlman_2018, Suresh2022}. The red diamond shows our BLPSR candidate. We note that we chose a cutoff of 0.1\,mJy\,kpc\(^2\) for our (\(L_{\rm min}(\mathrm{10\,GHz})\)) axis to zoom into our sensitivity curves.
    }
    \label{fig:Xband_sensitivity}
\end{figure*}

We estimated our survey's pulsar detection sensitivity using 2,167 known field pulsars from the Australian National Telescope Facility (ATNF) catalog 2.6.3 \citep{Manchester_2005}\footnote{https://www.atnf.csiro.au/research/pulsar/psrcat/}.
We derived their 1.4\,GHz luminosity using $S \times d^2$, where $S$ and $d$ are the catalogue parameters \texttt{S1400} and \texttt{DIST\_A}, respectively, corresponding to the mean flux density at 1.4\,GHz (in mJy) and the distance based on an dependent distance estimate. \texttt{DIST\_DM}, the distance based on the YMW16 electron density model, was used where \texttt{DIST\_A} was not available. An additional 2,176 pulsars were excluded from this analysis due to missing relevant parameter values. To compare these with our X-band observations at $\nu_{c} \approx 10$\,GHz, we scaled their luminosities using \(L_{1.4} = L_{\nu_{c}} \left({1.4}/{\nu_{c}}\right)^{\alpha}\), where $\alpha$ is the spectral index. We adopt \(\alpha=-1.76\) for CPs and \(\alpha=-1.95\) for MSPs, where MSPs are defined as those with $P<30$\,ms \citep{Torne2021}. The black points in Figure \ref{fig:Xband_sensitivity} show the distribution of these known pulsars in period--pseudo-luminosity space.  

We now define \(L_{\nu_{c}} = S_{\rm min}\, d_{GC}^2\), where $S_{\rm min}$ represents the minimum detectable flux density, and $d_{GC}$ the distance to the GC (8.18\,kpc; \citealt{Gravity_Collab_2019}). $S_{\rm min}$ can be expressed by the radiometer equation \citep{Lorimer_2004}
\begin{equation}
S_{\rm min} = \frac{S/N_{\rm min} \; T_{\rm sys}^{\rm GC}}{G \; \sqrt{n_{p} \; t_{\rm int} \; \Delta{\nu}}} \; \sqrt{\frac{\delta}{1-\delta}},
\label{eq:Smin}
\end{equation}
where \(T_{\rm sys}^{\rm GC}\) is the net system temperature toward the GC \citep{Oneil_2002,Suresh_2021}. \(T_{\rm sys}^{GC}\) is defined by 
\begin{equation}
\begin{split}
T_{\mathrm{\rm sys}}^{\mathrm{GC}}(\nu_{c}, t) =\; & T_{R_g}(\nu_{c}) 
+\ \bigl(T_{\mathrm{GC}}(\nu_{c}) + T_{\mathrm{CMB}}\bigr)\, e^{-A(t)\,\tau_{\nu_{c}}} \\
&+\ T_{\mathrm{atm}}\,\Bigl(1 - e^{-A(t)\,\tau_{\nu_{c}}}\Bigr),
\label{eq:GCsys}
\end{split}
\end{equation}
where $T_{\mathrm{GC}}(\nu_{c}) \approx 568\,\mathrm{K}\,({\nu_{c}}/1\,\mathrm{GHz})^{-1.13}$ is the background continuum temperature in the direction of the GC \citep{Rajwade_2017, Gajjar_2021}. $T_{\mathrm{CMB}} \approx 2.73\,\mathrm{K}$ is the isotropic cosmic microwave background (CMB) temperature \citep{Fixsen_2009}, and $T_{\mathrm{atm}} \approx 269\,\mathrm{K}$ is the estimated atmospheric temperature at $\nu_{c}$\,=\,10\,GHz\footnote{\url{http://www.greenbankobservatory.org/~rmaddale/WeatherGFS3/tatm.html}}. The contribution of the atmosphere to \(T_{\rm sys}^{\rm GC}\) at $\nu_{c}$\,=\,10\,GHz at our observed elevations is significantly lower, as it is modulated by the opacity term \(1 - e^{-A(t)\,\tau_{\nu_{c}}} \). \(\tau_{\nu_{c}} \approx 10^{-4}\,\left(80 \;+\; 1.25\, e^{\sqrt{\frac{\nu_{c}}{1\,\mathrm{GHz}}}}\right)\) represents the zenith atmospheric opacity\footnote{\label{fn:GBTidlcal}\url{https://www.gb.nrao.edu/GBT/DA/gbtidl/gbtidl_calibration.pdf}} and $A(t)=1/\sin\!\bigl(\theta_{\mathrm{GC}}(t)\bigr)$ measures the airmass at the elevation $\theta_{\mathrm{GC}}(t)$ of the GC. For brevity, we take the mean $\theta_{\mathrm{GC}}(t)$ for all observations at each epoch and find $\theta_{\mathrm{GC}}(t) = 19.7$\textdegree\, and 18.7\textdegree\, for Epoch 1 and 2, respectively. Using these elevation angles and a zenith opacity of $\tau_{\nu_c} \approx 0.011$ at 10\,GHz, we compute atmospheric attenuation factors of $e^{-A(t)\,\tau_{\nu_c}} \approx 0.968$ and $0.966$ for Epochs 1 and 2, respectively. $T_{R_g}(\nu_{c})$ denotes the receiver temperature of 17\,K for $\nu_{c} = 10$\,GHz (see the GBT Proposer's Guide\footnote{\url{http://www.gb.nrao.edu/scienceDocs/GBTog.pdf}}). We calculate a \(T_{\rm sys}^{\rm GC}\) of 72 and 70\,K for epoch 1 and 2, respectively. The GC and CMB terms are modulated by the frequency- and elevation-dependent atmospheric opacity (e.g., GBT Memo \#16, \#19, and \#302). Rear spillover contributes negligibly to the overall $T_{\mathrm{GC}}(\nu_{c})$ at $\nu_{c}$\,=\,10\,GHz for the GBT due to its offset Gregorian design and low spillover efficiency ($\eta_l$ = 0.99, see GBTIDL Calibration Guide)\footref{fn:GBTidlcal}, and is not included in Equation \ref{eq:GCsys}.

We use \((S/N)_{\rm min}=3\),
\(G=2\,\rm{K/J}\,\) for the telescope gain, and \(n_p=2\) for the number of polarizations. \(t_{\rm int}\) is the integration time and \(\Delta\nu\) is the frequency bandwidth, which varies for the observations (see Section \ref{sec:observations_description} for details). \(\delta = {W_{\rm eff}/}{P}\) is the pulsar duty cycle, which is estimated using the ratio \(W_{50}/P\), where \(W_{50}\) is the FWHM obtained from the ATNF catalogue. This yields median duty cycles of 2.58\% for CPs and 8.63\% for MSPs. The effective pulse width takes into account the effects of scattering, dispersion, and instrumentation, and is given by
\begin{equation}
W_{eff} =  \sqrt{W_{\rm int}^2 + \tau_{s}^2 + \Delta t_{\rm DM}^2 + \delta t^2 },
\label{eq:Weff}
\end{equation}
where \(W_{\rm int}=\delta P\) is the intrinsic pulse width, \(\tau_s\) is the scattering timescale, \(\Delta t_{\rm DM}\) is the intra-channel dispersion smearing \citep{Clarke_2013}, and \(\delta t\) is the time resolution of the digitized data \citep{Manchester_1996}. 

We consider the weak scattering model, as observed for PSR\,J1745--2900 \citep{Spitler_2014}, \(\tau_{ws} = {1.3\,\rm }{\nu^{-4}_{\rm GHz}}\)\,sec. Although earlier studies such as \citet{Macquart_2015} have explored the strong-scattering regime using a scattering screen located $\approx$130\,pc from the GC, they estimate that the optimal search frequency for detecting MSPs under such conditions is $\approx$25\,GHz, which is beyond the range of our survey.

The intra-channel smearing is calculated by
\begin{equation}
\Delta t_{\rm DM} = \frac{2\, DM\, \Delta\nu_{\text{sub-band}}}{\kappa\, \nu_{c}^3},
\label{eq:channel_smear}
\end{equation}
where \(\Delta\nu_{\text{sub-band}}\) is the sub-band  bandwidth, \(\kappa = 2.41\times10^{-16}\,\rm pc\,cm^{-3}\,s\), and \(DM=1778\,\rm pc\,cm^{-3}\) (for PSR\,J1745--2900) is chosen. Because smearing within each of the 4096 sub-bands is negligible (see Section \ref{sect:preprocessing}), we are sensitive to the fastest MSPs and their harmonics.

Our sensitivity measurements align with pulsar detection estimates toward the GC, demonstrating that our survey is capable of detecting the brightest MSPs under the assumption of weak scattering (Figure \ref{fig:Xband_sensitivity}), provided that the intrinsic GC pulsar population is comparable to that of the broader Milky Way. Our long pointings achieved the highest sensitivity of any GC survey to date, and the lower half of Table \ref{tab:past_surveys} summarizes the corresponding \(S_{\mathrm{\rm min}}\) and \(L_{\mathrm{\rm min}}\) values for MSPs (assuming $P_0\approx10$\,ms), which are the primary focus of this survey. For the BLPSR candidate, we assume the median MSP duty cycle for consistency ($\delta$=8.63\%), where its $\approx$\,3$\sigma$ detection corresponds to a flux density of $S_{\rm 10\, GHz}$ $\approx$ 0.007\,mJy and a radio luminosity of $L_{\rm 10\,GHz}$ $\approx$ 0.44 mJy\,kpc$^{2}$. For our survey threshold of \(S/N_{\rm min}=3\), we obtain corresponding $S_\mathrm{min}$ values of 0.02\,mJy (short pointings), 0.007\,mJy and 0.006\,mJy (1-hr long pointings for E1 and E2, respectively), and 0.004\,mJy (2-hr long pointings) for MSPs. Figure \ref{fig:Xband_sensitivity} also shows our sensitivity for CPs ($ > 30$\,ms), with the full 2\,hr Epoch 2 integrations reaching \(S_{\mathrm{\rm min}} \approx 0.002\,\mathrm{mJy}\) and \(L_{\mathrm{\rm min}} \approx 0.14\,\mathrm{mJy\,kpc^2}\)(assuming $P_0\approx1$\,s). In contrast, even our most sensitive observation (Epoch 2; 2-hr integrations) would only be sensitive to pulsars with $P_0$ $\gtrsim$ 0.2\,s under strong scattering conditions (dotted blue line).

Several other deep pulsar searches have also targeted the GC using various radio telescopes --- including Effelsberg, the GBT, and Parkes --- covering frequencies up to 156\,GHz \citep{Torne2021}. Table~\ref{tab:past_surveys} summarizes the key parameters of these prior surveys (top half). For cross-survey sensitivity comparison, we report their minimum detectable luminosities, $L_{\rm min}(\nu_c)$, using $d_{GC}$ = 8.18\,kpc, and we also scale them to $\nu_{c}=10\,\mathrm{GHz}$ (\(L_{\rm min}(\mathrm{10\,GHz})\)), assuming a power-law spectral index of $\alpha$=-1.95 for MSPs \citep{Torne2021}. We do not include the \(S/N_{\rm min}\) or $S_\mathrm{min}$ from previous surveys, as these are based on different detection criteria, making direct comparison inconsistent. As seen in Table~\ref{tab:past_surveys}, out of all previous surveys, the Parkes survey at $\nu_{c}=8.4$\,GHz with a 1.2\,hr integration time \citep{Johnston2006} is the most similar to our X-band observations but suffers from a reduced bandwidth and smaller effective area, which limits its sensitivity to MSPs unless their luminosity exceeds \(13.4\,\mathrm{mJy\,kpc^{2}}\). The final column lists the number of pulsars (or candidates) detected by each survey, most of which are outlined in Section \ref{sec:known_pulsars}. 

Figure \ref{fig:Xband_sensitivity} also shows the sensitivity curve for the previous BL-GC survey at C-band (4--8\,GHz) conducted with the GBT \citep{Suresh2022}. While a \(L_{\mathrm{\rm min}}(\mathrm{10\,GHz})\) of $\approx$\,1\,mJy\,kpc\(^2\) was achieved for MSPs, the shorter integration time of 30-min and lower frequency coverage in \cite{Suresh2022} likely compromised any detections due to the complex pulsar orbital dynamics and scattering effects expected in the GC. Additionally, the coverage in the 1.4--6\,GHz range likely also impeded detection of MSPs due to their inherent faintness \citep{Rajwade_2017} and extended eclipses in binaries at low frequencies (see Section \ref{sec:orbital_params}). Our BL--GC X--band survey with the GBT represents one of the most sensitive pulsar searches ever conducted toward the GC. 

\subsection{BLPSR Follow-up}
\label{sec:blpsr_followup}
BLPSR was detected only in Scan 1.1 of Epoch 1 during our blind searches. Following its initial detection, we conducted targeted follow-up searches on the subsequent four 1-hr scans of Epoch 1. These searches employed \texttt{prepfold} with 512 sub-bands, the corresponding \texttt{mask} file, and the ephemeris derived in Section~\ref{sec:confirm_blpsr}, utilizing both the \texttt{-coarse} and \texttt{-fine} flags. We searched both with and without (\texttt{-nosearch}) constraints in \(P\)–\(\dot{P}\) and DM–\(\phi\) space, allowing for up to \(\pm4\) full pulse phase wraps of error to accommodate expected pulse phase drift over the duration of each observation. The parameter space spanned a spin frequency range of $f_{s}$\,\(\pm\,0.00\overline{1}\)\,Hz with a step size of \(4.3402\overline{7}\times10^{-6}\)\,Hz, and a spin frequency derivative range of $\dot{f_{s}}$\,\(\pm\,3.086\times10^{-7}\)\,Hz/\,s with a step size of \(2.411\times10^{-9}\)\,Hz/\,s--searching over 513 different trial values for \(f_{s}\) and \(\dot{f}_{s}\). The \texttt{-coarse} flag examined 385 DMs (step size 3.47\,pc\,cm\(^{-3}\)), while the \texttt{-fine} flag covered 129 DMs (step size 1.74\,pc\,cm\(^{-3}\)), constituting an extensive parameter search. We also blindly searched scans 1.1 and 1.2 coherently but did not recover the signal, likely because the BLPSR had already begun fading during the latter half of scan 1.1 (see Section \ref{sec:segmentation_test}).

We applied the same targeted searches described above to the 2-hr scans from Epoch 2 and further subdivided them into 1-hr segments to accommodate possible variations in pulsar acceleration due to binary motion. This expanded our search parameter space, improving sensitivity to a wider range of orbital periods and accelerations (see Figure \ref{fig:contour_lines}), since some pulsars can be eclipsed or exhibit varying acceleration over portions of their orbit (see Section \ref{sec:orbital_params} for more details). Despite these thorough targeted searches, BLPSR was not re-detected. 

\subsection{Is BLPSR a real pulsar?}
\label{sec:real_unreal}
We have presented and discussed the observed properties of our BLPSR candidate and conducted a series of extensive tests to evaluate its significance, along with detailed follow-up observations, all of which resulted in non-detections. To summarize, BLPSR was identified as a top candidate in our acceleration search, exhibiting the highest coherent power and Gaussian significance  (see Sections \ref{sec:long_pointings} and \ref{sec:blpsr_cand}). It displayed pulsar-like properties across both time and frequency, and showed a relatively weak but noticeable peak in the S/N--DM plane (Figure~\ref{fig:MSP_Cand_origdetect}). However, it was not detected in any subsequent observations (see Section \ref{sec:blpsr_followup}).

In Section~\ref{sec:data_quality_test}, we examined the quality of the immediate follow-up data (Scans 1.2--1.5) and found no evidence of sensitivity loss or instrumental issues that could plausibly explain the non-detection. In Section~\ref{sec:peak_flux_with_dm}, we attempted to verify the expected variation of peak S/N with DM but were unable to draw a robust conclusion due to the low signal strength. When visualizing the candidate across both time and frequency, and measuring the KS-statistics for broadbandness and time-persistency (\(D_{\rm broad}^{\rm KS}\) and \(D_{\rm time}^{\rm KS}\)), we found evidence of significant structure, as would be expected for a true pulsar. 
 
To evaluate whether such features could arise from noise alone, we carried out a novel randomization test (Section~\ref{sec:randomtests}). When considering only the KS-statistics, we found that roughly one in a thousand noise-originating candidates could exhibit similar values which is likely due to a lower S/N folded profile. However, when factoring in BLPSR's unusually high coherent power, the likelihood of a noise fluctuation producing such a candidate dropped to approximately one in a million. Despite these results, we refrain from concluding that BLPSR is a real pulsar due to several key limitations. Its folded profile, while coherent, was relatively narrow and close to our detection threshold. More importantly, BLPSR was not re-detected in either the immediate follow-up scans or in subsequent observations conducted nearly a year later using a different, more sensitive receiver (see Section \ref{sec:blpsr_followup}). 

In light of these factors --- and given the extraordinary implications of detecting a pulsar near Sgr A* --- we remain highly skeptical of BLPSR and emphasize that a much stronger burden of proof is required before asserting its astrophysical origin. Nonetheless, in the following subsections, we explore the possible orbital constraints and astrophysical scenarios that could explain its non-detection in subsequent follow-up observations. In summary, we remain dubious of the astrophysical nature of BLPSR, but recognize the possibility that it may be a real, yet highly variable or intermittent pulsar ---  potentially influenced by extreme propagation effects, binary motion, or intrinsic emission variability. Given its anomalous nature, BLPSR merits continued scrutiny and targeted re-observation efforts, though we withhold any definitive classification pending further evidence.

\subsection{BLPSR Orbital Constraints and Astrophysical Scenarios for Non-Detections}
\label{sec:orbital_params}

In Section~\ref{sec:randomtests}, we explored the possibility that BLPSR may have arisen from noise fluctuations, which could explain its singular detection and absence in subsequent scans and epochs. The signal also appears to diminish in significance toward the end of the corresponding scan, prompting us to consider several astrophysical factors that could complicate re-detection, particularly for a pulsar in a binary system near the GC, such as complex orbital dynamics and interactions within the dense stellar environment. The strong variability in scattering screen properties and the poorly understood high-electron density environment in the GC is caused by fluctuations in the interstellar medium \citep{Lazio_1998, Dexter2017, Abbate2023}. These variations are dependent on the pulsar's orbital position, as well as the uniformity (uniform versus patchy), movement rate, and distance of the intervening scattering screen \citep{Lazio_1998,Dexter2017}. Such complex propagation effects could cause normally persistent pulsed signals into being highly intermittent, further complicating pulsar detection efforts near the GC. If BLPSR is indeed a genuine astrophysical signal, these factors could naturally lead to intermittent visibility or short-term detectability windows, complicating confirmation efforts from our follow-up observations. 

\begin{figure*}
  \centering 
  \includegraphics[width=0.49\linewidth]{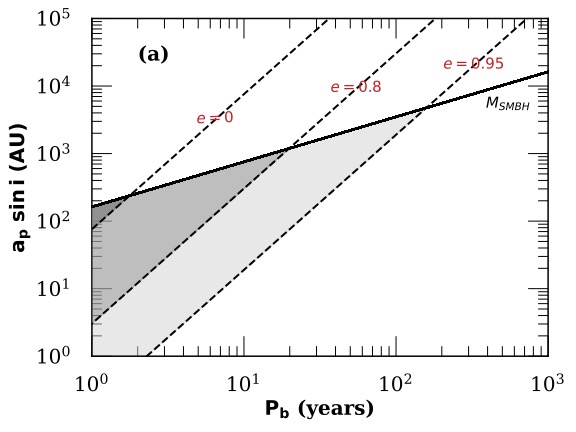}
  \includegraphics[width=0.49\linewidth]{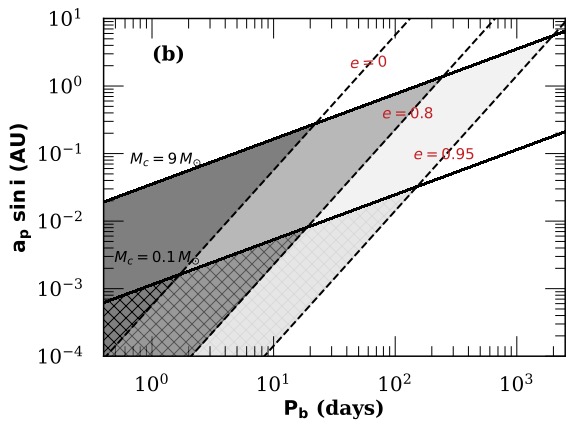}
  \caption{Allowed parameter space for the orbital period $P_b$ and projected semi-major axis $a_p\,\sin{i}$ of BLPSR, assuming it is orbiting \textbf{(a)} the SMBH Sgr A* ($M_{\rm Sgr\,A*} = 4.30 \times 10^{6}\,M_{\odot}$), or \textbf{(b)} a stellar-mass companion of either $M_c = 9\,M_{\odot}$ or $M_c = 0.1\,M_{\odot}$. The solid black lines represent Kepler's third law for the true semi-major axis $a_p$, which is the upper limit on $a_p\,\sin{i}$, while the dashed lines are the lower limits on $a_p\,\sin{i}$ for different orbital eccentricities ($e$ = 0, 0.8, 0.95). The shaded regions, bounded by a solid and dashed line, mark the allowed parameter space for each eccentricity, including the full ranges of $A_T$ and $\omega$. The intersection point of a solid and dashed line marks the overall maximum $P_b$ and $a_p$ permitted for each eccentricity. In (b) the cross-hatched region represents the allowed parameter space for $M_c = 0.1\,M_{\odot}$, which is nested within that of $M_c= 9\,M_{\odot}$.}
  \label{fig:Pb_vs_ap}
\end{figure*}

Even our single detection of the candidate pulsar BLPSR constrains its orbital parameters significantly. The candidate was blindly detected at $z=2.75$, corresponding to line-of-sight acceleration $a_{l} = -0.45\, \text{m}\,{\text{s}^{-2}}$ during Scan 1.1, as noted in Section \ref{sec:blpsr_cand}. This tentative non-zero acceleration may indicate that it is in a binary system.  We employ Equation \ref{eqn:bacghi_al} for $a_{l}$ as a function of orbital parameters \citep{Bagchi_2013, Liu_2021}, which determines the range of $a_p$ and $P_b$ that could produce the observed $a_l=-0.45$~m~s$^{-2}$ at some point along the orbit. For a given value of $e$, the angles that maximize the right side of Equation~\ref{eqn:bacghi_al}  are $A_{T}=0$, and $\omega=\pi/2$. By choosing these angles, Equation~\ref{eqn:bacghi_al} yields the minimum possible value of the projected semi-major axis $a_p\,\sin{i}$ as a function of $P_b$ for a particular $e$, illustrated by the dashed lines in Figure~\ref{fig:Pb_vs_ap}. Then, for a particular assumed mass, $a_p$ as a function of $P_b$ is given by Kepler's third law, which represents the maximum possible value of $a_p\,\sin{i}$ (solid line in Figure~\ref{fig:Pb_vs_ap}).  Finally, the intersection point of the solid and dashed line marks the overall maximum of $P_b$ and $a_p$ for the chosen $M$ and $e$.

We explore these orbital constraints for a pulsar mass of $2.14~M_{\odot}$, three different companion masses, and eccentricities $e=0$, 0.8, and 0.95. First, we consider the SMBH Sgr A* ($M_{\rm Sgr\,A*} = 4.30 \times 10^{6}\, M_{{\odot}}$) as the companion. Figure \ref{fig:Pb_vs_ap}a defines the allowed range of $a_{p}\,\sin{i}$ and $P_b$, which for each eccentricity is the shaded region between the corresponding dashed line and the solid line. At $e=0$, the apex of the darkest shaded triangle indicates a maximum $a_p = 237.89$ AU, corresponding to $P_b=1.77$ years. For $e=0.85$, the maximum $a_p=1189.78$ AU at $P_b=19.80$ years, while for $e=0.95$, the maximum $a_p=4759.22$ AU at $P_b=\,158.39$ years.

Similarly, Figure \ref{fig:Pb_vs_ap}b presents the same eccentricity cases, but for a stellar-mass BH companion ($M_c=9$\,M$_{{\odot}}$) and a representative redback or white dwarf companion ($M_c=0.1\,M_{\odot}$). Given the significantly lower companion masses compared to Sgr A*, the corresponding orbital periods $P_{b}$ are much shorter --- on the order of days. For the stellar-mass BH ($M_c=9$\,M$_{\odot}$), the maximum orbital period for $e=0$ is 22.0 days, which corresponds to $a_p=0.28$\,AU. For $e=0.8$, the maximum $P_b$ increases to 247.22 days with $a_p$\,=\,1.39\,AU, while for $e=0.95$, it extends to 1978.0 days with $a_p$\,=\,5.56\,AU. The allowed parameter space for $M_c$ = 0.1$M_{\odot}$ is nested within that of $M_c$ =9$M_{\odot}$, but with different maximum values, as indicated by the cross-hatched regions. For $M_c=0.1\,M_{\odot}$, the maximum $P_b$ for $e=0$ is 1.6 days, which corresponds to $a_p$\,=\,0.0016\,AU. For $e=0.85$, the maximum $P_b$ is 18.80 days with $a_p=0.008$\,AU, and for $e=0.95$, $P_b$ reaches 147.61 days with $a_p=0.03$ AU. 

Redback binaries are known to produce broad eclipses throughout their orbit, particularly around superior conjunction, when the companion star is between us and the pulsar. Although their orbital periods are shorter compared to more massive companions, they may experience extensive eclipses --- up to 50\% of the orbit --- across all phases, particularly at lower frequencies \citep{Archibald_2009}, although massive young companions have also been observed to be eclipsed, e.g., PSR\,B1259-63 binary with an OB star \citep{Chernyakoba_2021}. This is due to intrabinary gas driven off the companion by pulsar wind irradiation, which can enshroud the pulsar in a highly gaseous environment. Consequently, such pulsars are only detected when the system is in a favorable orbital phase, specifically when the companion is near inferior conjunction. Although these eclipses and the associated DM variations hinder detections at lower frequencies (1--2\,GHz), their behavior at higher frequencies, such as X-band, remains less certain. If BLPSR resides in such an eclipsing binary system, its non-detection in subsequent observations may be due to orbital phase-dependent occultation, with the initial detection having occurred during a more favorable orbital phase.

\subsubsection{Intrinsic Emission Variability}
Beyond external factors, several pulsars are known to exhibit pulsar nulling \citep{Gajjar_2012_Nulling}, an extreme form of emission suppression characterized by intermittent behavior where emission appears to intermittently switch off for several days \citep{Kramer_2006_Intermittent_PSRs}. Additionally, intermittency can significantly affect survey completeness; for example, if a pulsar is active for only 10\% of the time and its nulling periods exceed the duration of a typical survey integration, there is only a 1\% chance of detecting it both during the survey and in a follow-up confirmation observation \citep{Eatough_2013}. Although this intermittency phenomenon has been largely observed in canonical long-period isolated pulsars, it cannot be completely ruled out for MSPs. It is possible that BLPSR exhibits a similar intrinsic magnetospheric state-switching behavior, hindering its re-detection in our later follow-up observations.

\subsection{The ``missing pulsar'' problem}
\label{sec:nondetections_complexity}
Despite conducting one of the most sensitive and extensive pulsar surveys towards the GC to date, we did not conclusively detect any new pulsars, apart from the marginal BLPSR candidate, whose astrophysical origin remains uncertain. Under the assumption that the GC pulsar population resembles that observed in the rest of the Milky Way \citep{Macquart2010, Macquart_2015}, our survey --- particularly the longer scans --- should have been sensitive to a significant fraction of CPs ($\approx$ 47\%) and the brightest MSPs ($\approx$ 9\%) under weak scattering conditions and when accounting for a typical beaming fraction of $f_b = 0.7$ \citep{Kramer_1998} (see Section \ref{sec:survey_sensitivity}). Thus, the absence of such detections further deepens the longstanding ``missing pulsar'' problem in the GC \citep{Cordes_1997}. 

If BLPSR is not a pulsar, the implications are more severe: even the most luminous MSPs, expected to lie above our sensitivity thresholds, remain undetected. This suggests that the mechanisms potentially contributing to the marginal detection of BLPSR (see Section \ref{sec:orbital_params}), such as highly eccentric orbits, variable scattering from a patchy screen, or intrinsic emission variability (e.g., nulling or intermittent emission states), may be widespread among the GC pulsar population. Although our searches probed a broad range of accelerations and orbital parameters, pulsars may still evade detection due to the unique challenges posed by the GC environment such as frequent dynamical interactions due to the high stellar density. As MSPs migrate inward via dynamical friction, close encounters with compact objects, including NSs and stellar- or intermediate-mass BHs, can disrupt pulsar binaries and eject them into unbound hyperbolic orbits \citep{Jiale_2022}. Isolated pulsars can also be ejected. Additionally, NSs may be dynamically ejected from the few central parsecs due to either high natal kick velocities \citep{Boodram2022} or gravitational interactions with massive stars within the central $\sim 1$\,pc \citep{Abbate2018}. Testing whether the innermost region of the GC is truly devoid of pulsars would require deeper surveys covering a broader area; however, detailed calculations of the likelihoods associated with these alternative non-detection scenarios are beyond the scope of the present work.

While the GC shares its extreme stellar density with globular clusters --- where MSPs are abundant \citep{Schodel_2018} --- its environment differs in key ways. It hosts a SMBH, has much higher velocity dispersion, and contains a younger and more massive stellar population. These factors likely lead to more frequent, violent dynamical interactions, increasing the likelihood of binary disruption and NS ejection rather than MSP retention and recycling. Additionally, as mentioned in Section \ref{sec:orbital_params}, a pulsar's beam emission could precess away during its orbit, implying that caution must be used when it comes to dismissing possible pulsar candidates toward the GC, such as BLPSR \citep{Macquart2010}. Thus, our null results (or marginal candidate) highlight both the observational challenges of detecting (and confirming) pulsars near Sgr A*, as well as the limitations of relying solely on deeper sensitivity. 

Given that GC pulsars may only be briefly detectable during favorable scattering or orbital phases, we advocate for future surveys to prioritize regular, high-cadence monitoring over single ultra-deep integrations. Capturing transient windows of visibility, such as those arising from refractive scintillation or specific orbital configurations that minimize eclipsing, may prove more effective than improving sensitivity alone. Sustained temporal coverage, rather than deeper integrations, may be the key to finally resolving the missing pulsar problem in the GC. If these efforts also fail, it may indicate that the weak scattering case does not apply in the GC, or that the GC pulsar population near Sgr A* is very different from that of the Galactic field and globular clusters. This region might instead be dominated by magnetars, which are thought to account for 10--50\% of NS births in the Galaxy \citep{Macquart_2015, Keane_2008}.

\section{Future and Ongoing Work}
\label{sec:future_work}
We conducted an additional 12\,hr of follow-up observations on A00 with the GBT in 2024 January--February, using 2--3-hr scans to further investigate BLPSR; these data will be presented in a forthcoming publication. Ongoing analysis includes thorough acceleration and jerk searches, as well as the splitting of scans into 1-hr segments to account for a potential long orbital period (see Section \ref{sec:orbital_params}). Future work will include computationally intensive coherent searches across subsequent scans and full epochs, as well as complementary single-pulse searches. In addition, we have recently observed 8.37\,hr of Very Large Array (VLA) time in D-configuration at X-band (synthesized beam of 7\farcs2 and $\Delta\,\nu$\,=\,4\,GHz). With a two-fold sensitivity increase and 2-hr integration times, we expect a 6.5$\sigma$ detection for BLPSR (rms noise of 3.48 $\mu$Jy/beam) if this candidate is indeed a genuine pulsar and falls within the VLA synthesized beam. While these observations will initially focus on the innermost few arcseconds of the A00 region, future VLA proposals will aim to map the surrounding area. Additionally, the BL-GC survey with the GBT is continuing to collect data, with plans to expand to Ku--W bands (12--93\,GHz) (see \cite{Gajjar_2021}). This comprehensive frequency coverage will be essential for detecting the most luminous pulsars near the GC. Such efforts will lay the groundwork for future pulsar detections with next-generation interferometers that will ultimately resolve the longstanding missing pulsar problem.

\section{Conclusions}
\label{sec:conclusions}
We have conducted one of the most sensitive pulsar surveys to date targeting the innermost ($1\farcm4$) region of the GC with the GBT at X-band. Our observations, covering both short and long integrations of Sgr A* and its surrounding region ($\sim 8 \arcmin$ diameter), reached sensitivities low enough to detect CPs and MSPs down to \(L_{\mathrm{\rm min}} \approx 0.14\,\mathrm{mJy\,kpc^2}\) and \(L_{\mathrm{\rm min}} \approx 0.26\,\mathrm{mJy\,kpc^2}\), respectively, assuming weak scattering. No pulsar candidates emerged in the short pointings, but we report a promising 8.19\,ms MSP candidate, BLPSR, at a DM of 2775\,pc\,cm\(^{-3}\) detected in the Epoch 1 long integrations during our $z_{\rm max}$\,=\,200 acceleration searches. 

BLPSR exhibited pulsar-like properties and was selected on its detection significance, high coherent power, and its temporal and spectral characteristics after folding. We examined its characteristics by evaluating its S/N flux as a function of DM, and quantifying its signal persistence across both time and frequency using statistical Kolmogorov-Smirnov tests. Randomizing the dataset and applying these KS-tests yielded a false alarm probability of $\sim$$10^{-6}$, suggesting the candidate is unlikely to be a noise artifact. However, its low S/N, transient nature, and absence in subsequent scans and follow-up observations warrant caution. Further tests and observations are necessary to conclusively confirm or reject BLPSR as a pulsar.

The absence of a large pulsar population in our survey may point to observational biases --- such as interstellar scattering, unfavorable pulsar beaming geometry, or binary-induced acceleration smearing --- or suggest that pulsars near Sgr A* are intrinsically faint or exceptionally rare. Future surveys with greater sensitivity, longer integration times, and broader orbital phase coverage, such as those enabled by the mid-frequency array of the Square Kilometer Array Phase-1 (SKA1-MID) and the full SKA \citep{Macquart_2015, Schoedel_2024_whitepaper}, will be critical to resolve any pulsars in the GC. Detecting, confirming, and timing a pulsar in a close orbit around Sgr A* remains a major goal for testing general relativity, understanding the SMBH, and probing the dense and turbulent environment at the heart of our Galaxy. Such surveys will ultimately reveal the long-hypothesized pulsar population or further deepen the missing pulsar problem in the Galactic Center. 
 
\begin{acknowledgements}
We thank the anonymous referee for their review and for providing several suggestions and clarifications that improved the manuscript. We are also grateful to Prof. D. J. Helfand for a careful reading of the manuscript, and to Scott Ransom for his prompt and helpful responses on the PRESTO GitHub page. Breakthrough Listen is managed by the Breakthrough Initiatives, sponsored by the Breakthrough Prize Foundation. The Green Bank Observatory is a facility of the National Science Foundation operated under cooperative agreement by Associated Universities, Inc.  We acknowledge use of computing resources from Columbia University's Shared Research Computing Facility project, which is supported by NIH Research Facility Improvement Grant 1G20RR030893-01, and associated funds from the New York State Empire State Development, Division of Science Technology and Innovation (NYSTAR) Contract C090171, both awarded April 15, 2010. 
\end{acknowledgements}

\facilities{GBT}
\software{PRESTO \citep{2001PhDT.......123R}, DSPSR \citep{DSPSR_vanStraten:2010hy}, PSRCHIVE \citep{PSRCHIVE_Hotan_2004}}

\bibliography{references}

@ARTICLE{Ransom_2002,
       author = {{Ransom}, Scott M. and {Eikenberry}, Stephen S. and {Middleditch}, John},
        title = "{Fourier Techniques for Very Long Astrophysical Time-Series Analysis}",
      journal = {\aj},
         year = 2002,
        month = sep,
       volume = {124},
       number = {3},
        pages = {1788-1809},
          doi = {10.1086/342285},
archivePrefix = {arXiv},
       eprint = {astro-ph/0204349},
 primaryClass = {astro-ph},
       adsurl = {https://ui.adsabs.harvard.edu/abs/2002AJ....124.1788R}
}

@ARTICLE{Gajjar_2022,
       author = {{Gajjar}, Vishal and {LeDuc}, Dominic and {Chen}, Jiani and {Siemion}, Andrew P.~V. and {Sheikh}, Sofia Z. and {Brzycki}, Bryan and {Croft}, Steve and {Czech}, Daniel and {DeBoer}, David and {DeMarines}, Julia and {Drew}, Jamie and {Isaacson}, Howard and {Lacki}, Brian C. and {Lebofsky}, Matt and {MacMahon}, David H.~E. and {Ng}, Cherry and {de Pater}, Imke and {Perez}, Karen I. and {Price}, Danny C. and {Suresh}, Akshay and {Webb}, Claire and {Worden}, S. Pete},
        title = "{Searching for Broadband Pulsed Beacons from 1883 Stars Using Neural Networks}",
      journal = {\apj},
         year = 2022,
        month = jun,
       volume = {932},
       number = {2},
          eid = {81},
        pages = {81},
          doi = {10.3847/1538-4357/ac6dd5},
archivePrefix = {arXiv},
       eprint = {2205.02964},
 primaryClass = {astro-ph.HE},
       adsurl = {https://ui.adsabs.harvard.edu/abs/2022ApJ...932...81G}
}

@article{Cafardo_2021,
doi = {10.3847/1538-4357/ac0efe},
url = {https://dx.doi.org/10.3847/1538-4357/ac0efe},
year = {2021},
month = {sep},
publisher = {The American Astronomical Society},
volume = {918},
number = {1},
pages = {30},
author = {Fabio Cafardo and Rodrigo Nemmen and (Fermi LAT Collaboration)},
title = {Fermi-LAT Observations of Sagittarius A*: Imaging Analysis},
journal = {The Astrophysical Journal}
}

@ARTICLE{Lazarus_2015,
       author = {{Lazarus}, P. and {Brazier}, A. and {Hessels}, J.~W.~T. and {Karako-Argaman}, C. and {Kaspi}, V.~M. and {Lynch}, R. and {Madsen}, E. and {Patel}, C. and {Ransom}, S.~M. and {Scholz}, P. and {Swiggum}, J. and {Zhu}, W.~W. and {Allen}, B. and {Bogdanov}, S. and {Camilo}, F. and {Cardoso}, F. and {Chatterjee}, S. and {Cordes}, J.~M. and {Crawford}, F. and {Deneva}, J.~S. and {Ferdman}, R. and {Freire}, P.~C.~C. and {Jenet}, F.~A. and {Knispel}, B. and {Lee}, K.~J. and {van Leeuwen}, J. and {Lorimer}, D.~R. and {Lyne}, A.~G. and {McLaughlin}, M.~A. and {Siemens}, X. and {Spitler}, L.~G. and {Stairs}, I.~H. and {Stovall}, K. and {Venkataraman}, A.},
        title = "{Arecibo Pulsar Survey Using ALFA. IV. Mock Spectrometer Data Analysis, Survey Sensitivity, and the Discovery of 40 Pulsars}",
      journal = {\apj},
         year = 2015,
        month = oct,
       volume = {812},
       number = {1},
          eid = {81},
        pages = {81},
          doi = {10.1088/0004-637X/812/1/81},
archivePrefix = {arXiv},
       eprint = {1504.02294},
 primaryClass = {astro-ph.HE},
       adsurl = {https://ui.adsabs.harvard.edu/abs/2015ApJ...812...81L}
}

@ARTICLE{Eatough2021,
       author = {{Eatough}, R.~P. and {Torne}, P. and {Desvignes}, G. and {Kramer}, M. and {Karuppusamy}, R. and {Klein}, B. and {Spitler}, L.~G. and {Lee}, K.~J. and {Champion}, D.~J. and {Liu}, K. and {Wharton}, R.~S. and {Rezzolla}, L. and {Falcke}, H.},
        title = "{Multi-epoch searches for relativistic binary pulsars and fast transients in the Galactic Centre}",
      journal = {\mnras},
         year = 2021,
        month = nov,
       volume = {507},
       number = {4},
        pages = {5053-5068},
          doi = {10.1093/mnras/stab2344},
archivePrefix = {arXiv},
       eprint = {2108.05241},
 primaryClass = {astro-ph.HE},
       adsurl = {https://ui.adsabs.harvard.edu/abs/2021MNRAS.507.5053E}
}

@ARTICLE{Torne2021,
       author = {{Torne}, P. and {Desvignes}, G. and {Eatough}, R.~P. and {Kramer}, M. and {Karuppusamy}, R. and {Liu}, K. and {Noutsos}, A. and {Wharton}, R. and {Kramer}, C. and {Navarro}, S. and {Paubert}, G. and {Sanchez}, S. and {Sanchez-Portal}, M. and {Schuster}, K.~F. and {Falcke}, H. and {Rezzolla}, L.},
        title = "{Searching for pulsars in the Galactic centre at 3 and 2 mm}",
      journal = {\aap},
         year = 2021,
        month = jun,
       volume = {650},
          eid = {A95},
        pages = {A95},
          doi = {10.1051/0004-6361/202140775},
archivePrefix = {arXiv},
       eprint = {2103.16581},
 primaryClass = {astro-ph.HE},
       adsurl = {https://ui.adsabs.harvard.edu/abs/2021A&A...650A..95T}
}

@ARTICLE{Gajjar_2021,
       author = {{Gajjar}, Vishal and {Perez}, Karen I. and {Siemion}, Andrew P.~V. and {Foster}, Griffin and {Brzycki}, Bryan and {Chatterjee}, Shami and {Chen}, Yuhong and {Cordes}, James M. and {Croft}, Steve and {Czech}, Daniel and {DeBoer}, David and {DeMarines}, Julia and {Drew}, Jamie and {Gowanlock}, Michael and {Isaacson}, Howard and {Lacki}, Brian C. and {Lebofsky}, Matt and {MacMahon}, David H.~E. and {Morrison}, Ian S. and {Ng}, Cherry and {de Pater}, Imke and {Price}, Danny C. and {Sheikh}, Sofia Z. and {Suresh}, Akshay and {Webb}, Claire and {Pete Worden}, S.},
        title = "{The Breakthrough Listen Search For Intelligent Life Near the Galactic Center. I.}",
      journal = {\aj},
         year = 2021,
        month = jul,
       volume = {162},
       number = {1},
          eid = {33},
        pages = {33},
          doi = {10.3847/1538-3881/abfd36},
archivePrefix = {arXiv},
       eprint = {2104.14148},
 primaryClass = {astro-ph.HE},
       adsurl = {https://ui.adsabs.harvard.edu/abs/2021AJ....162...33G}
}

@ARTICLE{Mori_2021,
       author = {{Mori}, Kaya and {Hailey}, Charles J. and {Schutt}, Theo Y.~E. and {Mandel}, Shifra and {Heuer}, Keri and {Grindlay}, Jonathan E. and {Hong}, Jaesub and {Ponti}, Gabriele and {Tomsick}, John A.},
        title = "{The X-Ray Binary Population in the Galactic Center Revealed through Multi-decade Observations}",
      journal = {\apj},
         year = 2021,
        month = nov,
       volume = {921},
       number = {2},
          eid = {148},
        pages = {148},
          doi = {10.3847/1538-4357/ac1da5},
archivePrefix = {arXiv},
       eprint = {2108.07312},
 primaryClass = {astro-ph.HE},
       adsurl = {https://ui.adsabs.harvard.edu/abs/2021ApJ...921..148M}
}

@INPROCEEDINGS{Oneil_2002,
       author = {{O'Neil}, K.},
        title = "{Single-Dish Calibration Techniques at Radio Wavelengths}",
    booktitle = {Single-Dish Radio Astronomy: Techniques and Applications},
         year = 2002,
       editor = {{Stanimirovic}, Snezana and {Altschuler}, Daniel and {Goldsmith}, Paul and {Salter}, Chris},
       series = {Astronomical Society of the Pacific Conference Series},
       volume = {278},
        month = dec,
        pages = {293-311},
          doi = {10.48550/arXiv.astro-ph/0203001},
archivePrefix = {arXiv},
       eprint = {astro-ph/0203001},
 primaryClass = {astro-ph},
       adsurl = {https://ui.adsabs.harvard.edu/abs/2002ASPC..278..293O}
}

@ARTICLE{Suresh_2021,
       author = {{Suresh}, Akshay and {Cordes}, James M. and {Chatterjee}, Shami and {Gajjar}, Vishal and {Perez}, Karen I. and {Siemion}, Andrew P.~V. and {Price}, Danny C.},
        title = "{4-8 GHz Spectrotemporal Emission from the Galactic Center Magnetar PSR J1745-2900}",
      journal = {\apj},
         year = 2021,
        month = nov,
       volume = {921},
       number = {2},
          eid = {101},
        pages = {101},
          doi = {10.3847/1538-4357/ac1d45},
archivePrefix = {arXiv},
       eprint = {2108.05404},
 primaryClass = {astro-ph.HE},
       adsurl = {https://ui.adsabs.harvard.edu/abs/2021ApJ...921..101S}
}

@ARTICLE{Fixsen_2009,
       author = {{Fixsen}, D.~J.},
        title = "{The Temperature of the Cosmic Microwave Background}",
      journal = {\apj},
         year = 2009,
        month = dec,
       volume = {707},
       number = {2},
        pages = {916-920},
          doi = {10.1088/0004-637X/707/2/916},
archivePrefix = {arXiv},
       eprint = {0911.1955},
 primaryClass = {astro-ph.CO},
       adsurl = {https://ui.adsabs.harvard.edu/abs/2009ApJ...707..916F}
}

@ARTICLE{Jiale_2022,
       author = {{Li}, Kaye Jiale and {Wu}, Kinwah and {Leung}, Po Kin and {Singh}, Dinesh},
        title = "{Relativistic scattering of a fast spinning neutron star by a massive black hole}",
      journal = {\mnras},
         year = 2022,
        month = apr,
       volume = {511},
       number = {3},
        pages = {3602-3617},
          doi = {10.1093/mnras/stab2925},
archivePrefix = {arXiv},
       eprint = {2110.03494},
 primaryClass = {astro-ph.HE},
       adsurl = {https://ui.adsabs.harvard.edu/abs/2022MNRAS.511.3602L}
}

@article{Yan_2015,
doi = {10.1088/0004-637X/814/1/5},
url = {https://dx.doi.org/10.1088/0004-637X/814/1/5},
year = {2015},
month = {nov},
publisher = {The American Astronomical Society},
volume = {814},
number = {1},
pages = {5},
author = {Yan, Zhen and Shen, Zhi-Qiang and Wu, Xin-Ji and Manchester, R. N. and Weltevrede, P. and Wu, Ya-Jun and Zhao, Rong-Bing and Yuan, Jian-Ping and Lee, Ke-Jia and Fan, Qing-Yuan and Hong, Xiao-Yu and Jiang, Dong-Rong and Li, Bin and Liang, Shi-Guang and Ling, Quan-Bao and Liu, Qing-Hui and Qian, Zhi-Han and Zhang, Xiu-Zhong and Zhong, Wei-Ye and Ye, Shu-Hua},
title = {SINGLE-PULSE RADIO OBSERVATIONS OF THE GALACTIC CENTER MAGNETAR PSR J1745–2900},
journal = {The Astrophysical Journal}
}

@ARTICLE{Zhao_2020,
       author = {{Zhao}, Jun-Hui and {Morris}, Mark R. and {Goss}, W.~M.},
        title = "{A Population of Compact Radio Variables and Transients in the Radio-bright Zone at the Galactic Center Observed with the Jansky Very Large Array}",
      journal = {\apj},
         year = 2020,
        month = dec,
       volume = {905},
       number = {2},
          eid = {173},
        pages = {173},
          doi = {10.3847/1538-4357/abc75e},
archivePrefix = {arXiv},
       eprint = {2011.01368},
 primaryClass = {astro-ph.HE},
       adsurl = {https://ui.adsabs.harvard.edu/abs/2020ApJ...905..173Z}
}

@ARTICLE{Gravity_Collab_2019,
       author = {{Gravity Collaboration} and {Abuter}, R. and {Amorim}, A. and
         {Baub{\"o}ck}, M. and {Berger}, J.~P. and {Bonnet}, H. and {Brand
        ner}, W. and {Cl{\'e}net}, Y. and {Coud{\'e} Du Foresto}, V. and
         {de Zeeuw}, P.~T. and {Dexter}, J. and {Duvert}, G. and {Eckart}, A. and
         {Eisenhauer}, F. and {F{\"o}rster Schreiber}, N.~M. and {Garcia}, P. and
         {Gao}, F. and {Gendron}, E. and {Genzel}, R. and {Gerhard}, O. and
         {Gillessen}, S. and {Habibi}, M. and {Haubois}, X. and {Henning}, T. and
         {Hippler}, S. and {Horrobin}, M. and {Jim{\'e}nez-Rosales}, A. and
         {Jocou}, L. and {Kervella}, P. and {Lacour}, S. and
         {Lapeyr{\`e}re}, V. and {Le Bouquin}, J. -B. and {L{\'e}na}, P. and
         {Ott}, T. and {Paumard}, T. and {Perraut}, K. and {Perrin}, G. and
         {Pfuhl}, O. and {Rabien}, S. and {Rodriguez Coira}, G. and
         {Rousset}, G. and {Scheithauer}, S. and {Sternberg}, A. and
         {Straub}, O. and {Straubmeier}, C. and {Sturm}, E. and
         {Tacconi}, L.~J. and {Vincent}, F. and {von Fellenberg}, S. and
         {Waisberg}, I. and {Widmann}, F. and {Wieprecht}, E. and
         {Wiezorrek}, E. and {Woillez}, J. and {Yazici}, S.},
        title = "{A geometric distance measurement to the Galactic center black hole with 0.3\% uncertainty}",
      journal = {\aap},
         year = 2019,
        month = may,
       volume = {625},
          eid = {L10},
        pages = {L10},
          doi = {10.1051/0004-6361/201935656},
archivePrefix = {arXiv},
       eprint = {1904.05721},
 primaryClass = {astro-ph.GA},
       adsurl = {https://ui.adsabs.harvard.edu/abs/2019A&A...625L..10G}
}

@article{Nogueras_2019,
   title={Early formation and recent starburst activity in the nuclear disk of the Milky Way},
   volume={4},
   ISSN={2397-3366},
   url={http://dx.doi.org/10.1038/s41550-019-0967-9},
   DOI={10.1038/s41550-019-0967-9},
   number={4},
   journal={Nature Astronomy},
   publisher={Springer Science and Business Media LLC},
   author={Nogueras-Lara, Francisco and Schödel, Rainer and Gallego-Calvente, Aurelia Teresa and Gallego-Cano, Eulalia and Shahzamanian, Banafsheh and Dong, Hui and Neumayer, Nadine and Hilker, Michael and Najarro, Francisco and Nishiyama, Shogo and et al.},
   year={2019},
   month={Dec},
   pages={377–381}
}

@ARTICLE{Andersen_2018,
       author = {{Andersen}, Bridget C. and {Ransom}, Scott M.},
        title = "{A Fourier Domain {\textquotedblleft}Jerk{\textquotedblright} Search for Binary Pulsars}",
      journal = {\apjl},
         year = 2018,
        month = aug,
       volume = {863},
       number = {1},
          eid = {L13},
        pages = {L13},
          doi = {10.3847/2041-8213/aad59f},
archivePrefix = {arXiv},
       eprint = {1807.07900},
 primaryClass = {astro-ph.HE},
       adsurl = {https://ui.adsabs.harvard.edu/abs/2018ApJ...863L..13A}
}

@ARTICLE{Rajwade_2017,
   author = {{Rajwade}, K.~M. and {Lorimer}, D.~R. and {Anderson}, L.~D.},
    title = "{Detecting pulsars in the Galactic Centre}",
  journal = "MNRAS",
archivePrefix = "arXiv",
   eprint = {1611.06977},
 primaryClass = "astro-ph.HE",
     year = 2017,
    month = oct,
   volume = 471,
    pages = {730-739},
      doi = {10.1093/mnras/stx1661},
   adsurl = {http://adsabs.harvard.edu/abs/2017MNRAS.471..730R}
}

@ARTICLE{Dexter2014,
   author = {{Dexter}, J. and {O'Leary}, R.~M.},
    title = "{The Peculiar Pulsar Population of the Central Parsec}",
  journal = "The Astrophysical Journal",
archivePrefix = "arXiv",
   eprint = {1310.7022},
 primaryClass = "astro-ph.GA",
     year = 2014,
    month = mar,
   volume = 783,
      eid = {L7},
    pages = {L7},
      doi = {10.1088/2041-8205/783/1/L7},
   adsurl = {http://adsabs.harvard.edu/abs/2014ApJ...783L...7D}
}

@ARTICLE{Liu_2021,
       author = {{Liu}, Kuo and {Desvignes}, Gregory and {Eatough}, Ralph P. and {Karuppusamy}, Ramesh and {Kramer}, Michael and {Torne}, Pablo and {Wharton}, Robert and {Chatterjee}, Shami and {Cordes}, James M. and {Crew}, Geoffrey B. and {Goddi}, Ciriaco and {Ransom}, Scott M. and {Rottmann}, Helge and {Abbate}, Federico and {Bower}, Geoffrey C. and {Brinkerink}, Christiaan D. and {Falcke}, Heino and {Noutsos}, Aristeidis and {Hern{\'a}ndez-G{\'o}mez}, Antonio and {Jiang}, Wu and {Johnson}, Michael D. and {Lu}, Ru-Sen and {Pidopryhora}, Yurii and {Rezzolla}, Luciano and {Shao}, Lijing and {Shen}, Zhiqiang and {Wex}, Norbert},
        title = "{An 86 GHz Search for Pulsars in the Galactic Center with the Atacama Large Millimeter / submillimeter Array}",
      journal = {\apj},
         year = 2021,
        month = jun,
       volume = {914},
       number = {1},
          eid = {30},
        pages = {30},
          doi = {10.3847/1538-4357/abf9a2},
archivePrefix = {arXiv},
       eprint = {2104.08986},
 primaryClass = {astro-ph.HE},
       adsurl = {https://ui.adsabs.harvard.edu/abs/2021ApJ...914...30L}
}

@article{Bagchi_2013,
    author = {Bagchi, Manjari and Lorimer, Duncan R. and Wolfe, Spencer},
    title = "{On the detectability of eccentric binary pulsars}",
    journal = {Monthly Notices of the Royal Astronomical Society},
    volume = {432},
    number = {2},
    pages = {1303-1314},
    year = {2013},
    month = {04},
    issn = {0035-8711},
    doi = {10.1093/mnras/stt559},
    url = {https://doi.org/10.1093/mnras/stt559},
    eprint = {https://academic.oup.com/mnras/article-pdf/432/2/1303/18751713/stt559.pdf},
}

@article{Suresh2022,
doi = {10.3847/1538-4357/ac74c0},
url = {https://dx.doi.org/10.3847/1538-4357/ac74c0},
year = {2022},
month = {jul},
publisher = {The American Astronomical Society},
volume = {933},
number = {2},
pages = {121},
author = {Suresh, Akshay and Cordes, James M. and Chatterjee, Shami and Gajjar, Vishal and Perez, Karen I. and Siemion, Andrew P. V. and Lebofsky, Matt and MacMahon, David H. E. and Ng, Cherry},
title = {4–8 GHz Fourier-domain Searches for Galactic Center Pulsars},
journal = {The Astrophysical Journal}
}

@article{Perez_2023,
doi = {10.3847/1538-4357/acdc23},
url = {https://dx.doi.org/10.3847/1538-4357/acdc23},
year = {2023},
month = {jul},
publisher = {The American Astronomical Society},
volume = {952},
number = {2},
pages = {150},
author = {Karen I. Perez and Slavko Bogdanov and Jules P. Halpern and Vishal Gajjar},
title = {Green Bank Telescope Discovery of the Redback Binary Millisecond Pulsar PSR J0212+5321},
journal = {The Astrophysical Journal}
}

@ARTICLE{Lower_2024,
       author = {{Lower}, Marcus E. and {Dai}, Shi and {Johnston}, Simon and {Barr}, Ewan D.},
        title = "{A Millisecond Pulsar Binary Embedded in a Galactic Center Radio Filament}",
      journal = {\apjl},
         year = 2024,
        month = may,
       volume = {967},
       number = {1},
          eid = {L16},
        pages = {L16},
          doi = {10.3847/2041-8213/ad4866},
archivePrefix = {arXiv},
       eprint = {2404.09098},
 primaryClass = {astro-ph.HE},
       adsurl = {https://ui.adsabs.harvard.edu/abs/2024ApJ...967L..16L}
}

@ARTICLE{Frayer_2017,
       author = {{Frayer}, David T.},
        title = "{The GBT Beam Shape at 109 GHz}",
      journal = {arXiv e-prints},
         year = 2017,
        month = mar,
          eid = {arXiv:1704.00025},
        pages = {arXiv:1704.00025},
          doi = {10.48550/arXiv.1704.00025},
archivePrefix = {arXiv},
       eprint = {1704.00025},
 primaryClass = {astro-ph.IM},
       adsurl = {https://ui.adsabs.harvard.edu/abs/2017arXiv170400025F}
}

@ARTICLE{Cromartie_2020,
       author = {{Cromartie}, H.~T. and {Fonseca}, E. and {Ransom}, S.~M. and {Demorest}, P.~B. and {Arzoumanian}, Z. and {Blumer}, H. and {Brook}, P.~R. and {DeCesar}, M.~E. and {Dolch}, T. and {Ellis}, J.~A. and {Ferdman}, R.~D. and {Ferrara}, E.~C. and {Garver-Daniels}, N. and {Gentile}, P.~A. and {Jones}, M.~L. and {Lam}, M.~T. and {Lorimer}, D.~R. and {Lynch}, R.~S. and {McLaughlin}, M.~A. and {Ng}, C. and {Nice}, D.~J. and {Pennucci}, T.~T. and {Spiewak}, R. and {Stairs}, I.~H. and {Stovall}, K. and {Swiggum}, J.~K. and {Zhu}, W.~W.},
        title = "{Relativistic Shapiro delay measurements of an extremely massive millisecond pulsar}",
      journal = {Nature Astronomy},
         year = 2020,
        month = jan,
       volume = {4},
        pages = {72-76},
          doi = {10.1038/s41550-019-0880-2},
archivePrefix = {arXiv},
       eprint = {1904.06759},
 primaryClass = {astro-ph.HE},
       adsurl = {https://ui.adsabs.harvard.edu/abs/2020NatAs...4...72C}
}

@ARTICLE{Heywood_2022,
       author = {{Heywood}, I. and {Rammala}, I. and {Camilo}, F. and {Cotton}, W.~D. and {Yusef-Zadeh}, F. and {Abbott}, T.~D. and {Adam}, R.~M. and {Adams}, G. and {Aldera}, M.~A. and {Asad}, K.~M.~B. and {Bauermeister}, E.~F. and {Bennett}, T.~G.~H. and {Bester}, H.~L. and {Bode}, W.~A. and {Botha}, D.~H. and {Botha}, A.~G. and {Brederode}, L.~R.~S. and {Buchner}, S. and {Burger}, J.~P. and {Cheetham}, T. and {de Villiers}, D.~I.~L. and {Dikgale-Mahlakoana}, M.~A. and {du Toit}, L.~J. and {Esterhuyse}, S.~W.~P. and {Fanaroff}, B.~L. and {February}, S. and {Fourie}, D.~J. and {Frank}, B.~S. and {Gamatham}, R.~R.~G. and {Geyer}, M. and {Goedhart}, S. and {Gouws}, M. and {Gumede}, S.~C. and {Hlakola}, M.~J. and {Hokwana}, A. and {Hoosen}, S.~W. and {Horrell}, J.~M.~G. and {Hugo}, B. and {Isaacson}, A.~I. and {J{\'o}zsa}, G.~I.~G. and {Jonas}, J.~L. and {Joubert}, A.~F. and {Julie}, R.~P.~M. and {Kapp}, F.~B. and {Kenyon}, J.~S. and {Kotz{\'e}}, P.~P.~A. and {Kriek}, N. and {Kriel}, H. and {Krishnan}, V.~K. and {Lehmensiek}, R. and {Liebenberg}, D. and {Lord}, R.~T. and {Lunsky}, B.~M. and {Madisa}, K. and {Magnus}, L.~G. and {Mahgoub}, O. and {Makhaba}, A. and {Makhathini}, S. and {Malan}, J.~A. and {Manley}, J.~R. and {Marais}, S.~J. and {Martens}, A. and {Mauch}, T. and {Merry}, B.~C. and {Millenaar}, R.~P. and {Mnyandu}, N. and {Mokone}, O.~J. and {Monama}, T.~E. and {Mphego}, M.~C. and {New}, W.~S. and {Ngcebetsha}, B. and {Ngoasheng}, K.~J. and {Ockards}, M.~T. and {Oozeer}, N. and {Otto}, A.~J. and {Passmoor}, S.~S. and {Patel}, A.~A. and {Peens-Hough}, A. and {Perkins}, S.~J. and {Ramaila}, A.~J.~T. and {Ramanujam}, N.~M.~R. and {Ramudzuli}, Z.~R. and {Ratcliffe}, S.~M. and {Robyntjies}, A. and {Salie}, S. and {Sambu}, N. and {Schollar}, C.~T.~G. and {Schwardt}, L.~C. and {Schwartz}, R.~L. and {Serylak}, M. and {Siebrits}, R. and {Sirothia}, S.~K. and {Slabber}, M. and {Smirnov}, O.~M. and {Sofeya}, L. and {Taljaard}, B. and {Tasse}, C. and {Tiplady}, A.~J. and {Toruvanda}, O. and {Twum}, S.~N. and {van Balla}, T.~J. and {van der Byl}, A. and {van der Merwe}, C. and {Van Tonder}, V. and {Van Wyk}, R. and {Venter}, A.~J. and {Venter}, M. and {Wallace}, B.~H. and {Welz}, M.~G. and {Williams}, L.~P. and {Xaia}, B.},
        title = "{The 1.28 GHz MeerKAT Galactic Center Mosaic}",
      journal = {\apj},
         year = 2022,
        month = feb,
       volume = {925},
       number = {2},
          eid = {165},
        pages = {165},
          doi = {10.3847/1538-4357/ac449a},
archivePrefix = {arXiv},
       eprint = {2201.10541},
 primaryClass = {astro-ph.GA},
       adsurl = {https://ui.adsabs.harvard.edu/abs/2022ApJ...925..165H}
}

@ARTICLE{FermiLAT_2017,
       author = {{Ackermann}, M. and {Ajello}, M. and {Albert}, A. and {Atwood}, W.~B. and {Baldini}, L. and {Ballet}, J. and {Barbiellini}, G. and {Bastieri}, D. and {Bellazzini}, R. and {Bissaldi}, E. and {Blandford}, R.~D. and {Bloom}, E.~D. and {Bonino}, R. and {Bottacini}, E. and {Brandt}, T.~J. and {Bregeon}, J. and {Bruel}, P. and {Buehler}, R. and {Burnett}, T.~H. and {Cameron}, R.~A. and {Caputo}, R. and {Caragiulo}, M. and {Caraveo}, P.~A. and {Cavazzuti}, E. and {Cecchi}, C. and {Charles}, E. and {Chekhtman}, A. and {Chiang}, J. and {Chiappo}, A. and {Chiaro}, G. and {Ciprini}, S. and {Conrad}, J. and {Costanza}, F. and {Cuoco}, A. and {Cutini}, S. and {D'Ammando}, F. and {de Palma}, F. and {Desiante}, R. and {Digel}, S.~W. and {Di Lalla}, N. and {Di Mauro}, M. and {Di Venere}, L. and {Drell}, P.~S. and {Favuzzi}, C. and {Fegan}, S.~J. and {Ferrara}, E.~C. and {Focke}, W.~B. and {Franckowiak}, A. and {Fukazawa}, Y. and {Funk}, S. and {Fusco}, P. and {Gargano}, F. and {Gasparrini}, D. and {Giglietto}, N. and {Giordano}, F. and {Giroletti}, M. and {Glanzman}, T. and {Gomez-Vargas}, G.~A. and {Green}, D. and {Grenier}, I.~A. and {Grove}, J.~E. and {Guillemot}, L. and {Guiriec}, S. and {Gustafsson}, M. and {Harding}, A.~K. and {Hays}, E. and {Hewitt}, J.~W. and {Horan}, D. and {Jogler}, T. and {Johnson}, A.~S. and {Kamae}, T. and {Kocevski}, D. and {Kuss}, M. and {La Mura}, G. and {Larsson}, S. and {Latronico}, L. and {Li}, J. and {Longo}, F. and {Loparco}, F. and {Lovellette}, M.~N. and {Lubrano}, P. and {Magill}, J.~D. and {Maldera}, S. and {Malyshev}, D. and {Manfreda}, A. and {Martin}, P. and {Mazziotta}, M.~N. and {Michelson}, P.~F. and {Mirabal}, N. and {Mitthumsiri}, W. and {Mizuno}, T. and {Moiseev}, A.~A. and {Monzani}, M.~E. and {Morselli}, A. and {Negro}, M. and {Nuss}, E. and {Ohsugi}, T. and {Orienti}, M. and {Orlando}, E. and {Ormes}, J.~F. and {Paneque}, D. and {Perkins}, J.~S. and {Persic}, M. and {Pesce-Rollins}, M. and {Piron}, F. and {Principe}, G. and {Rain{\`o}}, S. and {Rando}, R. and {Razzano}, M. and {Razzaque}, S. and {Reimer}, A. and {Reimer}, O. and {S{\'a}nchez-Conde}, M. and {Sgr{\`o}}, C. and {Simone}, D. and {Siskind}, E.~J. and {Spada}, F. and {Spandre}, G. and {Spinelli}, P. and {Suson}, D.~J. and {Tajima}, H. and {Tanaka}, K. and {Thayer}, J.~B. and {Tibaldo}, L. and {Torres}, D.~F. and {Troja}, E. and {Uchiyama}, Y. and {Vianello}, G. and {Wood}, K.~S. and {Wood}, M. and {Zaharijas}, G. and {Zimmer}, S. and {Fermi LAT Collaboration}},
        title = "{The Fermi Galactic Center GeV Excess and Implications for Dark Matter}",
      journal = {\apj},
         year = 2017,
        month = may,
       volume = {840},
       number = {1},
          eid = {43},
        pages = {43},
          doi = {10.3847/1538-4357/aa6cab},
archivePrefix = {arXiv},
       eprint = {1704.03910},
 primaryClass = {astro-ph.HE},
       adsurl = {https://ui.adsabs.harvard.edu/abs/2017ApJ...840...43A}
}

@article{Liu_2014,
    author = {Liu, K. and Eatough, R. P. and Wex, N. and Kramer, M.},
    title = "{Pulsar–black hole binaries: prospects for new gravity tests with future radio telescopes}",
    journal = {Monthly Notices of the Royal Astronomical Society},
    volume = {445},
    number = {3},
    pages = {3115-3132},
    year = {2014},
    month = {10},
    issn = {0035-8711},
    doi = {10.1093/mnras/stu1913},
    url = {https://doi.org/10.1093/mnras/stu1913},
    eprint = {https://academic.oup.com/mnras/article-pdf/445/3/3115/3547559/stu1913.pdf},
}

@ARTICLE{Spitler_2014,
   author = {{Spitler}, L.~G. and {Lee}, K.~J. and {Eatough}, R.~P. and {Kramer}, M. and 
	{Karuppusamy}, R. and {Bassa}, C.~G. and et al.},
    title = "{Pulse Broadening Measurements from the Galactic Center Pulsar J1745-2900}",
  journal = "The Astrophysical Journal Letters",
archivePrefix = "arXiv",
   eprint = {1309.4673},
 primaryClass = "astro-ph.HE",
     year = 2014,
    month = jan,
   volume = 780,
      eid = {L3},
    pages = {L3},
      doi = {10.1088/2041-8205/780/1/L3},
   adsurl = {http://adsabs.harvard.edu/abs/2014ApJ...780L...3S}
}

@ARTICLE{Eatough_2013a,
   author = {{Eatough}, R. and {Karuppusamy}, R. and {Kramer}, M. and {Klein}, B. and 
	{Champion}, D. and {Kraus}, A. et al.},
    title = "{Detection of radio pulsations from the direction of the NuSTAR 3.76 second X-ray pulsar at 8.35 GHz}",
  journal = {The Astronomer's Telegram},
     year = 2013,
    month = may,
   volume = 5040,
   adsurl = {http://adsabs.harvard.edu/abs/2013ATel.5040....1E}
}

@ARTICLE{Eatough_2013b,
   author = {{Eatough}, R.~P. and {Falcke}, H. and {Karuppusamy}, R. and 
	{Lee}, K.~J. and {Champion}, D.~J. and {Keane}, E.~F. and et al.},
    title = "{A strong magnetic field around the supermassive black hole at the centre of the Galaxy}",
  journal = "Nature",
archivePrefix = "arXiv",
   eprint = {1308.3147},
 primaryClass = "astro-ph.GA",
     year = 2013,
    month = sep,
   volume = 501,
    pages = {391-394},
      doi = {10.1038/nature12499},
   adsurl = {http://adsabs.harvard.edu/abs/2013Natur.501..391E}
}

@article{Keane_2008,
    author = {Keane, E. F. and Kramer, M.},
    title = {On the birthrates of Galactic neutron stars},
    journal = {Monthly Notices of the Royal Astronomical Society},
    volume = {391},
    number = {4},
    pages = {2009-2016},
    year = {2008},
    month = {12},
    issn = {0035-8711},
    doi = {10.1111/j.1365-2966.2008.14045.x},
    url = {https://doi.org/10.1111/j.1365-2966.2008.14045.x},
    eprint = {https://academic.oup.com/mnras/article-pdf/391/4/2009/4905977/mnras0391-2009.pdf},
}

@ARTICLE{Schodel_2018,
       author = {{Sch{\"o}del}, R. and {Gallego-Cano}, E. and {Dong}, H. and {Nogueras-Lara}, F. and {Gallego-Calvente}, A.~T. and {Amaro-Seoane}, P. and {Baumgardt}, H.},
        title = "{The distribution of stars around the Milky Way's central black hole. II. Diffuse light from sub-giants and dwarfs}",
      journal = {\aap},
         year = 2018,
        month = jan,
       volume = {609},
          eid = {A27},
        pages = {A27},
          doi = {10.1051/0004-6361/201730452},
archivePrefix = {arXiv},
       eprint = {1701.03817},
 primaryClass = {astro-ph.GA},
       adsurl = {https://ui.adsabs.harvard.edu/abs/2018A&A...609A..27S}
}

@ARTICLE{Genzel_1996,
       author = {{Genzel}, R. and {Thatte}, N. and {Krabbe}, A. and {Kroker}, H. and {Tacconi-Garman}, L.~E.},
        title = "{The Dark Mass Concentration in the Central Parsec of the Milky Way}",
      journal = {\apj},
         year = 1996,
        month = nov,
       volume = {472},
        pages = {153},
          doi = {10.1086/178051},
       adsurl = {https://ui.adsabs.harvard.edu/abs/1996ApJ...472..153G}
}

@ARTICLE{Schoedel_2024_whitepaper,
       author = {{Schoedel}, Rainer and {Alberdi}, Antxon and {Jimenez-Serra}, Izaskun and {Yusef-Zadeh}, Farhad and {Gardini}, Angela and {Kramer}, Michael and {Perez Torres}, Miguel and {Morris}, Mark R. and {Forbrich}, Jan and {Ingallinera}, Adriano and {Nogueras-Lara}, Francisco and {Henshaw}, Jonathan D. and {Longmore}, Steven N. and {Moldon}, Javier and {Heywood}, Ian and {Rammala}, Isabella and {Verdes Montenegro}, Lourdes and {Sanchez Exposito}, Susana},
        title = "{The SKA Galactic Centre Survey -- A White Paper}",
      journal = {arXiv e-prints},
         year = 2024,
        month = jun,
          eid = {arXiv:2406.04022},
        pages = {arXiv:2406.04022},
          doi = {10.48550/arXiv.2406.04022},
archivePrefix = {arXiv},
       eprint = {2406.04022},
 primaryClass = {astro-ph.GA},
       adsurl = {https://ui.adsabs.harvard.edu/abs/2024arXiv240604022S}
}

@ARTICLE{Yao_2017_ymw16,
       author = {{Yao}, J.~M. and {Manchester}, R.~N. and {Wang}, N.},
        title = "{A New Electron-density Model for Estimation of Pulsar and FRB Distances}",
      journal = {\apj},
         year = 2017,
        month = jan,
       volume = {835},
       number = {1},
          eid = {29},
        pages = {29},
          doi = {10.3847/1538-4357/835/1/29},
archivePrefix = {arXiv},
       eprint = {1610.09448},
 primaryClass = {astro-ph.GA},
       adsurl = {https://ui.adsabs.harvard.edu/abs/2017ApJ...835...29Y}
}

@ARTICLE{Price_2021,
       author = {{Price}, D.~C. and {Flynn}, C. and {Deller}, A.},
        title = "{A comparison of Galactic electron density models using PyGEDM}",
      journal = {\pasa},
         year = 2021,
        month = aug,
       volume = {38},
          eid = {e038},
        pages = {e038},
          doi = {10.1017/pasa.2021.33},
archivePrefix = {arXiv},
       eprint = {2106.15816},
 primaryClass = {astro-ph.GA},
       adsurl = {https://ui.adsabs.harvard.edu/abs/2021PASA...38...38P}
}

@article{Kramer_1998,
doi = {10.1086/305790},
url = {https://dx.doi.org/10.1086/305790},
year = {1998},
month = {jul},
publisher = {},
volume = {501},
number = {1},
pages = {270},
author = {Kramer, Michael and Xilouris, Kiriaki M. and Lorimer, Duncan R. and Doroshenko, Oleg and Jessner, Axel and Wielebinski, Richard and Wolszczan, Alexander and Camilo, Fernando},
title = {The Characteristics of Millisecond Pulsar Emission. I. Spectra, Pulse Shapes, and the Beaming Fraction},
journal = {The Astrophysical Journal}
}

@ARTICLE{Chernyakoba_2021,
       author = {{Chernyakova}, Maria and {Malyshev}, Denys and {van Soelen}, Brian and {O'Sullivan}, Shane and {Sobey}, Charlotte and {Tsygankov}, Sergey and {Mc Keague}, Samuel and {Green}, Jacob and {Kirwan}, Matthew and {Santangelo}, Andrea and {P{\"u}hlhofer}, Gerd and {Monageng}, Itumeleng M.},
        title = "{Multi-Wavelength Properties of the 2021 Periastron Passage of PSR B1259-63}",
      journal = {Universe},
         year = 2021,
        month = jul,
       volume = {7},
       number = {7},
          eid = {242},
        pages = {242},
          doi = {10.3390/universe7070242},
archivePrefix = {arXiv},
       eprint = {2106.03759},
 primaryClass = {astro-ph.HE},
       adsurl = {https://ui.adsabs.harvard.edu/abs/2021Univ....7..242C}
}

@ARTICLE{Macquart_2015,
       author = {{Macquart}, Jean-Pierre and {Kanekar}, Nissim},
        title = "{On Detecting Millisecond Pulsars at the Galactic Center}",
      journal = {\apj},
         year = 2015,
        month = jun,
       volume = {805},
       number = {2},
          eid = {172},
        pages = {172},
          doi = {10.1088/0004-637X/805/2/172},
archivePrefix = {arXiv},
       eprint = {1504.02492},
 primaryClass = {astro-ph.HE},
       adsurl = {https://ui.adsabs.harvard.edu/abs/2015ApJ...805..172M}
}

@ARTICLE{Clarke_2013,
       author = {{Clarke}, Nathan and {Macquart}, Jean-Pierre and {Trott}, Cathryn},
        title = "{Performance of a Novel Fast Transients Detection System}",
      journal = {\apjs},
         year = 2013,
        month = mar,
       volume = {205},
       number = {1},
          eid = {4},
        pages = {4},
          doi = {10.1088/0067-0049/205/1/4},
archivePrefix = {arXiv},
       eprint = {1301.3968},
 primaryClass = {astro-ph.IM},
       adsurl = {https://ui.adsabs.harvard.edu/abs/2013ApJS..205....4C}
}

@article{Burgay_2013,
   title={The High Time Resolution Universe Pulsar Survey – VII. Discovery of five millisecond pulsars and the different luminosity properties of binary and isolated recycled pulsars},
   volume={433},
   ISSN={1365-2966},
   url={http://dx.doi.org/10.1093/mnras/stt721},
   DOI={10.1093/mnras/stt721},
   number={1},
   journal={Monthly Notices of the Royal Astronomical Society},
   publisher={Oxford University Press (OUP)},
   author={Burgay, M. and Bailes, M. and Bates, S. D. and Bhat, N. D. R. and Burke-Spolaor, S. and Champion, D. J. and Coster, P. and D’Amico, N. and Johnston, S. and Keith, M. J. and et al.},
   year={2013},
   month={May},
   pages={259–269}
}

@article{Pearlman_2018,
doi = {10.3847/1538-4357/aade4d},
url = {https://dx.doi.org/10.3847/1538-4357/aade4d},
year = {2018},
month = {oct},
publisher = {The American Astronomical Society},
volume = {866},
number = {2},
pages = {160},
author = {Pearlman, Aaron B. and Majid, Walid A. and Prince, Thomas A. and Kocz, Jonathon and Horiuchi, Shinji},
title = {Pulse Morphology of the Galactic Center Magnetar PSR J1745–2900},
journal = {The Astrophysical Journal}
}

@ARTICLE{Wharton_2012,
       author = {{Wharton}, R.~S. and {Chatterjee}, S. and {Cordes}, J.~M. and {Deneva}, J.~S. and {Lazio}, T.~J.~W.},
        title = "{Multiwavelength Constraints on Pulsar Populations in the Galactic Center}",
      journal = {\apj},
         year = 2012,
        month = jul,
       volume = {753},
       number = {2},
          eid = {108},
        pages = {108},
          doi = {10.1088/0004-637X/753/2/108},
archivePrefix = {arXiv},
       eprint = {1111.4216},
 primaryClass = {astro-ph.HE},
       adsurl = {https://ui.adsabs.harvard.edu/abs/2012ApJ...753..108W}
}

@ARTICLE{Liu_2012,
       author = {{Liu}, K. and {Wex}, N. and {Kramer}, M. and {Cordes}, J.~M. and {Lazio}, T.~J.~W.},
        title = "{Prospects for Probing the Spacetime of Sgr A* with Pulsars}",
      journal = {\apj},
         year = 2012,
        month = mar,
       volume = {747},
       number = {1},
          eid = {1},
        pages = {1},
          doi = {10.1088/0004-637X/747/1/1},
archivePrefix = {arXiv},
       eprint = {1112.2151},
 primaryClass = {astro-ph.HE},
       adsurl = {https://ui.adsabs.harvard.edu/abs/2012ApJ...747....1L}
}

@ARTICLE{Macquart2010,
   author = {{Macquart}, J.-P. and {Kanekar}, N. and {Frail}, D.~A. and {Ransom}, S.~M.
	},
    title = "{A High-frequency Search for Pulsars within the Central Parsec of Sgr A*}",
  journal = "The Astrophysical Journal",
archivePrefix = "arXiv",
   eprint = {1004.1643},
     year = 2010,
    month = jun,
   volume = 715,
    pages = {939-946},
      doi = {10.1088/0004-637X/715/2/939},
   adsurl = {http://adsabs.harvard.edu/abs/2010ApJ...715..939M}
}

@article{Deneva2009,
        doi = {10.1088/0004-637x/702/2/l177},
        url = {https://doi.org/10.1088/0004-637x/702/2/l177},
        year = 2009,
        month = {aug},
        publisher = {American Astronomical Society},
        volume = {702},
        number = {2},
        pages = {L177--L181},
        author = {J. S. Deneva and J. M. Cordes and T. J. W. Lazio},
        title = {{DISCOVERY} {OF} {THREE} {PULSARS} {FROM} A {GALACTIC} {CENTER} {PULSAR} {POPULATION}},
        journal = {The Astrophysical Journal}}

@ARTICLE{Archibald_2009,
       author = {{Archibald}, Anne M. and {Stairs}, Ingrid H. and {Ransom}, Scott M. and {Kaspi}, Victoria M. and {Kondratiev}, Vladislav I. and {Lorimer}, Duncan R. and {McLaughlin}, Maura A. and {Boyles}, Jason and {Hessels}, Jason W.~T. and {Lynch}, Ryan and {van Leeuwen}, Joeri and {Roberts}, Mallory S.~E. and {Jenet}, Frederick and {Champion}, David J. and {Rosen}, Rachel and {Barlow}, Brad N. and {Dunlap}, Bart H. and {Remillard}, Ronald A.},
        title = "{A Radio Pulsar/X-ray Binary Link}",
      journal = {Science},
         year = 2009,
        month = jun,
       volume = {324},
       number = {5933},
        pages = {1411},
          doi = {10.1126/science.1172740},
archivePrefix = {arXiv},
       eprint = {0905.3397},
 primaryClass = {astro-ph.HE},
       adsurl = {https://ui.adsabs.harvard.edu/abs/2009Sci...324.1411A}
}

@ARTICLE{Wang_2006,
       author = {{Wang}, Q.~D. and {Lu}, F.~J. and {Gotthelf}, E.~V.},
        title = "{G359.95-0.04: an energetic pulsar candidate near Sgr A*}",
      journal = {\mnras},
         year = 2006,
        month = apr,
       volume = {367},
       number = {3},
        pages = {937-944},
          doi = {10.1111/j.1365-2966.2006.09998.x},
archivePrefix = {arXiv},
       eprint = {astro-ph/0512643},
 primaryClass = {astro-ph},
       adsurl = {https://ui.adsabs.harvard.edu/abs/2006MNRAS.367..937W}
}

@ARTICLE{Johnston2006,
       author = {{Johnston}, Simon and {Kramer}, M. and {Lorimer}, D.~R. and {Lyne}, A.~G. and {McLaughlin}, M. and {Klein}, B. and {Manchester}, R.~N.},
        title = "{Discovery of two pulsars towards the Galactic Centre}",
      journal = {\mnras},
         year = 2006,
        month = nov,
       volume = {373},
       number = {1},
        pages = {L6-L10},
          doi = {10.1111/j.1745-3933.2006.00232.x},
archivePrefix = {arXiv},
       eprint = {astro-ph/0606465},
 primaryClass = {astro-ph},
       adsurl = {https://ui.adsabs.harvard.edu/abs/2006MNRAS.373L...6J}
}

@ARTICLE{Muno_2005,
       author = {{Muno}, M.~P. and {Pfahl}, E. and {Baganoff}, F.~K. and {Brandt}, W.~N. and {Ghez}, A. and {Lu}, J. and {Morris}, M.~R.},
        title = "{An Overabundance of Transient X-Ray Binaries within 1 Parsec of the Galactic Center}",
      journal = {\apjl},
         year = 2005,
        month = apr,
       volume = {622},
       number = {2},
        pages = {L113-L116},
          doi = {10.1086/429721},
archivePrefix = {arXiv},
       eprint = {astro-ph/0412492},
 primaryClass = {astro-ph},
       adsurl = {https://ui.adsabs.harvard.edu/abs/2005ApJ...622L.113M}
}

@ARTICLE{Ghez_2005,
       author = {{Ghez}, A.~M. and {Salim}, S. and {Hornstein}, S.~D. and {Tanner}, A. and {Lu}, J.~R. and {Morris}, M. and {Becklin}, E.~E. and {Duch{\^e}ne}, G.},
        title = "{Stellar Orbits around the Galactic Center Black Hole}",
      journal = {\apj},
         year = 2005,
        month = feb,
       volume = {620},
       number = {2},
        pages = {744-757},
          doi = {10.1086/427175},
archivePrefix = {arXiv},
       eprint = {astro-ph/0306130},
 primaryClass = {astro-ph},
       adsurl = {https://ui.adsabs.harvard.edu/abs/2005ApJ...620..744G}
}

@ARTICLE{Manchester_2005,
       author = {{Manchester}, R.~N. and {Hobbs}, G.~B. and {Teoh}, A. and {Hobbs}, M.},
        title = "{The Australia Telescope National Facility Pulsar Catalogue}",
      journal = {\aj},
         year = 2005,
        month = apr,
       volume = {129},
       number = {4},
        pages = {1993-2006},
          doi = {10.1086/428488},
archivePrefix = {arXiv},
       eprint = {astro-ph/0412641},
 primaryClass = {astro-ph},
       adsurl = {https://ui.adsabs.harvard.edu/abs/2005AJ....129.1993M}
}

@ARTICLE{Manchester_1996,
       author = {{Manchester}, R.~N. and {Lyne}, A.~G. and {D'Amico}, N. and {Bailes}, M. and {Johnston}, S. and {Lorimer}, D.~R. and {Harrison}, P.~A. and {Nicastro}, L. and {Bell}, J.~F.},
        title = "{The Parkes Southern Pulsar Survey. I. Observing and data analysis systems and initial results.}",
      journal = {\mnras},
         year = 1996,
        month = apr,
       volume = {279},
       number = {4},
        pages = {1235-1250},
          doi = {10.1093/mnras/279.4.1235},
       adsurl = {https://ui.adsabs.harvard.edu/abs/1996MNRAS.279.1235M}
}

@BOOK{Lorimer_2004,
       author = {{Lorimer}, D.~R. and {Kramer}, M.},
        title = "{Handbook of Pulsar Astronomy}",
         year = 2004,
       volume = {4},
       adsurl = {https://ui.adsabs.harvard.edu/abs/2004hpa..book.....L}
}

@article{Figer_2004,
        doi = {10.1086/380392},
        url = {https://doi.org/10.1086/380392},
        year = 2004,
        month = {jan},
        publisher = {American Astronomical Society},
        volume = {601},
        number = {1},
        pages = {319--339},
        author = {Donald F. Figer and R. Michael Rich and Sungsoo S. Kim and Mark Morris and Eugene Serabyn},
        title = {An Extended Star Formation History for the Galactic Center {fromHubble} Space {TelescopeNICMOS} Observations},
        journal = {The Astrophysical Journal}
}

@ARTICLE{Pfahl_2004,
       author = {{Pfahl}, Eric and {Loeb}, Abraham},
        title = "{Probing the Spacetime around Sagittarius A* with Radio Pulsars}",
      journal = {\apj},
         year = 2004,
        month = nov,
       volume = {615},
       number = {1},
        pages = {253-258},
          doi = {10.1086/423975},
archivePrefix = {arXiv},
       eprint = {astro-ph/0309744},
 primaryClass = {astro-ph},
       adsurl = {https://ui.adsabs.harvard.edu/abs/2004ApJ...615..253P}
}

@ARTICLE{Cordes_2003_NE2001model,
       author = {{Cordes}, J.~M. and {Lazio}, T.~J.~W.},
        title = "{NE2001. II. Using Radio Propagation Data to Construct a Model for the Galactic Distribution of Free Electrons}",
      journal = {arXiv e-prints},
         year = 2003,
        month = jan,
          eid = {astro-ph/0301598},
        pages = {astro-ph/0301598},
archivePrefix = {arXiv},
       eprint = {astro-ph/0301598},
 primaryClass = {astro-ph},
       adsurl = {https://ui.adsabs.harvard.edu/abs/2003astro.ph..1598C}
}

@ARTICLE{Spiewak_2018,
       author = {{Spiewak}, R. and {Bailes}, M. and {Barr}, E.~D. and {Bhat}, N.~D.~R. and {Burgay}, M. and {Cameron}, A.~D. and {Champion}, D.~J. and {Flynn}, C.~M.~L. and {Jameson}, A. and {Johnston}, S. and {Keith}, M.~J. and {Kramer}, M. and {Kulkarni}, S.~R. and {Levin}, L. and {Lyne}, A.~G. and {Morello}, V. and {Ng}, C. and {Possenti}, A. and {Ravi}, V. and {Stappers}, B.~W. and {van Straten}, W. and {Tiburzi}, C.},
        title = "{PSR J2322-2650 - a low-luminosity millisecond pulsar with a planetary-mass companion}",
      journal = {\mnras},
         year = 2018,
        month = mar,
       volume = {475},
       number = {1},
        pages = {469-477},
          doi = {10.1093/mnras/stx3157},
archivePrefix = {arXiv},
       eprint = {1712.04445},
 primaryClass = {astro-ph.HE},
       adsurl = {https://ui.adsabs.harvard.edu/abs/2018MNRAS.475..469S}
}

@article{Knispel_2013,
   author = {Knispel, B. and Eatough, R.~P. and Kim, H. and Keane, E.~F. and Allen, B. and Anderson, D. and Aulbert, C. and et al.},
   year = {2013},
   title = {{Einstein@Home Discovery of 24 Pulsars in the Parkes Multi-beam Pulsar Survey}},
   journal = {Astrophysical Journal},
   volume = {774},
   number = {2},
   pages = {93}
}

@ARTICLE{Sengar_2024,
       author = {{Sengar}, R. and {Bailes}, M. and {Balakrishnan}, V. and {Barr}, E.~D. and {Bhat}, N.~D.~R. and {Burgay}, M. and {Bernadich}, M.~C.~I. and {Cameron}, A.~D. and {Champion}, D.~J. and {Chen}, W. and {Flynn}, C.~M.~L. and {Jameson}, A. and {Johnston}, S. and {Keith}, M.~J. and {Kramer}, M. and {Morello}, V. and {Ng}, C. and {Possenti}, A. and {Stevenson}, S. and {Shannon}, R.~M. and {van Straten}, W. and {Wongphechauxsorn}, J.},
        title = "{The High Time Resolution Universe Pulsar Survey - XIX. A coherent GPU-accelerated reprocessing and the discovery of 71 pulsars in the Southern Galactic plane}",
      journal = {\mnras},
         year = 2025,
        month = feb,
       volume = {536},
       number = {4},
        pages = {3159-3176},
          doi = {10.1093/mnras/stae2716},
archivePrefix = {arXiv},
       eprint = {2412.07104},
 primaryClass = {astro-ph.HE},
       adsurl = {https://ui.adsabs.harvard.edu/abs/2025MNRAS.536.3159S}
}

@ARTICLE{Eatough_2013,
       author = {{Eatough}, R.~P. and {Kramer}, M. and {Lyne}, A.~G. and {Keith}, M.~J.},
        title = "{A coherent acceleration search of the Parkes multibeam pulsar survey - techniques and the discovery and timing of 16 pulsars}",
      journal = {\mnras},
         year = 2013,
        month = may,
       volume = {431},
       number = {1},
        pages = {292-307},
          doi = {10.1093/mnras/stt161},
archivePrefix = {arXiv},
       eprint = {1301.6346},
 primaryClass = {astro-ph.IM},
       adsurl = {https://ui.adsabs.harvard.edu/abs/2013MNRAS.431..292E}
}

@PHDTHESIS{2001PhDT.......123R,
       author = {{Ransom}, Scott Mitchell},
        title = "{New search techniques for binary pulsars}",
       school = {Harvard University},
         year = 2001,
        month = jan,
       adsurl = {https://ui.adsabs.harvard.edu/abs/2001PhDT.......123R}
}

@article{Cordes_1997,
        doi = {10.1086/303569},
        url = {https://doi.org/10.1086/303569},
        year = 1997,
        month = {feb},
        publisher = {American Astronomical Society},
        volume = {475},
        number = {2},
        pages = {557--564},
        author = {James M. Cordes and T. Joseph W. Lazio},
        title = {Finding Radio Pulsars in and beyond the Galactic Center},
        journal = {The Astrophysical Journal}
}

@ARTICLE{Mezger_1999,
       author = {{Mezger}, P.~G. and {Zylka}, R. and {Philipp}, S. and {Launhardt}, R.},
        title = "{The nuclear bulge of the Galaxy. II. The K band luminosity function of the central 30 PC}",
      journal = {\aap},
         year = 1999,
        month = aug,
       volume = {348},
        pages = {457-465},
       adsurl = {https://ui.adsabs.harvard.edu/abs/1999A&A...348..457M}
}

@article{Wex_Kopeikin_1999,
        doi = {10.1086/306933},
        url = {https://doi.org/10.1086/306933},
        year = 1999,
        month = {mar},
        publisher = {American Astronomical Society},
        volume = {514},
        number = {1},
        pages = {388--401},
        author = {N. Wex and S. M. Kopeikin},
        title = {Frame Dragging and Other Precessional Effects in Black Hole Pulsar Binaries},
        journal = {The Astrophysical Journal}
}

@article{Lazio_1998,
        doi = {10.1086/306174},
        url = {https://doi.org/10.1086/306174},
        year = 1998,
        month = {oct},
        publisher = {American Astronomical Society},
        volume = {505},
        number = {2},
        pages = {715--731},
        author = {T. Joseph W. Lazio and James M. Cordes},
        title = {Hyperstrong Radio-Wave Scattering in the Galactic Center. {II}. A Likelihood Analysis of Free Electrons in the Galactic Center},
        journal = {The Astrophysical Journal}
}

@article{Lorimer_1993,
    author = {Lorimer, D. R. and Bailes, M. and Dewey, R. J. and Harrison, P. A.},
    title = "{Pulsar statistics: the birthrate and initial spin periods of radio pulsars}",
    journal = {Monthly Notices of the Royal Astronomical Society},
    volume = {263},
    number = {2},
    pages = {403-415},
    year = {1993},
    month = {07},
    issn = {0035-8711},
    doi = {10.1093/mnras/263.2.403},
    url = {https://doi.org/10.1093/mnras/263.2.403},
    eprint = {https://academic.oup.com/mnras/article-pdf/263/2/403/18539280/mnras263-0403.pdf},
}

@ARTICLE{Cordes_1991,
       author = {{Cordes}, James M. and {Lazio}, T.~J.},
        title = "{Interstellar Scattering Effects on the Detection of Narrow-Band Signals}",
      journal = {\apj},
         year = 1991,
        month = jul,
       volume = {376},
        pages = {123},
          doi = {10.1086/170261},
       adsurl = {https://ui.adsabs.harvard.edu/abs/1991ApJ...376..123C}
}

@ARTICLE{MacMahon_BLDR,
       author = {{MacMahon}, David H.~E. and {Price}, Danny C. and {Lebofsky}, Matthew and {Siemion}, Andrew P.~V. and {Croft}, Steve and {DeBoer}, David and {Enriquez}, J. Emilio and {Gajjar}, Vishal and {Hellbourg}, Gregory and {Isaacson}, Howard and {Werthimer}, Dan and {Abdurashidova}, Zuhra and {Bloss}, Marty and {Brandt}, Joe and {Creager}, Ramon and {Ford}, John and {Lynch}, Ryan S. and {Maddalena}, Ronald J. and {McCullough}, Randy and {Ray}, Jason and {Whitehead}, Mark and {Woody}, Dave},
        title = "{The Breakthrough Listen Search for Intelligent Life: A Wideband Data Recorder System for the Robert C. Byrd Green Bank Telescope}",
      journal = {\pasp},
         year = 2018,
        month = apr,
       volume = {130},
       number = {986},
        pages = {044502},
          doi = {10.1088/1538-3873/aa80d2},
archivePrefix = {arXiv},
       eprint = {1707.06024},
 primaryClass = {astro-ph.IM},
       adsurl = {https://ui.adsabs.harvard.edu/abs/2018PASP..130d4502M}
}

@ARTICLE{Lebofsky_2019,
       author = {{Lebofsky}, Matthew and {Croft}, Steve and {Siemion}, Andrew P.~V. and {Price}, Danny C. and {Enriquez}, J. Emilio and {Isaacson}, Howard and {MacMahon}, David H.~E. and {Anderson}, David and {Brzycki}, Bryan and {Cobb}, Jeff and {Czech}, Daniel and {DeBoer}, David and {DeMarines}, Julia and {Drew}, Jamie and {Foster}, Griffin and {Gajjar}, Vishal and {Gizani}, Nectaria and {Hellbourg}, Greg and {Korpela}, Eric J. and {Lacki}, Brian and {Sheikh}, Sofia and {Werthimer}, Dan and {Worden}, Pete and {Yu}, Alex and {Zhang}, Yunfan Gerry},
        title = "{The Breakthrough Listen Search for Intelligent Life: Public Data, Formats, Reduction, and Archiving}",
      journal = {\pasp},
         year = 2019,
        month = dec,
       volume = {131},
       number = {1006},
        pages = {124505},
          doi = {10.1088/1538-3873/ab3e82},
archivePrefix = {arXiv},
       eprint = {1906.07391},
 primaryClass = {astro-ph.IM},
       adsurl = {https://ui.adsabs.harvard.edu/abs/2019PASP..131l4505L}
}

@ARTICLE{Cordes_McLaughlin_2003,
       author = {{Cordes}, J.~M. and {McLaughlin}, M.~A.},
        title = "{Searches for Fast Radio Transients}",
      journal = {\apj},
         year = 2003,
        month = oct,
       volume = {596},
       number = {2},
        pages = {1142-1154},
          doi = {10.1086/378231},
archivePrefix = {arXiv},
       eprint = {astro-ph/0304364},
 primaryClass = {astro-ph},
       adsurl = {https://ui.adsabs.harvard.edu/abs/2003ApJ...596.1142C}
}

@ARTICLE{Heimdall_Barsdell_2012,
       author = {{Barsdell}, B.~R. and {Bailes}, M. and {Barnes}, D.~G. and {Fluke}, C.~J.},
        title = "{Accelerating incoherent dedispersion}",
      journal = {\mnras},
         year = 2012,
        month = may,
       volume = {422},
       number = {1},
        pages = {379-392},
          doi = {10.1111/j.1365-2966.2012.20622.x},
archivePrefix = {arXiv},
       eprint = {1201.5380},
 primaryClass = {astro-ph.IM},
       adsurl = {https://ui.adsabs.harvard.edu/abs/2012MNRAS.422..379B}
}

@software{Presto_Ransom_2011,
       author = {{Ransom}, Scott},
        title = "{PRESTO: PulsaR Exploration and Search TOolkit}",
 howpublished = {Astrophysics Source Code Library, record ascl:1107.017},
         year = 2011,
        month = jul,
          eid = {ascl:1107.017},
       adsurl = {https://ui.adsabs.harvard.edu/abs/2011ascl.soft07017R}
}

@article{DSPSR_vanStraten:2010hy,
    author = "van Straten, W. and Bailes, M.",
    title = "{DSPSR: Digital Signal Processing Software for Pulsar Astronomy}",
    eprint = "1008.3973",
    archivePrefix = "arXiv",
    primaryClass = "astro-ph.IM",
    doi = "10.1071/AS10021",
    journal = "Publ. Astron. Soc. Austral.",
    volume = "28",
    pages = "1",
    year = "2011"
}

@ARTICLE{PSRCHIVE_Hotan_2004,
       author = {{Hotan}, A.~W. and {van Straten}, W. and {Manchester}, R.~N.},
        title = "{PSRCHIVE and PSRFITS: An Open Approach to Radio Pulsar Data Storage and Analysis}",
      journal = {\pasa},
         year = 2004,
        month = jan,
       volume = {21},
       number = {3},
        pages = {302-309},
          doi = {10.1071/AS04022},
archivePrefix = {arXiv},
       eprint = {astro-ph/0404549},
 primaryClass = {astro-ph},
       adsurl = {https://ui.adsabs.harvard.edu/abs/2004PASA...21..302H}
}

@ARTICLE{Keith_2009_MNRAS,
       author = {{Keith}, M.~J. and {Eatough}, R.~P. and {Lyne}, A.~G. and {Kramer}, M. and {Possenti}, A. and {Camilo}, F. and {Manchester}, R.~N.},
        title = "{Discovery of 28 pulsars using new techniques for sorting pulsar candidates}",
      journal = {\mnras},
         year = 2009,
        month = may,
       volume = {395},
       number = {2},
        pages = {837-846},
          doi = {10.1111/j.1365-2966.2009.14543.x},
archivePrefix = {arXiv},
       eprint = {0901.3570},
 primaryClass = {astro-ph.SR},
       adsurl = {https://ui.adsabs.harvard.edu/abs/2009MNRAS.395..837K}
}

@ARTICLE{Brzycki_2024_GC_SETI,
       author = {{Brzycki}, Bryan and {Siemion}, Andrew P.~V. and {de Pater}, Imke and {Choza}, Carmen and {Croft}, Steve and {Gajjar}, Vishal and {Drew}, Jamie and {Lacki}, Brian C. and {Price}, Danny C. and {Sheikh}, Sofia Z.},
        title = "{The Breakthrough Listen Search for Intelligent Life: Galactic Center Search for Scintillated Technosignatures}",
      journal = {\aj},
         year = 2024,
        month = dec,
       volume = {168},
       number = {6},
          eid = {284},
        pages = {284},
          doi = {10.3847/1538-3881/ad7e18},
       adsurl = {https://ui.adsabs.harvard.edu/abs/2024AJ....168..284B}
}

@ARTICLE{Dexter2017,
       author = {{Dexter}, J. and {Deller}, A. and {Bower}, G.~C. and {Demorest}, P. and {Kramer}, M. and {Stappers}, B.~W. and {Lyne}, A.~G. and {Kerr}, M. and {Spitler}, L.~G. and {Psaltis}, D. and {Johnson}, M. and {Narayan}, R.},
        title = "{Locating the intense interstellar scattering towards the inner Galaxy}",
      journal = {\mnras},
         year = 2017,
        month = nov,
       volume = {471},
       number = {3},
        pages = {3563-3576},
          doi = {10.1093/mnras/stx1777},
archivePrefix = {arXiv},
       eprint = {1707.03842},
 primaryClass = {astro-ph.GA},
       adsurl = {https://ui.adsabs.harvard.edu/abs/2017MNRAS.471.3563D}
}

@ARTICLE{Abbate2018,
       author = {{Abbate}, F. and {Mastrobuono-Battisti}, A. and {Colpi}, M. and {Possenti}, A. and {Sippel}, A.~C. and {Dotti}, M.},
        title = "{Probing the formation history of the nuclear star cluster at the Galactic Centre with millisecond pulsars}",
      journal = {\mnras},
         year = 2018,
        month = jan,
       volume = {473},
       number = {1},
        pages = {927-936},
          doi = {10.1093/mnras/stx2364},
archivePrefix = {arXiv},
       eprint = {1708.01627},
 primaryClass = {astro-ph.GA},
       adsurl = {https://ui.adsabs.harvard.edu/abs/2018MNRAS.473..927A}
}

@ARTICLE{Abbate2023,
       author = {{Abbate}, F. and {Noutsos}, A. and {Desvignes}, G. and {Wharton}, R.~S. and {Torne}, P. and {Kramer}, M. and {Eatough}, R.~P. and {Karuppusamy}, R. and {Liu}, K. and {Shao}, L. and {Wongphechauxsorn}, J.},
        title = "{Rotation measure variations in Galactic Centre pulsars}",
      journal = {\mnras},
         year = 2023,
        month = sep,
       volume = {524},
       number = {2},
        pages = {2966-2977},
          doi = {10.1093/mnras/stad2047},
archivePrefix = {arXiv},
       eprint = {2307.03230},
 primaryClass = {astro-ph.HE},
       adsurl = {https://ui.adsabs.harvard.edu/abs/2023MNRAS.524.2966A}
}

@ARTICLE{DellaMonica2023,
       author = {{Della Monica}, Riccardo and {De Martino}, Ivan and {De Laurentis}, Mariafelicia},
        title = "{Testing space-time geometries and theories of gravity at the Galactic centre with pulsar's time delay}",
      journal = {\mnras},
         year = 2023,
        month = sep,
       volume = {524},
       number = {3},
        pages = {3782-3796},
          doi = {10.1093/mnras/stad2125},
archivePrefix = {arXiv},
       eprint = {2305.18178},
 primaryClass = {gr-qc},
       adsurl = {https://ui.adsabs.harvard.edu/abs/2023MNRAS.524.3782D}
}

@ARTICLE{Wang_2009a,
       author = {{Wang}, Yan and {Creighton}, Teviet and {Price}, Richard H. and {Jenet}, Frederick A.},
        title = "{Strong Field Effects on Pulsar Arrival Times: General Orientations}",
      journal = {\apj},
         year = 2009,
        month = nov,
       volume = {705},
       number = {2},
        pages = {1252-1259},
          doi = {10.1088/0004-637X/705/2/1252},
archivePrefix = {arXiv},
       eprint = {0909.2709},
 primaryClass = {astro-ph.GA},
       adsurl = {https://ui.adsabs.harvard.edu/abs/2009ApJ...705.1252W}
}

@ARTICLE{Wang_2009b,
       author = {{Wang}, Yan and {Jenet}, Frederick A. and {Creighton}, Teviet and {Price}, Richard H.},
        title = "{Strong Field Effects on Pulsar Arrival Times: Circular Orbits and Equatorial Beams}",
      journal = {\apj},
         year = 2009,
        month = may,
       volume = {697},
       number = {1},
        pages = {237-246},
          doi = {10.1088/0004-637X/697/1/237},
archivePrefix = {arXiv},
       eprint = {0812.2302},
 primaryClass = {astro-ph},
       adsurl = {https://ui.adsabs.harvard.edu/abs/2009ApJ...697..237W}
}

@ARTICLE{Stovall2012,
       author = {{Stovall}, Kevin and {Creighton}, Teviet and {Price}, Richard H. and {Jenet}, Fredrick A.},
        title = "{Observability of Pulsar Beam Bending by the Sgr A* Black Hole}",
      journal = {\apj},
         year = 2012,
        month = jan,
       volume = {744},
       number = {2},
          eid = {143},
        pages = {143},
          doi = {10.1088/0004-637X/744/2/143},
archivePrefix = {arXiv},
       eprint = {1102.5470},
 primaryClass = {astro-ph.GA},
       adsurl = {https://ui.adsabs.harvard.edu/abs/2012ApJ...744..143S}
}

@ARTICLE{FaucherGiguere2011,
       author = {{Faucher-Gigu{\`e}re}, Claude-Andr{\'e} and {Loeb}, Abraham},
        title = "{Pulsar-black hole binaries in the Galactic Centre}",
      journal = {\mnras},
         year = 2011,
        month = aug,
       volume = {415},
       number = {4},
        pages = {3951-3961},
          doi = {10.1111/j.1365-2966.2011.19019.x},
archivePrefix = {arXiv},
       eprint = {1012.0573},
 primaryClass = {astro-ph.HE},
       adsurl = {https://ui.adsabs.harvard.edu/abs/2011MNRAS.415.3951F}
}

@ARTICLE{Johannsen2016,
       author = {{Johannsen}, Tim},
        title = "{Sgr A* and general relativity}",
      journal = {Classical and Quantum Gravity},
         year = 2016,
        month = jun,
       volume = {33},
       number = {11},
          eid = {113001},
        pages = {113001},
          doi = {10.1088/0264-9381/33/11/113001},
archivePrefix = {arXiv},
       eprint = {1512.03818},
 primaryClass = {astro-ph.GA},
       adsurl = {https://ui.adsabs.harvard.edu/abs/2016CQGra..33k3001J}
}

@ARTICLE{Boodram2022,
       author = {{Boodram}, Oliver and {Heinke}, Craig O.},
        title = "{Millisecond pulsar kicks cause difficulties in explaining the Galactic Centre gamma-ray excess}",
      journal = {\mnras},
         year = 2022,
        month = may,
       volume = {512},
       number = {3},
        pages = {4239-4247},
          doi = {10.1093/mnras/stac702},
archivePrefix = {arXiv},
       eprint = {2205.03479},
 primaryClass = {astro-ph.HE},
       adsurl = {https://ui.adsabs.harvard.edu/abs/2022MNRAS.512.4239B}
}

@ARTICLE{Psaltis2016,
       author = {{Psaltis}, Dimitrios and {Wex}, Norbert and {Kramer}, Michael},
        title = "{A Quantitative Test of the No-hair Theorem with Sgr A* Using Stars, Pulsars, and the Event Horizon Telescope}",
      journal = {\apj},
         year = 2016,
        month = feb,
       volume = {818},
       number = {2},
          eid = {121},
        pages = {121},
          doi = {10.3847/0004-637X/818/2/121},
archivePrefix = {arXiv},
       eprint = {1510.00394},
 primaryClass = {astro-ph.HE},
       adsurl = {https://ui.adsabs.harvard.edu/abs/2016ApJ...818..121P}
}

@ARTICLE{Gajjar_2012_Nulling,
       author = {{Gajjar}, Vishal and {Joshi}, B.~C. and {Kramer}, M.},
        title = "{A survey of nulling pulsars using the Giant Meterwave Radio Telescope}",
      journal = {\mnras},
         year = 2012,
        month = aug,
       volume = {424},
       number = {2},
        pages = {1197-1205},
          doi = {10.1111/j.1365-2966.2012.21296.x},
archivePrefix = {arXiv},
       eprint = {1205.2550},
 primaryClass = {astro-ph.SR},
       adsurl = {https://ui.adsabs.harvard.edu/abs/2012MNRAS.424.1197G}
}

@ARTICLE{Kramer_2006_Intermittent_PSRs,
       author = {{Kramer}, M. and {Lyne}, A.~G. and {O'Brien}, J.~T. and {Jordan}, C.~A. and {Lorimer}, D.~R.},
        title = "{A Periodically Active Pulsar Giving Insight into Magnetospheric Physics}",
      journal = {Science},
         year = 2006,
        month = apr,
       volume = {312},
       number = {5773},
        pages = {549-551},
          doi = {10.1126/science.1124060},
archivePrefix = {arXiv},
       eprint = {astro-ph/0604605},
 primaryClass = {astro-ph},
       adsurl = {https://ui.adsabs.harvard.edu/abs/2006Sci...312..549K}
}

@ARTICLE{Samsing2017,
       author = {{Samsing}, Johan and {MacLeod}, Morgan and {Ramirez-Ruiz}, Enrico},
        title = "{Formation of Tidal Captures and Gravitational Wave Inspirals in Binary-single Interactions}",
      journal = {\apj},
         year = 2017,
        month = sep,
       volume = {846},
       number = {1},
          eid = {36},
        pages = {36},
          doi = {10.3847/1538-4357/aa7e32},
archivePrefix = {arXiv},
       eprint = {1609.09114},
 primaryClass = {astro-ph.HE},
       adsurl = {https://ui.adsabs.harvard.edu/abs/2017ApJ...846...36S}
}

@book{Scott1992Multivariate,
  author    = {Scott, David W.},
  title     = {Multivariate Density Estimation: Theory, Practice, and Visualization},
  publisher = {John Wiley \& Sons},
  year      = {1992},
  address   = {New York},
  isbn      = {0471547700}
}

@article{Virtanen2020SciPy,
  author    = {Virtanen, Pauli and Gommers, Ralf and Oliphant, Travis E. and others},
  title     = {{SciPy} 1.0: Fundamental Algorithms for Scientific Computing in Python},
  journal   = {Nature Methods},
  year      = {2020},
  volume    = {17},
  pages     = {261--272},
  doi       = {10.1038/s41592-019-0686-2}
}
\bibliographystyle{aasjournal}



\end{document}